\documentclass[11pt,reqno]{article}

\usepackage{amsmath}
\usepackage{amssymb}
\usepackage{amsthm}
\usepackage{bm}
\usepackage{booktabs}
\usepackage[bf,footnotesize]{caption}
\usepackage{changepage}
\usepackage{color}
\usepackage{eucal}
\usepackage{fullpage}
\usepackage{graphicx}
\usepackage{hhline}
\usepackage{lineno}
\usepackage{makecell}
\usepackage[sc]{mathpazo}
\usepackage{mathrsfs}
\usepackage{mathtools}
\usepackage{multirow}
\usepackage[numbers,sort&compress,square]{natbib}
\usepackage{setspace}
\usepackage{stmaryrd}
\usepackage[all,cmtip]{xy}

\newcommand*\patchAmsMathEnvironmentForLineno[1]{%
	\expandafter\let\csname old#1\expandafter\endcsname\csname #1\endcsname
	\expandafter\let\csname oldend#1\expandafter\endcsname\csname end#1\endcsname
	\renewenvironment{#1}%
	{\linenomath\csname old#1\endcsname}%
	{\csname oldend#1\endcsname\endlinenomath}}% 
	\newcommand*\patchBothAmsMathEnvironmentsForLineno[1]{%
	\patchAmsMathEnvironmentForLineno{#1}%
	\patchAmsMathEnvironmentForLineno{#1*}}%
	\AtBeginDocument{%
	\patchBothAmsMathEnvironmentsForLineno{equation}%
	\patchBothAmsMathEnvironmentsForLineno{align}%
	\patchBothAmsMathEnvironmentsForLineno{flalign}%
	\patchBothAmsMathEnvironmentsForLineno{alignat}%
	\patchBothAmsMathEnvironmentsForLineno{gather}%
	\patchBothAmsMathEnvironmentsForLineno{multline}%
}

\newcommand{\delhat}{\widehat{\Delta}}
\newcommand{\delhatsel}{\widehat{\Delta}_{\mathrm{sel}}}

\newcommand{\vx}{\mathbf{x}}
\newcommand{\vy}{\mathbf{y}}

\newcommand{\vA}{\mathbf{A}}
\newcommand{\vB}{\mathbf{B}}
\newcommand{\MSS}{\circlearrowright\left(\bm{\xi}\right)}
\newcommand{\RMC}{\mathrm{RMC}\left(\bm{\xi}\right)}
\newcommand{\E}{\mathbb{E}}

\theoremstyle{definition}

\newtheorem{lemma}{Lemma}

\newtheorem*{fixation}{Fixation Axiom}

\newcommand{\eq}[1]{Equation~\ref{eq:#1}}

\newcommand{\lem}[1]{Lemma~\ref{lem:#1}}

\title{\begin{center} \bfseries\singlespacing
Evolution of prosocial behavior in multilayer populations
\end{center}}
\author{\parbox[c]{16cm}{\onehalfspacing \normalsize \centering ~\\[-0.4cm] Qi Su$^{1,2}$\, , Alex McAvoy$^{2,3,4}$ ,  Yoichiro Mori$^{1,2,4}$, and Joshua B. Plotkin$^{1,2,4}$ \\ \quad\\ \footnotesize
$^{1}$Department of Biology, University of Pennsylvania, Philadelphia, PA 19104, USA \\
$^{2}$Center for Mathematical Biology, University of Pennsylvania, Philadelphia, PA 19104, USA \\
$^{3}$Department of Organismic and Evolutionary Biology, Harvard University, Cambridge, MA 02138, USA\\
$^{4}$Department of Mathematics, University of Pennsylvania, Philadelphia, PA 19104, USA\\[0.2cm]}
\date{}
}

\begin{document}

\allowdisplaybreaks

\maketitle

\begin{abstract}
Human societies include diverse social relationships. Friends, family, business colleagues, and online contacts can all contribute to one’s social life. Individuals may behave differently in different domains, but success in one domain may engender success in another. 
Here, we study this problem using multilayer networks to model multiple domains of social interactions, in which individuals experience different environments and may express different behaviors. We provide a mathematical analysis and find that coupling between layers tends to promote prosocial behavior. Even if prosociality is disfavored in each layer alone, multilayer coupling can promote its proliferation in all layers simultaneously. We apply this analysis to six real-world multilayer networks, ranging from the socio-emotional and professional relationships in a Zambian community, to the online and offline relationships within an academic University.  We discuss the implications of our results, which suggest that small modifications to interactions in one domain may catalyze prosociality in a different domain.
\end{abstract}

\section{Introduction}
The scale and sophistication of global human societies are due in no small part to cooperation. Altruistic behavior that benefits the collective, and entails personal costs to the individual, has long been recognized as an important aspect of both human and non-human societies \cite{1859-Darwin}. Just as prosocial behaviors have unquestionably shaped the past, they will also play a major role in shaping the present and future. From the collective action necessary to prevent the spread of infectious diseases \cite{block:NHB:2020,lopez:NHB:2020}, to efforts to combat climate change \cite{jacquet:NCC:2013,keohane:NCC:2016}, cooperation is a critical precursor to social prosperity.

At the same time, the emergence and stability of prosocial behaviors is perplexing in light of Darwin's notion of ``survival of the fittest'' \cite{Hamilton1963,Trivers1971}. Several mechanisms have been proposed to explain their widespread abundance \cite{nowak:Science:2006}, most notably spatial structure, which constrains interaction and dispersal patterns within a population \cite{1992-Nowak-p826-829,2005-Lieberman-p312-316,2009-Tarnita-p570-581,2009-Tarnita-p8601-8604,2013-Chen-p637-664,2014-Debarre-p3409-3409,2014-Allen-p113-151,2019-Qi-p20190041-20190041,Li2020}. The effects of population structure on cooperation have been studied theoretically, using computer simulations \cite{2005-Santos-p098104}, by approximation techniques \cite{2006-Ohtsuki-p502-505}, and by direct analysis of special cases \cite{2011-Hadjichrysanthou-p386-386,2014-Maciejewski-p1003567-1003567}; and they have been tested empirically in laboratory experiments \cite{2014-Rand-p17093-17098}. The latest mathematical results allow for extensive analysis of large families of heterogeneous population structures \cite{2017-Allen-p227-230,2019-Allen-p1147-1210,mcavoy-2020-nhb} and arbitrary initial configurations of individuals \cite{McAvoy2021}. A large portion of population structures favor antisocial traits, such as spite \cite{Fotouhi2018}, which is simultaneously intriguing and concerning.

Nonetheless, a single network cannot capture the complexity of social structures in human societies. Individuals typically form many different types of social relationships. They enjoy leisure time with friends and encounter colleagues in the workplace. They have physical contact with those who are nearby and participate in online social networks to keep in touch with friends or strangers who are more distant \cite{magnani-2013-arxiv,Heath1974,Krackhardt1987b,emmanuel-2001-oxford,Padgett1993}. Each type of relationship forms a domain in which interactions take place, and individuals may behave differently in different domains. Success in one domain, such as wealth accumulated in business settings, may nonetheless have an impact on success in other domains, such as influence and trustworthiness of opinions expressed on social media. The tendency of an individual's behavior to spread is therefore often dependent on their aggregate success across the domains in which they interact -- which introduces a form of coupling between different social domains.

Altruistic acts in different domains often involve different costs and benefits, such as donating a dollar to someone in person versus sharing a useful tip on social media. 
As a result, an individual is likely to exhibit different behaviors in distinct domains. 
These complexities of human social life violate the classic assumptions made in most prior game-theoretic studies of prosocial behavior, which typically focus on a single domain of interaction or assume that individuals use the same strategy against all opponents \cite{1992-Nowak-p826-829,2005-Lieberman-p312-316,2005-Santos-p098104,2006-Ohtsuki-p502-505,2009-Tarnita-p570-581,2009-Tarnita-p8601-8604,2013-Chen-p637-664,2014-Debarre-p3409-3409,2014-Allen-p113-151,Fotouhi2018,2019-Qi-p20190041-20190041,Li2020,2011-Hadjichrysanthou-p386-386,2014-Maciejewski-p1003567-1003567,2017-Allen-p227-230,2019-Allen-p1147-1210,McAvoy2021,mcavoy-2020-nhb}.
Compared with a growing literature on the dynamics and structural analysis of multiple-domain coupling \cite{2014-Kivelae-p203-271,Boccaletti2014}, the evolution of prosocial behavior has received much less attention and has been investigated only through numerical simulations in specific cases \cite{2012-Wang-p48001-48001,2012-Gomez-Gardenes-p620-620,Santos2014,2015-Wang-p124-124,Kleineberg2018}.
The general question of how coupling between domains influences behavior in a population, for an arbitrary number of domains each with arbitrary spatial structure and potentially different payoffs, remains unresolved and outside the scope of simulations studies \cite{2012-Wang-p48001-48001,2012-Gomez-Gardenes-p620-620,Santos2014,2015-Wang-p124-124,Kleineberg2018}.
Although numerical simulations are useful for rapid exploration within a set of parameters, the notion of ``generalizability", which is important for progress in the social and behavioral sciences \cite{Munafo2017}, demands that theoretical results be established mathematically so that the extent of their generality is known. However, mathematical results on this topic remain absent, so far, even for the simplest cases. 

In this study, we use a multilayer network to describe a population with multiple domains of strategic interactions. Each layer describes the network of interactions that occur in given domain, and the players can adopt different behavioral strategies in different domains. An individual's behavior in a given domain is preferentially copied by others in that domain, based on the individual's aggregate success across domains. We provide mathematical results applicable to any multilayer structure (i.e.~the number of layers and connections within each layer), any initial strategy configuration, and any strategy update rule in each layer. A thorough analysis of all two-layer networks with small size, a sample of large two-layer random networks, and six empirical multilayer social networks, demonstrates that coupling layers tends to strongly promote cooperation. If cooperation is disfavored in each layer alone, or even if layers individually favor spite, coupling layers can often promote cooperation in all layers. The multiple domains that structure human societies thus serve as a natural breeding ground for cooperation to flourish.

\section{Results}

\subsection{Model}
We model a population of $N$ individuals engaged in pairwise social interactions in multiple domains, or layers. Each individual uses separate strategies and plays distinct games in each layer. An individual's accumulated payoff over all layers governs how much influence she has on her peers' strategy updates in each layer.

In our model, nodes represent individuals and edges describe their social interactions. The population structure is described by a two-layer network, so that each individual corresponds to a node in layer one and an associated node in layer two (see Supplementary Information section 2.3.2 for analysis of more than two layers). Interactions within layer one occur along weighted edges $w_{ij}^{[1]}$ ($w_{ij}^{[1]}>0$); and interactions in layer two occur along weighted edges $w_{ij}^{[2]}$ ($w_{ij}^{[2]}>0$). 
The degree of node $i$ in layer one is $w_i^{[1]}=\sum_{j=1}^N w_{ij}^{[1]}$, whereas it is $w_i^{[2]}=\sum_{j=1}^N w_{ij}^{[2]}$ in layer two.

\begin{figure*}
\centering
\includegraphics[width=0.6\textwidth]{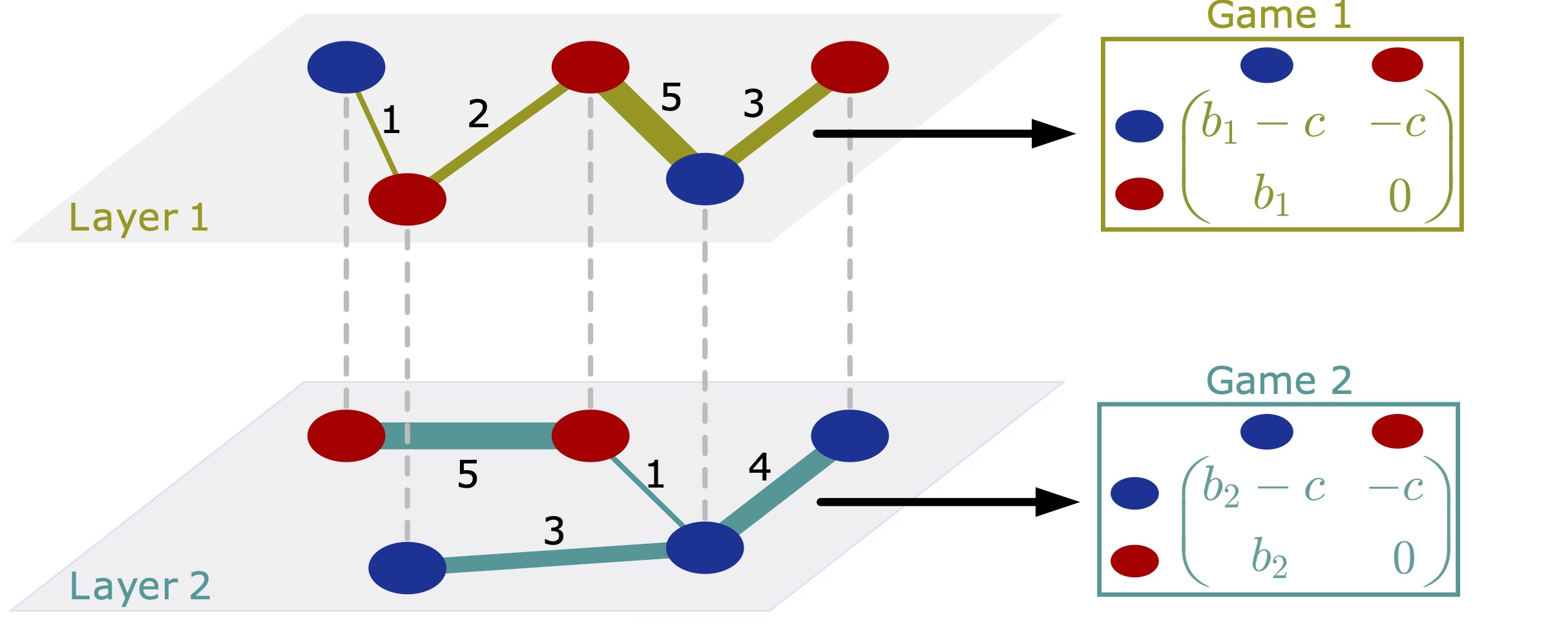}
\caption{\label{fig:1} \textbf{Evolutionary games in multilayer populations}.
	A population with two domains of social interaction is described by a two-layer network, with edge weights $w_{ij}^{[1]}$ in layer one and $w_{ij}^{[2]}$ 
	in layer two (see numbers next to edges for this example).
	Each player occupies a node in layer one and an associated node in layer two, as indicated by dashed lines. 
	Each player adopts a (possibly different) strategy in each layer, such as cooperation (blue) or defection (red). In each successive time step, each player $i$ plays game one with all her neighbors in layer one and derives an average payoff $u_i^{[1]}$ in layer one; the player also plays game two with all her neighbors in layer two and obtains average payoff $u_i^{[2]}$. Player $i$'s total payoff is the sum across layers, $u_i=u_i^{[1]}+u_i^{[2]}$, which determines her reproductive rate, $f_i = \exp\left(\delta u_i\right)$. 
	After all social interactions occur, a random player $i$ is selected to update her strategy in layer one by copying that of a random neighbor $j$ with probability proportional to $j$'s total fitness $w_{ij}^{[1]}f_j$ (i.e.~preferential copying of successful individuals). At the same time, a (possibly different) player $k$ updates his strategy in layer two, by copying that of a random neighbor $h$ proportional to $w_{kh}^{[2]}f_h$. We focus our analysis on donation games, in which each player chooses whether to pay a cost ($c$) to provide a benefit to her neighbor. The benefit may be different in layer one ($b_1$) than in layer two ($b_2$). 
}
\end{figure*}

Players engage in a donation game in every domain. In each layer, a player must choose either to cooperate ($C$) or defect ($D$) with her neighbors in that layer. A cooperative act means paying a cost of $c$ to provide the opponent with a benefit. The size of the benefit may differ across layers: $b_1$ in layer one and $b_2$ in layer two. Defection incurs no cost and provides no benefit to the opponent. A player's strategy may differ across layers, and so we let $s_i^{[1]}\in \{0,1\}$ denote player $i$'s strategy in layer one and $s_i^{[2]}\in \{0,1\}$ in layer two, where 1 denotes cooperation and 0 defection. This multilayer donation game is depicted in Fig.~\ref{fig:1}.

In each successive time step, each individual plays game one with all her neighbors in layer one, and she plays game two with all her neighbors in layer two. Each player $i$ obtains edge-weighted average payoff $u_i^{[1]}$ in layer one and $u_i^{[2]}$ in layer two, given by
\begin{equation}
	\begin{split}
		u_i^{[1]} &=-cs_i^{[1]}+b_1\sum_{j=1}^N p_{ij}^{[1]}s_j^{[1]}, \\
		u_i^{[2]} &=-cs_i^{[2]}+b_2\sum_{j=1}^N p_{ij}^{[2]}s_j^{[2]},
	\end{split}
\end{equation}
where $p_{ij}^{[1]}=w_{ij}^{[1]}/w_i^{[1]}$ and $p_{ij}^{[2]}=w_{ij}^{[2]}/w_i^{[2]}$.
Player $i$'s total payoff is the sum of those obtained in each layer, namely $u_i = u_i^{[1]}+u_i^{[2]}$. The total payoff across layers determines the rate at which a player's strategy spreads (i.e.~ its ``reproductive rate''), $f_i = \exp\left(\delta u_i\right)$, where $0<\delta<1$ is the intensity of selection \cite{2004-Nowak-p646-650}. The regimes $\delta\ll 1$ corresponds to weak selection \cite{2010-Wu-p46106-46106,Wu2013} and $\delta =0$ corresponds to neutral drift.

At the end of one time step, a random player $i$ is selected to update her strategy in layer one. With probability proportional to $w_{ij}^{[1]}f_j$, player $i$'s strategy in layer one is replaced by player $j$'s strategy in layer one. This update rule ensures that a player preferentially copies the strategy of successful individuals. At the same time, a random player $k$ is selected to update his strategy in layer two. With probability proportional to $w_{kh}^{[2]}f_h$, player $k$'s strategy in layer two is replaced by $h$'s strategy in layer two. 
We focus on this form of ``death-birth'' updating \cite{2006-Ohtsuki-p502-505}, and we also analyze other mechanisms such as pairwise-comparison updating, birth-death updating, and a mixture of the two (i.e. different update rules for different layers; see Supplementary Information section 2.1).

\subsection{General rule for the evolution of cooperation in multilayer populations}
In the absence of innovation (mutation), the population eventually settles into an absorbing state in which all players either cooperate or defect, in each layer. The absorbing state in the two layers may be different, e.g.~cooperation in layer one and defection in layer two. In general, selection can favor cooperation provided the benefit-to-cost ratio $b/c$ is sufficiently large \cite{2006-Ohtsuki-p502-505}. Here, we analyze how the critical benefit-to-cost ratio to support cooperation in layer one, $\left(b_1/c\right)^\ast$, depends on coupling with a second layer.

Let $\rho^{[1]}_{C}$ denote the probability that all players eventually cooperate in layer one, starting from some fixed configuration of cooperators and defectors. We use $\left(\rho^{[1]}_{C}\right)^{\circ}$ to denote this probability under neutral drift, i.e.~when $\delta=0$.
Selection is said to favor the emergence and fixation of cooperation (or cooperation replacing defection) in layer one when the inequality $\rho^{[1]}_{C}>\left(\rho^{[1]}_{C}\right)^{\circ}$ holds \cite{2004-Nowak-p646-650,2005-Lieberman-p312-316,2006-Ohtsuki-p502-505}. 
We focus primarily on the probability that cooperation will fix under weak selection, compared to neutral drift.
We also compare the fixation probability of cooperation to the fixation probability of defection, and we find qualitatively similar results using this relative measure (Supplementary Information section 1).

\begin{figure}
	\centering
	\includegraphics[width=0.9\textwidth]{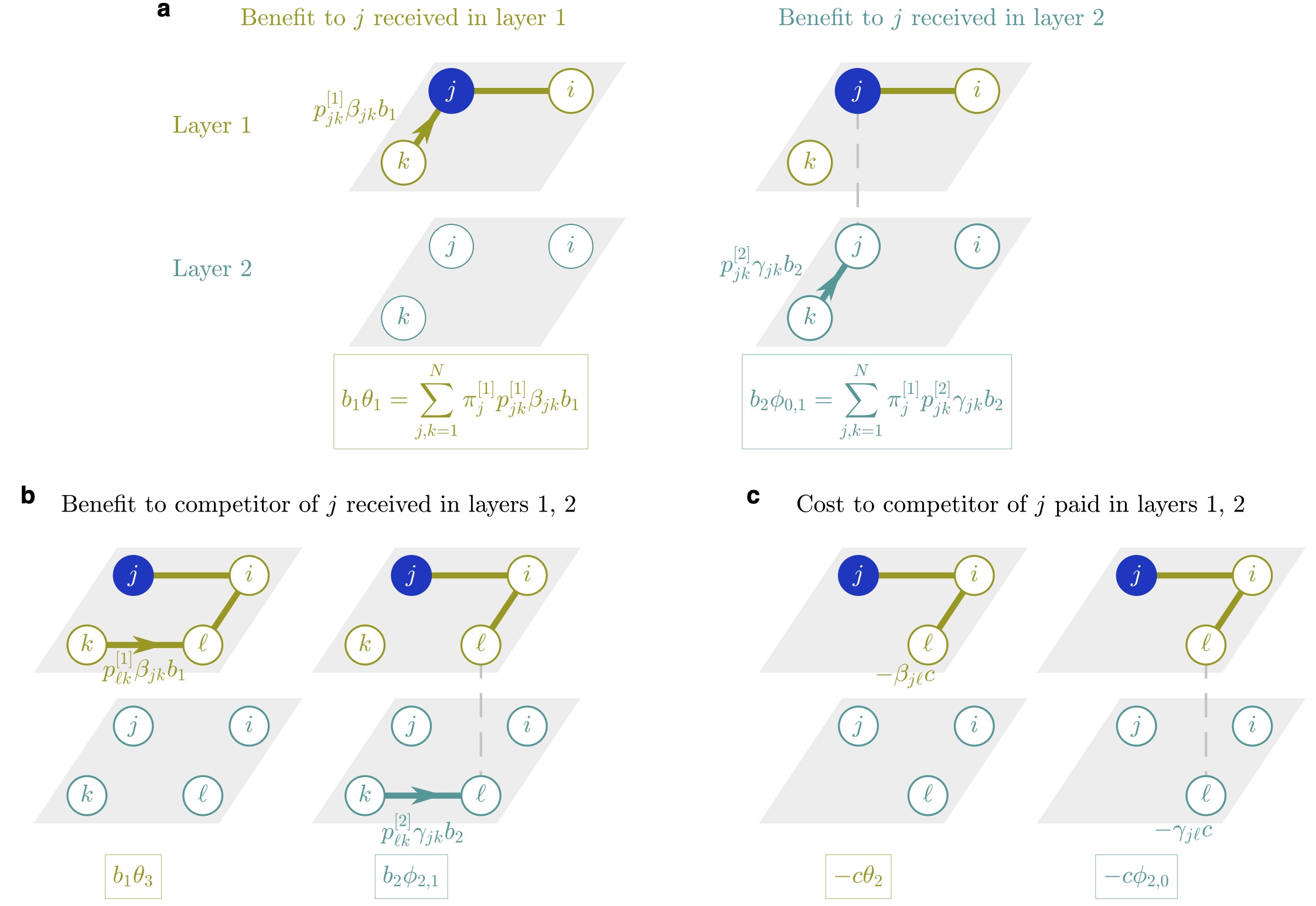}
	\caption{\textbf{General rule for the evolution of cooperation in multilayer populations}. We consider what happens when individual $i$ is chosen to update her strategy in layer one, and her neighbors compete to have their strategy copied. Cooperation will be selectively favored in layer one if a cooperative neighbor, node $j$, has greater expected payoff than a random neighbor, node $\ell$. Node $j$ receives an average benefit $b_1\theta_1$ from its own one-step neighbors in layer one (panel \textbf{a}, left). Node $j$ also receives an average benefit $b_2\phi_{0,1}$ from its own one-step neighbors in layer two (panel \textbf{a}, right). The expression for $\theta_1$ (respectively $\phi_{0,1}$) accounts for the probability $p_{jk}^{[1]}$ ($p_{jk}^{[2]}$) that a random walk moves from node $j$ to $k$ in layer one (layer two); and for the probability $\beta_{jk}$ ($\gamma_{jk}$) that node $k$ is cooperative in layer one (layer two) as node $j$ in layer one (see also Supplementary Information section 2.1.1). Node $j$ pays the cost $c\theta_0$ as a cooperator in layer one and $c\phi_{0,0}$ in layer two. Node $j$'s net payoff is therefore $\theta_1 b_1+\phi_{0,1}b_2-(\theta_0 c+\phi_{0,0}c)$. Any competitor of $j$, such as node $\ell$, is also vying to have its strategy copied. Note that in layer one, node $\ell$ is two steps away from node $j$. Node $\ell$ receives an average benefit $b_1\theta_3$ (respectively $b_2\phi_{2,1}$) from its one-step neighbors in layer one (layer two), who are three steps away in layer one (two steps away in layer one and one step away in layer two) from node $j$, as shown in panel \textbf{b}. Whenever $\ell$ is a cooperator she pays cost $c$, leading to an average cost $\theta_2c$ in layer one and $\phi_{2,0}c$ in layer two (panel \textbf{c}). Node $\ell$'s net payoff is therefore $\theta_3 b_1 + \phi_{2,1} b_2-(\theta_2+\phi_{2,0})c$. Selection will favor cooperation only if $\theta_1 b_1+\phi_{0,1}b_2-\theta_0 c-\phi_{0,0}c > \theta_3 b_1 + \phi_{2,1} b_2-(\theta_2+\phi_{2,0})c$.
	}
	\label{fig:2}
\end{figure}

To analyze the evolution of cooperation in multilayer networks, we adapt techniques from the study of strategy assortment in single-layer networks \cite{2017-Allen-p227-230,McAvoy2021,mcavoy-2020-nhb}, based on random walks within the network. It is necessary to first understand what a random walk in a multilayer network looks like. In a two-layer network, we define a random walk as follows: a step from node $i$ to $j$ in layer one (respectively layer two) occurs with probability $p_{ij}^{[1]}$ ($p_{ij}^{[2]}$). An $\left(n,m\right)$-step random walk in the network means an $n$-step random walk in layer one followed by an $m$-step random walk in layer two, where the beginning of the second random walk corresponds to the end of the first (e.g.~Fig.~\ref{fig:2}\textbf{b}).

We let $\theta_n$ denote the probability that the starting and ending nodes of an $n$-step random walk in layer one both employ the same strategy. For example, $\theta_1$ quantifies the correlation, or assortment, of strategies between neighboring nodes in layer one. Similarly, we let $\phi_{n,m}$ denote the probability that the starting and ending nodes of an $\left(n,m\right)$-step random employ the same strategy. For example, $\phi_{0,1}$ quantifies the strategy assortment between a node in layer one and a random neighbor in layer two. We can obtain $\theta_n$ and $\phi_{n,m}$ by solving systems of $O\left(N^{2}\right)$ linear equations (see Methods).

For any two-layer population structure and any initial strategy configuration, we have derived a general condition for when cooperation in layer one is favored by selection:
\begin{equation}
	\theta_1 b_1+\phi_{0,1}b_2-\theta_0 c-\phi_{0,0}c > \theta_3 b_1 + \phi_{2,1} b_2-\theta_2c-\phi_{2,0}c .
	\label{eq:db_formula_main}
\end{equation}
Informally, this condition states that a cooperative neighbor of a node in layer one must have a higher payoff than a random neighbor.
The four terms on the left side quantify the benefits and costs to a cooperative neighbor,
where $\theta_1b_1$ and $\theta_0 c$ denote the benefits and costs from layer one, and $\phi_{0,1}b_2$ and $\phi_{0,0}c$ denote the benefits and costs from layer two. 
The four terms on the right quantify the benefits and costs to a random neighbor,
where $\theta_3b_1$ and $\theta_2 c$ (respectively $\phi_{2,1}b_2$ and $\phi_{2,0}c$) denote the benefits and costs from layer one (layer two). These eight quantities collectively govern the fate of cooperation in multilayer networks, as depicted in Fig.~\ref{fig:2}. 
A special case of equation~(\ref{eq:db_formula_main}) is when layer one evolves independently from layer two, so that there are no benefits and costs arising from layer two, in which case selection favors cooperation whenever $\theta_1 b_1-\theta_0 c> \theta_3 b_1-\theta_2c$.

\subsection{Coupled ring networks}
\begin{figure}
	\centering
	\includegraphics[width=0.9\textwidth]{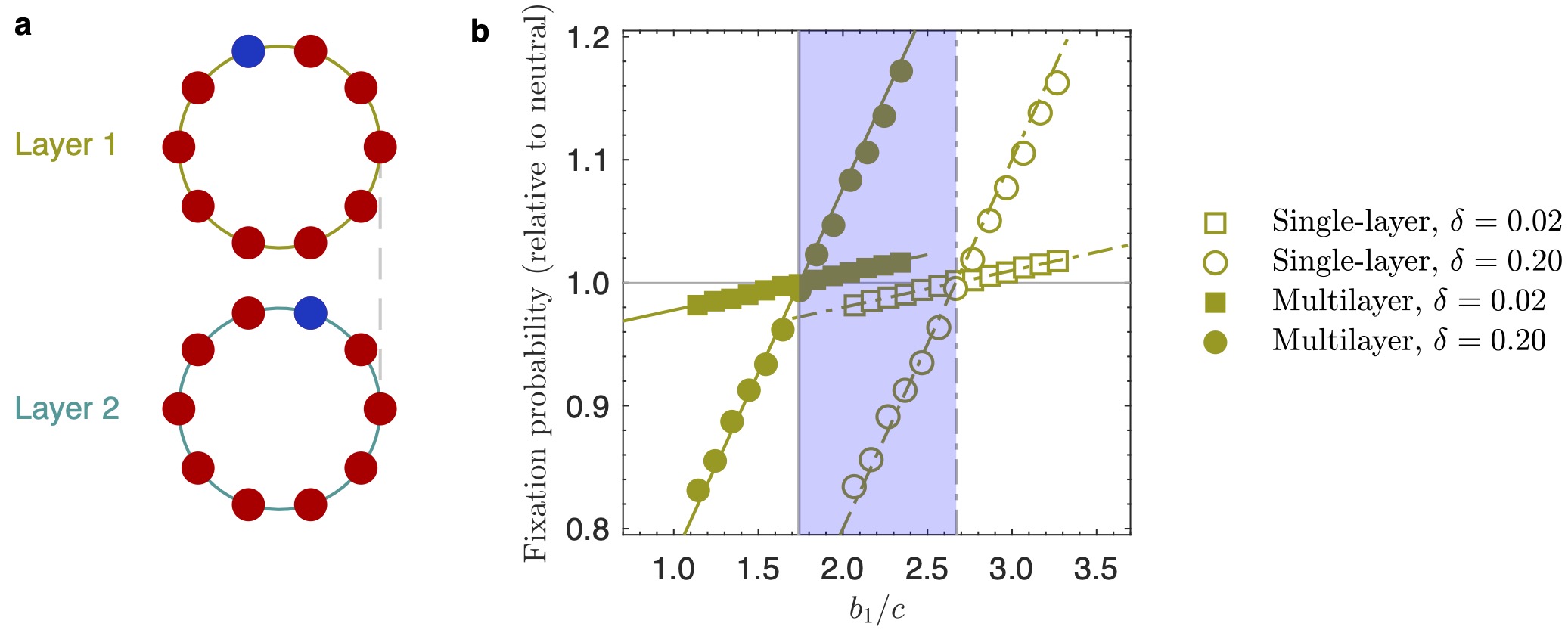}
	\caption{\label{fig:3} \textbf{Multilayer games can promote cooperation}. \textbf{a}, We consider a ``ring network" in each layer, with each node connected to two neighboring nodes. Nodes that occupy the same position in both layers represent the same individual, as indicated by the dashed line. The initial strategy configuration contains one cooperative individual in layer one (blue) and one cooperative individual in layer two (blue).
		\textbf{b}, The probability that cooperation will eventually fix in layer one, $\rho_C^{\left[1\right]}$, relative to the fixation probability under neutral drift, $\left(\rho^{[1]}_{C}\right)^{\circ}$. 
		We compare two scenarios: when the layers operate independently versus when the two layers are coupled. 
		Cooperation in layer one is favored by selection if it fixes with a greater probability than in the absence of selection (horizontal line). 
		Squares (for $\delta=0.02$) and circles (for $\delta=0.20$) indicate fixation probabilities estimated from $10^7$ replicate Monte Carlo simulations, and lines indicate analytical predictions.
		Our analysis under weak selection predicts that cooperation will be favored whenever the benefit-to-cost ratio ($b_1/c$) exceeds a critical value, indicated by the solid vertical line (for coupled layers) and by the dashed vertical line (for independent layers). 
		For the benefit-to-cost ratios indicated in light blue, coupling between layers promotes cooperation in layer one even though it would be disfavored by selection under evolution in layer one alone. 
		Parameters: $b_2=10$, $c=1$.}
\end{figure}

The general rule derived above allows us to study how multiple domains of social interactions influence the prospects for cooperation, in arbitrary interaction networks. 
In the following, we focus on unweighted networks.
We start with an illustrative example based on a two-layer ring network. We consider $N=10$ individuals are arranged in a ring, each with two neighbors in each layer. Initially, a single individual in each layer is cooperative, and the cooperator in layer one is connected to the cooperator in layer two (see Fig.~\ref{fig:3}\textbf{a}). When the two layers evolve independently, or in the absence of layer two, cooperation is favored by selection in layer one only if the benefit-to-cost ratio, $b_1/c$, exceeds a critical value, $(b_1/c)^*=8/3$ (dashed vertical line in Fig.~\ref{fig:3}\textbf{b}). But when the two layers are coupled and $b_2/c=10$, then critical value $(b_1/c)^*$ is reduced to 1.74 (solid vertical line in Fig.~\ref{fig:3}\textbf{b}). In other words, coupling games between layers promotes cooperation in layer one, making it far easier to evolve than in the absence of layer two. The reason is that, when layers are coupled, a player's success in one layer depends not only on her payoffs obtained in that layer, but also on her interactions in the other layer. In this case, the cooperator in layer one is being exploited by two neighboring defectors, as seen in Fig.~\ref{fig:3}\textbf{a}, but nonetheless she receives an extra benefit from a cooperative neighbor in layer two, who increases her fitness and promotes the spread of her (cooperative) strategy in layer one (see also Supplementary Fig.~1 for further details).

Coupling layers can have a significant effect on the probability that cooperation will spread and overtake a population, even in regimes where selection disfavors cooperation in the absence of coupling. For the example shown in Fig.~\ref{fig:3}, when the selection intensity is very small, e.g.~$\delta=0.02$, the fixation probability of cooperation can be increased by a small amount ($3\%$) relative to the case of independent layers; but when the selection intensity is moderate, such as $\delta=0.20$, the effect size can be as large as 27.76\% (Fig.~\ref{fig:3}\textbf{b}).
Although the absolute increase in fixation probability is always small, for weak selection, it makes sense to quantify the effect size relative to neutrality.

\begin{figure}
	\centering
	\includegraphics[width=0.5\textwidth]{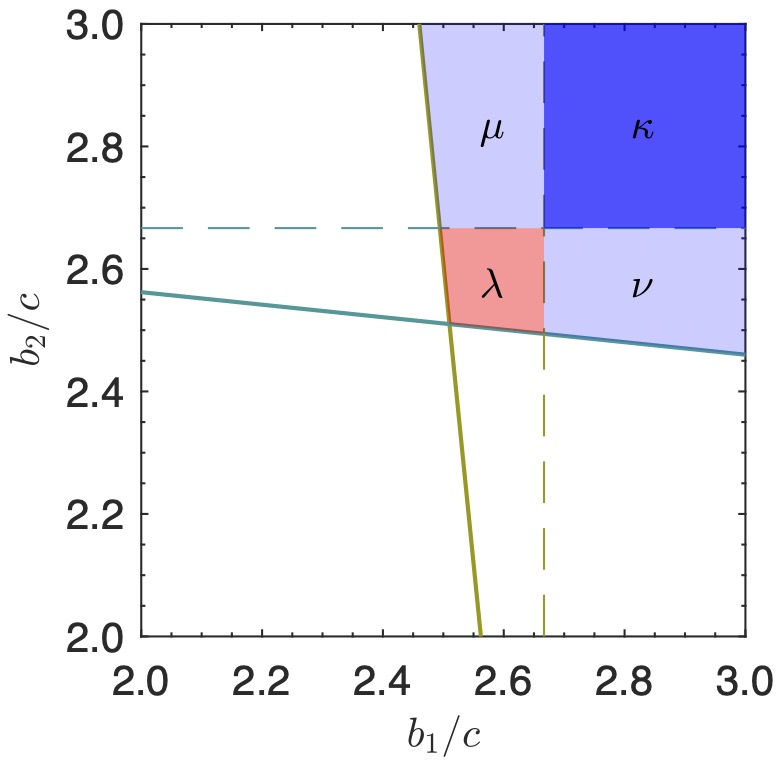}
	\caption{\label{fig:4} \textbf{When coupling promotes cooperation}. We analyze a two-layer ring network with the initial strategy configuration shown in Fig.~\ref{fig:3}\textbf{a}. If the population evolves in layer one alone, then cooperation is favored by selection only when $b_1/c$ exceeds the olive dashed line. Coupling with layer two facilitates the evolution of cooperation in layer one, decreasing the required benefit-to-cost ratio from the olive dashed line to the olive solid line. If the population evolves in layer two alone, cooperation is favored by selection only when $b_2/c$ exceeds the blue dashed line. Coupling with layer one facilitates the evolution of cooperation in layer two, decreasing the required benefit-to-cost ratio to the blue solid line. Without coupling, selection favors cooperation in both layers only in region $\kappa$. But coupling extends that region to $\kappa\mu\lambda\nu$. Note that in region $\lambda$, cooperation is disfavored in each layer on its own, but it is favored in both layers when they are coupled.
	}
\end{figure}

Figure~\ref{fig:4} illustrates more generally how multilayer coupling affects evolutionary dynamics in ring networks. When the two layers evolve separately, cooperation is favored in layer one only if $b_1/c$ exceeds the olive dashed line; and cooperation is favored in layer two only if $b_2/c$ exceeds the blue dashed line. Selection thus favors cooperation in both layers only when $b_1/c$ and $b_2/c$ lie in region $\kappa$. Coupling layers moves the benefit-to-cost ratio required for cooperation in layer one to olive solid line, and it moves the benefit-to-cost ratio required in layer two to blue solid line -- in both cases expanding the parameter range of costs and benefits that favor cooperation. 
In particular, the region $\lambda$ reveals the remarkable fact that even if cooperation is disfavored by selection in each layer alone, cooperation can nonetheless be favored in both layers simultaneously when they are coupled.

In the two-layer ring network, for any configuration with only one cooperator in layer one and one cooperator in layer two, we have derived a simple formula to calculate the critical benefit-to-cost ratio $\left(b_1/c\right)^\ast$ required to favor cooperation (see Methods). For more complicated initial configurations we can still resort to the general condition (equation~(\ref{eq:db_formula_main})) to obtain theoretical predictions, although the expressions are more complicated. Even among these simple graphs we find a diverse range of scenarios in which multilayer coupling promotes cooperation (see Supplementary Fig.~2).

\subsection{Coupled heterogeneous networks}
For ring networks, cooperation is favored in each layer alone provided the benefit-to-cost ratio exceeds some critical value. Coupling between layers can reduce the critical value and thereby promote cooperation. However, the prospects for cooperation may be far worse in other population structures. In fact, there are many single-layer population structures in which cooperation is never favored in a social dilemma, no matter how large the benefit-to-cost ratio is \cite{2009-Tarnita-p570-581,2017-Allen-p227-230,2019-Allen-p1147-1210}. 

\begin{figure}
	\centering
	\includegraphics[width=1\textwidth]{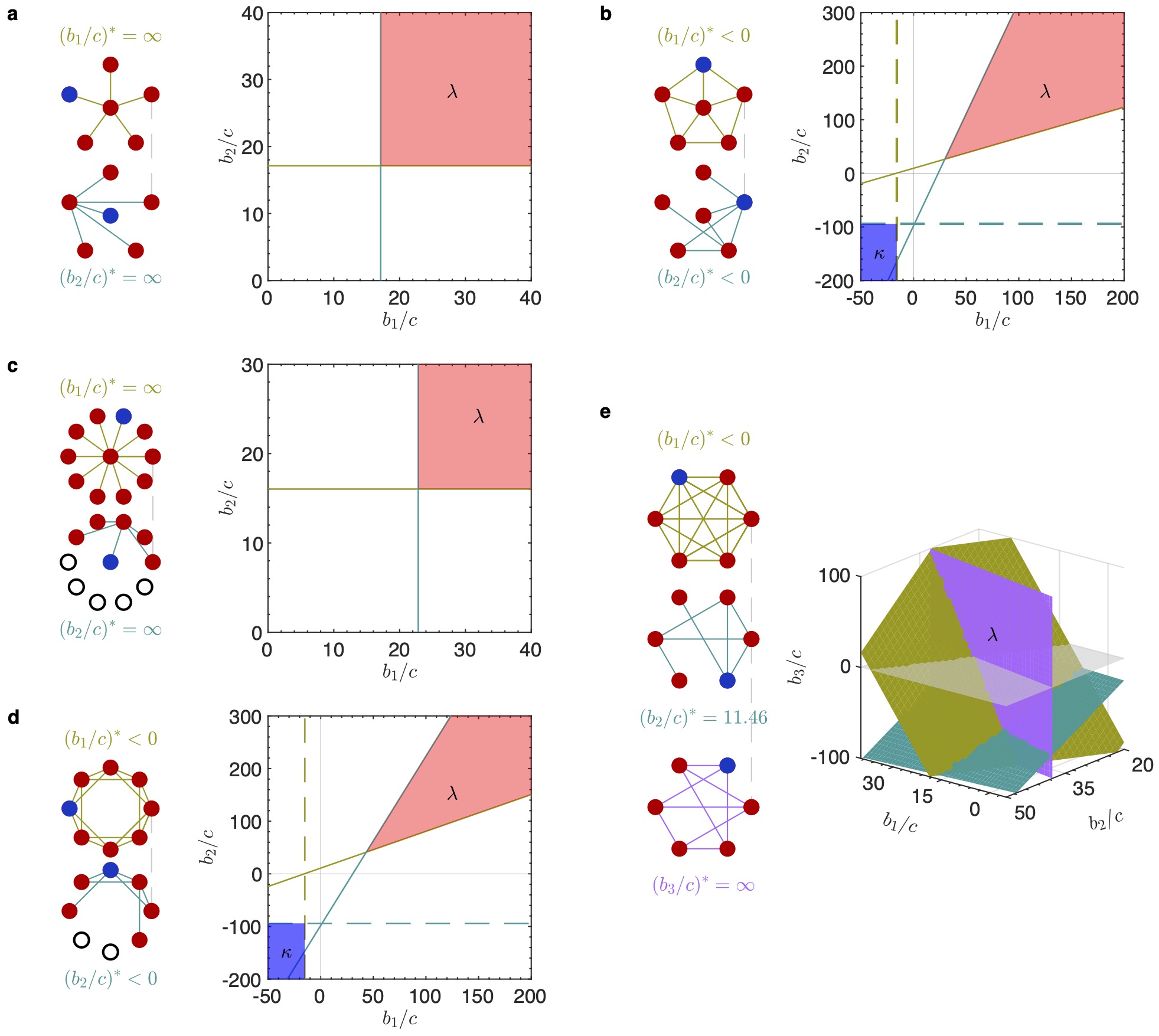}
	\caption{\label{fig:5} \textbf{Multilayer coupling can promote cooperation even when cooperation is disfavored in individual layers}. We present five representative examples. \textbf{a}, In each layer alone, the critical benefit-to-cost ratio is infinite, i.e. $\left(b_1/c\right)^\ast =\left(b_2/c\right)^\ast =\infty$. As a result, cooperation is never favored by selection, regardless of how large the benefit-to-cost ratio is. Nevertheless, when the two layers are coupled, selection then favors cooperation in both layers, provided $b_1/c$ and $b_2/c$ fall within the region $\lambda$. \textbf{b}, In each layer alone, the critical benefit-to-cost ratio is negative, i.e. $\left(b_1/c\right)^\ast ,\left(b_2/c\right)^\ast <0$. These negative ratios indicate that selection can favor the fixation of spite in each layer alone---so that an individual will pay a cost of $c>0$ to \textit{decrease} his partner's payoff. Nevertheless, when the two layers are coupled, selection then favors cooperation in both layers, provided $b_1/c$ and $b_2/c$ fall within the region $\lambda$. Multilayer networks can also rescue cooperation when there are different population sizes in different layers (\textbf{c},\textbf{d}), or for populations with more than two layers (\textbf{e}). In \textbf{c} and \textbf{d}, 
		open circles indicate absence of a node in that layer.
	}
\end{figure}

The star graph is an example of a population structure that always suppresses cooperation. The graph consists of a central hub and $N-1$ leaf nodes. Regardless of the initial strategy configuration, no finite value of the benefit-to-cost ratio can selectively favor cooperation (i.e.~$\left(b_1/c\right)^\ast = \infty$). Nonetheless, if we couple two stars in a certain way (Fig.~\ref{fig:5}\textbf{a}) then selection favors cooperation in both stars simultaneously provided $b_1/c$ and $b_2/c$ exceed $\left(18N^4-55N^3+64N^2-33N+6\right) /\left(4N^3-2N^2\right)$ (see Supplementary Information section 2.2.2 for detailed derivations). The region $\lambda$ in Fig.~\ref{fig:5}\textbf{a} depicts the benefit-to-cost ratios that favor cooperation in these two-layer graphs.

An even more striking example occurs on the wheel network, shown in Fig.~\ref{fig:5}\textbf{b}. For any initial strategy configuration on such networks, the critical benefit-to-cost ratio is negative, $\left(b_1/c\right)^\ast<0$ -- meaning that selection actually favors spite, an antisocial behavior where an individual pays a cost to decrease her neighbor's payoff. But if we couple one wheel network with another, as shown in Fig.~\ref{fig:5}\textbf{b}, cooperation can be favored on both layers, provided $b_1/c$ and $b_2/c$ lie in region $\lambda$. Together with the star network, this example shows that coupling can promote cooperation in multiple layers, even if selection always disfavors cooperation in each layer alone.

Our framework also applies to multilayer populations with different population sizes in different layers. That is, a player may have social interactions in layer one, but no social interactions in layer two (see examples in Fig.~\ref{fig:5}\textbf{c},\textbf{d}) -- corresponding, for example, to an individual who forgoes online social networking altogether. 
Figure~\ref{fig:5}\textbf{c} and \textbf{d} confirms that in such cases coupling can still allow cooperation to be favored in both layers, even if cooperation is disfavored in each layer alone for any benefit-to-cost ratio. In such populations with different population sizes in different layers the general rule for the evolution of cooperation is analogous to equation~(\ref{eq:db_formula_main}) (see Supplementary Information section 2.3.1).

Our framework also applies to multilayer populations with an arbitrary number of layers. 
Figure~\ref{fig:5}\textbf{e} illustrates an example of three-layer population. When the three layers evolve independently, cooperation is favored neither in layer one ($\left(b_1/c\right)^\ast <0$) nor in layer three ($\left(b_3/c\right)^\ast =\infty$). Coupling the three layers allows selection to favor cooperation, provided benefit-to-cost ratios lie in the three-dimensional region $\lambda$. In particular, coupling not only makes it possible for cooperation to be favored in layer one and layer three, but it also reduces the value of $b_2/c$ required for cooperation being favored in layer two. In Supplementary Information section 2.3.2, we derive the general condition for selection to favor cooperation on population structures with an arbitrary number of layers. Although coupling of layers can provide more opportunities for the evolution of cooperation, some choices of benefits and costs in layers may lead to negative effects. 
In the example shown in Fig.~\ref{fig:5}\textbf{e}, if $b_1/c$ and $b_3/c$ are selected beyond the region $\lambda$, then coupling domains may increase the critical benefit-to-cost ratio $\left(b_2/c\right)^\ast$, making it harder for cooperation to evolve in layer two.

\subsection{Small multilayer populations}
\begin{figure}
	\centering
	\includegraphics[width=0.8\textwidth]{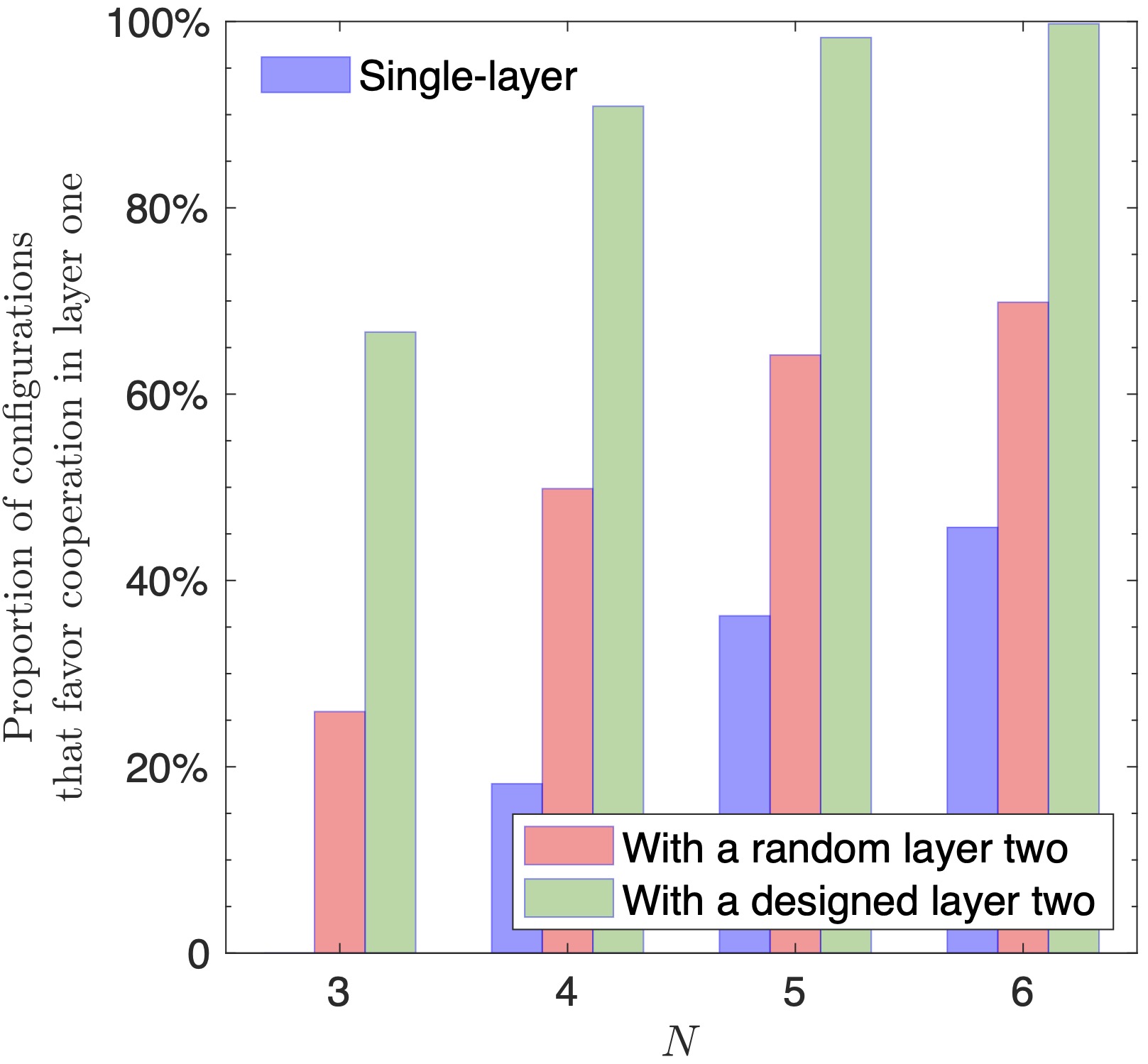}
	\caption{\label{fig:6} \textbf{Proportion of small networks that permit the evolution of cooperation.} We systematically analyzed all networks of size $N=3,4,5,6$, including all initial configurations containing a single cooperator. Blue bars indicate the proportion of single-layer networks and mutant configurations in which selection can favor cooperation in layer one for some benefit-to-cost ratio, i.e. $\left(b_1/c\right)^\ast >0$. For $N=3$, selection does not favor cooperation for any network and configuration, for any value of $b_1/c$. Coupling layer one with a randomly chosen network and strategy configuration in layer two increases the frequency of selection for cooperation (i.e. selection favors cooperation in layer one for some choice of $b_1/c>0$ and $b_2/c>0$, shown in red). Coupling layer one with a deliberately designed network and strategy configuration in layer two further increases the frequency of cooperation in layer one (green). In a majority of these cases, coupling to either a random or a designed network in layer two, selection actually favors cooperation in both layers simultaneously (see Supplementary Fig.~5).
	}
\end{figure}

To study behavioral dynamics across a variety of structures, we systematically analyzed all two-layer networks of size $N=3,4,5,6$ and all initial configurations of a single cooperator in each layer (see Methods for details). We first report the proportion of single-layer networks and strategy configurations in which cooperation can be favored in layer one alone for some choice of benefit-to-cost ratio (i.e. $\left(b_1/c\right)^\ast >0$, blue bars in Fig.~\ref{fig:6}). Coupling layer one with a randomly chosen network and strategy configuration in layer two can significantly increases the frequency of structures on which selection favors cooperation in layer one, for some values $b_1/c>0$ and $b_2/c>0$ (red bar). Coupling layer one with a deliberately designed network and configuration in layer two can further increase the frequency of cooperation (green bar). 
In a large proportion of these cases, coupling to either a random or a designed network in layer two, selection actually favors cooperation in both layers simultaneously (Supplementary Fig.~3). Therefore, in a systematic analysis of all small structures, multilayer networks have a significant positive impact on prospects for cooperation.

\subsection{Larger multilayer populations}
\begin{figure}
	\centering
	\includegraphics[width=0.7\textwidth]{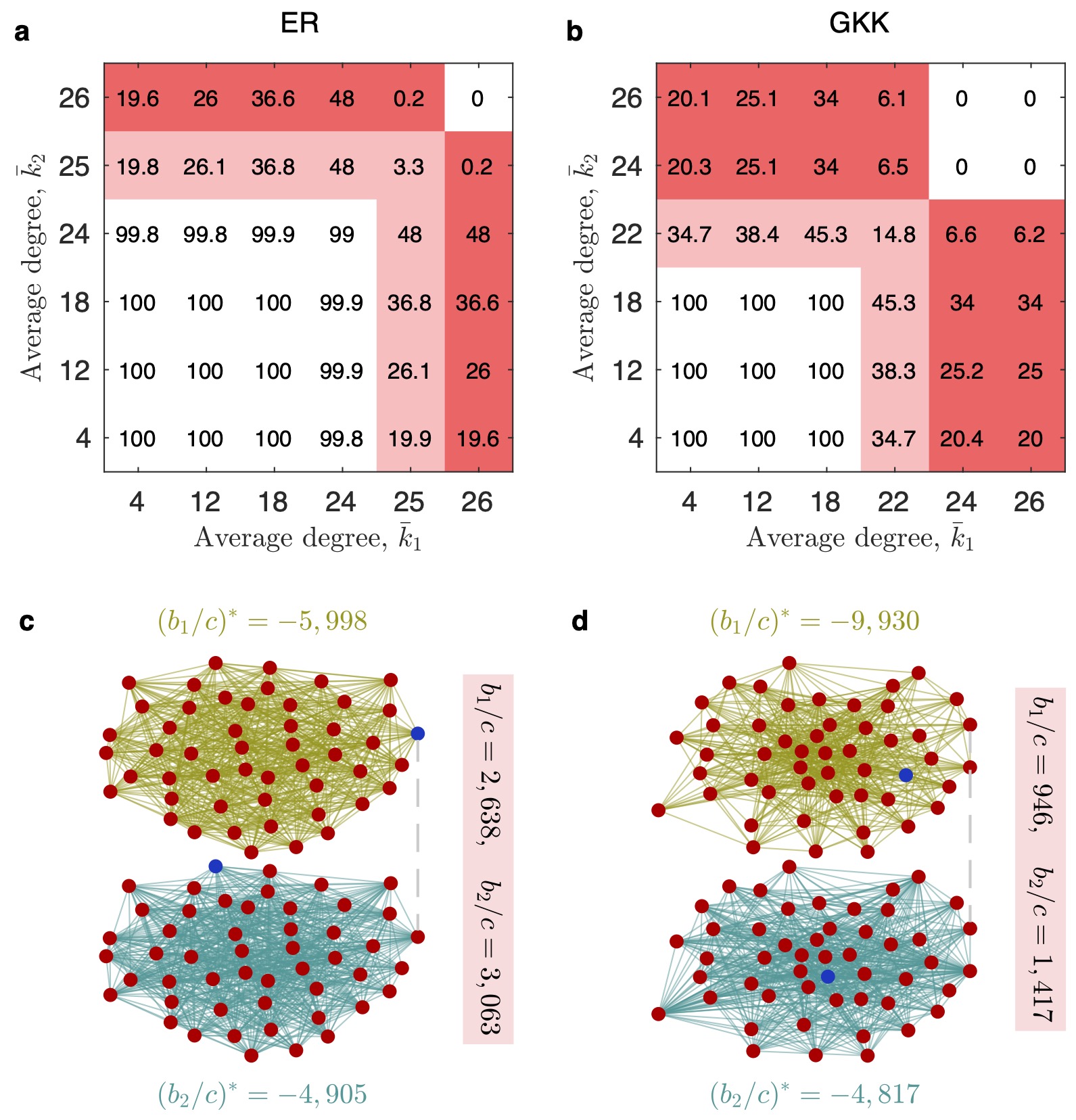}
	\caption{\label{fig:7} \textbf{Multilayer coupling can catalyze the evolution of cooperation in random networks}. We sampled 100 two-layer Erd\"{o}s-R\'{e}nyi (ER) random networks of size $N=50$, and 100 two-layer Goh-Kahng-Kim (GKK) networks generated by the Goh-Kahng-Kim algorithm \cite{2001-Goh-p278701-278701} of size $N=50$, for each pair of average node degrees, $\bar{k}_1$ and $\bar{k}_2$, in layers one and two, respectively. For each two-layer network we analyzed all $2{,}500$ initial configurations consisting of a single mutant cooperator in each layer. \textbf{a}, The proportion (percentage) of sampled two-layer ER networks and initial configurations in which selection can favor cooperation in both layers, for some positive values of $b_1/c$ and $b_2/c$. Highlighted entries indicate regimes when coupling increases the frequency of selection for cooperation in both layers compared to independent evolution in each layer. Coupling can have a dramatic effect---e.g.~favoring cooperation in both layers for nearly 50\% of sampled networks, compared to virtually never favoring cooperation without coupling (see Supplementary Fig.~6). For some regimes, coupling permits selection for cooperation in both layers even though one or both layers oppose its selection in the absence of coupling (dark red). \textbf{b}, The proportion (percentage) of sampled two-layer GKK networks and initial configurations in which selection can favor cooperation in both layers; highlighted entries indicate regimes when coupling increases the frequency of selection for cooperation in both layers compared to independent evolution in each layer. \textbf{c}, \textbf{d}, Examples of two-layer ER and GKK networks, respectively, in which spite is favored on each layer evolving independently, but cooperation is favored in both layers when coupled.
	}
\end{figure}

The networks explored above are all relatively small, but they nonetheless exhibit a diverse range of behavioral dynamics and surprising effects induced by multilayer coupling. 
To study behavior on larger networks, of size $N=50$, we sampled many two-layer Erd\"{o}s-R\'{e}nyi (ER) random networks\cite{1960-Erdoes-p17-61} and many two-layer Goh-Kahng-Kim (GKK) networks \cite{2001-Goh-p278701-278701} generated with exponent $\gamma=2.5$.
We sampled these networks across a diverse range of average node degrees in layer one and in layer two (Fig.~\ref{fig:7}\textbf{a}). 
The two classes of networks differ in their node degree distribution. 
For example, for average degree 4, the maximum node degree is 10 in ER random networks and up to 28 in GKK networks we study.
In each two-layer network we placed a single mutant cooperator in each layer and analyzed all $50\times 50=2{,}500$ initial strategy configurations. Figure~\ref{fig:7}\textbf{a} and \textbf{b} reports the frequency of structures for which selection can favor cooperation in both layers for some positive values of $b_1/c$ and $b_2/c$. Compared with the corresponding frequencies when the two layers evolve separately (see Supplementary Fig.~4), we find that coupling two layers is broadly conducive to cooperation, as shown in the highlighted area in Fig.~\ref{fig:7}\textbf{a},\textbf{b}. In particular, in the random networks with average degree greater than 26, cooperation is never favored for any benefit-to-cost ratio; whereas coupling such networks to a random network in layer two can often rescue cooperation (dark red area in Fig.~\ref{fig:7}\textbf{a}). Figure~\ref{fig:7}\textbf{c},\textbf{d} shows examples of random two-layer networks that favor the evolution of spite on each layer alone, but that can favor cooperation on both layers when coupled (see also Supplementary Figs.~5 and 6 for further analysis and examples).

We also investigated larger networks, with size up to $N=300$ and average degree $\bar{k}_1=
\bar{k}_2=4$, generated by the Goh-Kahng-Kim algorithm with exponent $\gamma=2.5$ and, alternatively, by the Barab\'{a}si-Albert algorithm \cite{1999-Barabasi-p509-512}. These networks exhibit broad distributions of node degree (Supplementary Fig.~7).
For each two-layer network, we randomly sampled $500$ initial strategy configurations. 
Among the Goh-Kahng-Kim networks, in $99.23\%$ of cases coupling layers decreases the benefit-to-cost ratio required for cooperation in layer one; furthermore, in $10.15\%$ of cases, coupling promotes cooperation in both layers simultaneously. Among the Barab\'{a}si-Albert networks, in $99.26\%$ of cases coupling layers decreases the benefit-to-cost ratio required for cooperation in layer one; and in $11.24\%$ of cases, coupling promotes cooperation in both layers simultaneously.

\subsection{Empirical multilayer populations}
\begin{figure}
	\centering
	\includegraphics[width=1\textwidth]{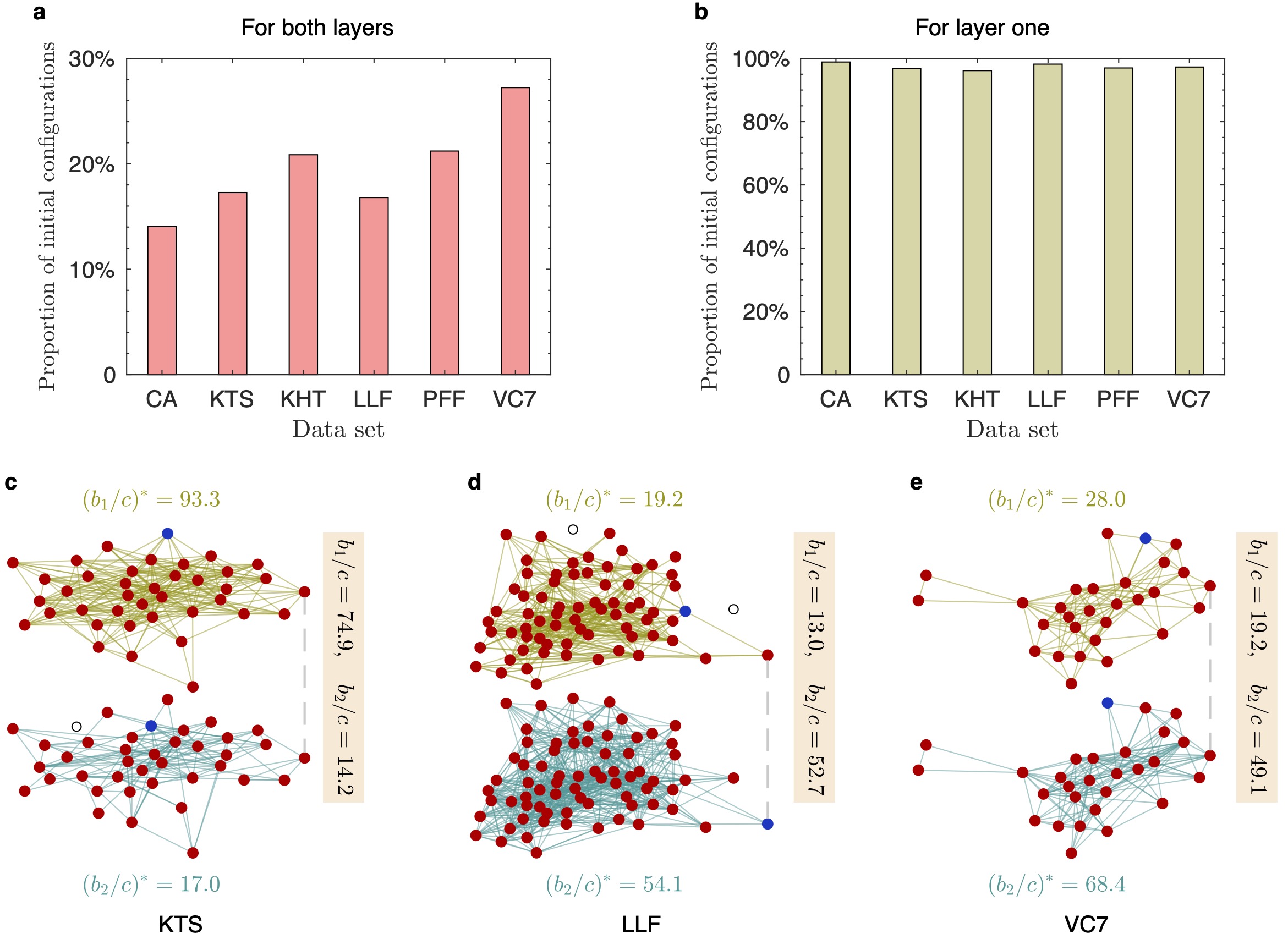}
	\caption{\label{fig:8} \textbf{Evolution of cooperation in six real-world two-layer networks}. We analyzed networks of online and offline relationships among 61 employees of the Computer Science Department at Aarhus University (CA) \cite{magnani-2013-arxiv}; social-emotional and professional relationships among 39 customers surveyed in a Zambian tailor shop (KTS) \cite{Heath1974}; friendship and professional relationships among 21 managers at a high-tech company (KHT) \cite{Krackhardt1987b}; friendship and professional relationships among 71 partners at the Lazega Law Firm (LLF) \cite{emmanuel-2001-oxford}; marriage and business relationships among 16 families in renaissance Florence (PFF) \cite{Padgett1993}; and friendship and scholastic relationships among 29 seventh-grade students in Victoria, Australia (VC7). 
		We considered all initial configurations with a single mutant cooperator in each layer, where individuals play the donation game. \textbf{a}, Proportion of configurations in which coupling layers reduces benefit-to-cost ratios required for cooperation to be favored in both layers, relative to when layers evolve independently. \textbf{b}, Proportion of initial configurations in which coupling layers reduces the benefit-to-cost ratio required for cooperation to be favored in layer one. \textbf{c}-\textbf{e}, Three example configurations with a single mutant cooperator (blue) among defectors (red), where open circles indicate isolated individuals. In these examples, selection favors cooperation in each layer alone provided the benefit-to-cost ratio exceeds a critical value, e.g. $\left(b_1/c\right)^\ast = 93.3$ in KTS layer one. Coupling layers reduces the benefit-to-cost ratio required for cooperation to evolve in one or both layers. For example, when $b_1/c=74.9$ and $b_2/c=14.2$, selection favors cooperation in both layers of the coupled KTS network. 
	}
\end{figure}

We also studied six real-world examples of communities engaged in multiple domains of social interaction. The six empirical two-layer networks \cite{magnani-2013-arxiv,Heath1974,Krackhardt1987b,emmanuel-2001-oxford,Padgett1993} range from online and offline relationships among members of the Computer Science Department at Aarhus University, to the marriage and business relationships among prominent families in renaissance Florence, and they range in population size from $N=21$ to $N=71$ (Fig.~\ref{fig:8}). We analyzed the prospects for cooperation when individuals play donation games in each layer, including all initial configurations with a single cooperator in each layer. In all of these empirical networks, even if two layers evolve separately cooperation can be favored in each layer provided the benefit-to-cost ratios are sufficiently large. Coupling the two layers can nonetheless reduce the benefit-to-cost ratios required to support cooperation. Figure~\ref{fig:8}\textbf{a} shows the proportions of initial configurations for which coupling facilitates cooperation in this way. Figure~\ref{fig:8}\textbf{c} shows an example of this phenomenon, using the two-layer network of socio-emotional and professional relationships among customers surveyed in a Zambian tailor shop; coupling these two domains of social interaction facilitates cooperation in both domains, by reducing the benefit-to-cost ratios required to favor prosocial behavior.

In practice, the behavioral outcome in one layer may be more important than in another layer, such as when more individuals interact in one layer, or when prosociality in one domain is more important for the overall welfare of a society. To study this in the context of real-world multilayer networks, we analyzed to what degree the benefit-to-cost ratio for cooperation to be favored in layer one alone can be reduced. In these analyses the prospect for cooperation in the second layer is left uncontrolled, and so cooperation might be disfavored in layer two. We find that in all six empirical two-layer networks, and for nearly all initial configurations, a proper choice of benefits and costs in layer two can serve to lower the critical benefit-to-cost ratio required for the evolution of cooperation in layer one (Fig.~\ref{fig:8}\textbf{b}).

The effect size of one layer on another can be substantial. In the case of the empirical networks of social and professional interactions in a Zambian tailor shop, for example, if interactions occur in a single layer (social interactions only), then the benefit-to-cost ratio required for cooperation to spread is unreasonably large: $(b_1/c)^*=93.3$. And yet, when behavior is coupled with professional interactions, by setting $b_2/c=30$ the benefit-to-cost ratio to favor cooperation in social interactions is dramatically reduced to $(b_1/c)=53.6$; at the same time the fixation probability of cooperation in that layer is increased by $135.2\%$ relative to neutrality (for selection intensity $\delta=0.2$), which is a measure of the effect size of coupling.

Remarkably, the critical benefit-to-cost ratio in layer one can sometimes be reduced to zero by coupling to a second layer (Supplementary Fig.~8), which indicates that cooperation can be favored in layer one despite providing no 
immediate benefit in that domain at all. This dramatic effect of coupling occurs 
for more than $25\%$ initial configurations in the six empirical networks. The spatial arrangement of cooperators strongly affects whether the required benefit-to-cost ratio can be reduced all the way to zero by coupling. In general, the closer two initial cooperators, one in each layer, the more likely that coupling can catalyze cooperation in layer one even without providing any immediate layer-one benefit (Supplementary Fig.~9). Aside from analyzing six empirical networks, we also illustrate this phenomenon in two-layer random networks with different degree distributions (Supplementary Figs.~10 and 11).
So far, we have assumed that individuals in each layer use averaged (edge-weighted) payoffs.
We find similar, cooperation-promoting effects of coupling layers when payoffs are accumulated across interactions (see Supplementary Information section 2.1.7).

\section{Discussion}

One of the many complexities of human societies is the structure of our social interactions. Structure is not confined to a single type of interaction, but includes the distinct domains of relationships in which we interact. This feature would not complicate the problem of understanding behavior if interactions and standing in one domain had no influence on other domains. But that is emphatically not the case. A person with a large online following, for example, can leverage this for success and appeal in professional relationships; and someone with success in business can garner support in politics or even religion. The empirical impact of coupling between domains can be dramatic, as exemplified by the famous Medici family of renaissance Florence \cite{Padgett1993}, but also in modern times. Understanding coupling between domains of social interaction is therefore critical to understanding what drives prosocial and selfish behavior in societies.

We have modelled the evolution of prosocial behaviors across domains using multilayer networks, where each individual uses separate strategies and plays distinct games in different layers. An individual's total payoff across domains determines his or her influence over peers. We find that the threshold for selection to favor cooperation in a multilayer population can be much lower than it is in a single-layer population \cite{2006-Ohtsuki-p502-505,2017-Allen-p227-230}. For a large portion of multilayer populations, coupling can promote cooperation in all layers, even when cooperation is disfavored in each layer alone. 
And so the prospects for cooperation are fundamentally changed when social interactions occur in distinct, but coupled, domains.

Our work has several potential implications for the evolution of prosocial behavior. The first noteworthy implication is that coupling between layers can often facilitate cooperation by proper coordination of the benefit-to-cost ratios between the two layers (equation~(\ref{eq:db_formula_main})). In practice, the benefit-to-cost ratio required for cooperation to spread in a single-layer network may be unreasonably large, as exemplified by the social interaction network measured in a Zambian tailor shop. But when coupled to the layer of professional interactions (layer two), an appropriate choice of the benefit-cost ratio in layer two can reduce the ratio required to support cooperation in layer one by as much as 40\%, while also increases the probability that cooperation fixes in layer one by over 130\%. More generally, we find that in up to $40\%$ of the two-layer networks we examined, cooperation can be favored in layer one even when there is no immediate benefit of cooperation in that layer ($b_1/c$ near zero), provided the benefits in layer two are sufficiently large.

Another potential implication concerns how interactions may be engineered or modified in one domain in order to promote cooperation in another, or in both. 
Indeed, not every multilayer structure is beneficial for cooperation; and even if the structure can favor cooperation, the benefit-to-cost ratio required may be unreasonably large.
So one can ask whether it is possible to slightly modify interactions in one layer to promote cooperation in both layers. 
Although modifying in-person interactions may be unfeasible, online interactions are often amenable to oversight or control.
Although this question is quite deep and difficult for full mathematical analysis, we have analyzed it systematically in all two-layer networks of size $6$ (See Supplementary Fig.~12). In these cases we find that adding or severing a small number of connections in one layer, if chosen properly, can rescue cooperation in both layers (see Supplementary Fig.~12 for intuition). Investigating this question in greater generality is a worthwhile avenue for future study.

Several prior studies have demonstrated that selection cannot favor cooperation in a single-layer structured population under birth-death or pairwise-comparison updating \cite{2006-Ohtsuki-p502-505,taylor:Nature:2007,traulsen:JTB:2007,allen:NC:2019}. More recent studies have found that game transitions \cite{Su2019pnas} and heterogeneous distributions of social goods \cite{mcavoy-2020-nhb} can catalyze cooperation under these update rules. Here, too, we find that a simple coupling of layers works efficiently to make cooperation favored by selection under birth-death or pairwise-comparison updating (see Supplementary Fig.~13). In practice, there may be considerable cultural differences between social domains, and it is not unreasonable to expect that the mechanisms of imitation and learning differ between layers. The multilayer approach allows for such a mixture of update rules in different layers (see Supplementary Information section 2.1).

As our aim has been to analyze multilayer populations in a mathematically rigorous manner, our study has several limitations. Since the population structures are fixed as traits evolve, there is an implicit assumption that networks change much more slowly than behaviors. Although this is a common assumption in the literature, it does exclude interesting cases involving dynamic topologies. Our analysis also requires weak selection. Stronger selection can complicate the formal analysis of evolutionary models in structured populations \cite{2015_lbsen_pnas}, but it is nonetheless an important aspect of natural populations and should be considered in future models of multilayer populations. The method we have employed for weak selection is computationally feasible for populations of moderate size, but calculations become more cumbersome in large populations (at least when allowing for arbitrarily complicated network topologies).
Generally, for an $L$-layer network of size $N$, the complexity of computing fixation probabilities is bounded by solving a linear system of size $O\left(L^2N^2\right)$. Furthermore, our metric for evolutionary success, fixation probability, is a long-term measure and does not capture the timescale of evolutionary processes as the population sojourns through transient states. Fixation probabilities themselves are relevant only when mutations appear sufficiently infrequently, which may or may not be true--especially in settings of cultural evolution in which ``mutation'' is interpreted as ``exploration.'' So while our analysis reveals many interesting properties of multilayer populations, we view this area as fertile ground for future theoretical investigations.

\section{Methods}
Here we briefly summarize our theoretical results on weak selection in multilayer populations, and we refer to Supplementary Information section 1 for detailed derivations. We consider a population structure described by a two-layer network of size $N$, with edge weights $w_{ij}^{[1]}$ in layer one and $w_{ij}^{[2]}$ in layer two. All edges are symmetric, i.e.~$w_{ij}^{[1]}=w_{ji}^{[1]}$ and $w_{ij}^{[2]}=w_{ji}^{[2]}$, and self loops are not allowed. The weighted degree of node $i$ is $w_i^{[1]}=\sum_{j=1}^N w_{ij}^{[1]}$ in layer one and $w_i^{[2]}=\sum_{j=1}^N w_{ij}^{[2]}$ in layer two. The relative weighted degree of node $i$ is thus $\pi_i^{[1]}=w_i^{[1]}/\sum_{j=1}^N w_{j}^{[1]}$ in layer one and $\pi_i^{[2]}=w_i^{[2]}/\sum_{j=1}^N w_{j}^{[2]}$ in layer two. Under death-birth updating, the relative weighted degree of $i$ in a given layer corresponds to the so-called reproductive value of $i$ in that layer \cite{taylor:AN:1990,Maciejewski-2014-jtb,2019-Allen-p1147-1210}, which represents the contribution of $i$ to future generations, in the absence of selection.

The evolutionary dynamics of death-birth updating in network-structured populations can be described in terms of random walks on networks \cite{2017-Allen-p227-230}. Here, too, random walks come into play, but since we are dealing with multilayer networks we need to be clear about their definitions. In a two-layer network, we define a random walk as follows. In layer one (resp. two), starting at node $i$, a one-step walk terminates at node $j$ with probability $p_{ij}^{[1]}=w_{ij}^{[1]}/w_{i}^{[1]}$ (resp. $p_{ij}^{[2]}=w_{ij}^{[2]}/w_{i}^{[2]})$. Let $\left(p^{[1]}\right)_{ij}^{(n)}$ denote the probability that a walker starting at node $i$ terminates at node $j$ after an $n$-step random walk in layer one. We define an $\left(n,m\right)$-step random walk to be an $n$-step walk in layer one followed by an $m$-step walk in layer two, where the beginning of the second random walk corresponds to the end of the first. Let $\left(p^{[1,2]}\right)_{ij}^{(n,m)}$ denote the probability that a walker starting at node $i$ terminates at node $j$ after an $\left(n,m\right)$-step walk.

The effects of selection depend on the assortment of strategies within the network. In a two-layer network, the spatial assortment involves not only strategies within the same layer but also those in the other layer. Let $\beta_{ij}$ denote the probability that, in layer one, both nodes $i$ and $j$ are cooperators under neutral drift. Similarly, let $\gamma_{ij}$ be the probability that both nodes $i$ in layer one and node $j$ in layer two are cooperators. When $i=j$, we let $\beta_{i}$ denote $\beta_{ij}$ and $\gamma_{i}$ denote $\gamma_{ij}$. For a formal mathematical description of the underlying distribution, see Supplementary Information section 1.

If $\bm{\xi}$ is any initial strategy configuration, then $\xi_i^{[L]}$ denotes is the strategy of node $i$ in layer $L$. The quantity then $\widehat{\bm{\xi}}^{\left[L\right]}=\sum_{i=1}^N \pi_i^{[L]}\xi_i^{[L]}$ represents the fixation probability of cooperators in layer $L$ under neutral drift ($\delta =0$) \cite{2019-Allen-p1147-1210}. In Supplementary Information section 1, we show that one can obtain $\beta_{ij}$ and $\gamma_{ij}$ by solving the following linear system of equations,
\begin{align}
	\begin{cases}
		\beta_{ij} =& \frac{N}{2}\left(\xi_{i}^{\left[1\right]}\xi_{j}^{\left[1\right]} - \widehat{\bm{\xi}}^{\left[1\right]}\right) 
		+ \frac{1}{2} \sum_{k=1}^{N} p_{ik}^{[1]} \beta_{kj} + \frac{1}{2} \sum_{k=1}^{N} p_{jk}^{[1]} \beta_{ik}, 
		\\  \\
		\beta_{i} =& N\left(\xi_{i}^{\left[1\right]} - \widehat{\bm{\xi}}^{\left[1\right]}\right) + \sum_{k=1}^{N} p_{ik}^{\left[1\right]} \beta_{k}, 
		\\  \\
		\gamma_{ij} =& \frac{N^{2}}{2N-1}\left(\xi_{i}^{\left[1\right]}\xi_{j}^{\left[2\right]} - \widehat{\bm{\xi}}^{\left[1\right]}\widehat{\bm{\xi}}^{\left[2\right]}\right) + \frac{1}{2N-1}\sum_{k_{1},k_{2}=1}^{N} p_{ik_{1}}^{\left[1\right]}p_{jk_{2}}^{\left[2\right]}\gamma_{k_{1}k_{2}} \\
		& +\frac{N-1}{2N-1}\sum_{k_{1}=1}^{N} p_{ik_{1}}^{\left[1\right]} \gamma_{k_{1}j} +\frac{N-1}{2N-1} \sum_{k_{2}=1}^{N} p_{jk_{2}}^{\left[2\right]} \gamma_{ik_{2}} ,
	\end{cases}
\end{align}
together with the additional constraints $\sum_{i=1}^N \pi_i^{[1]}\beta_i=0$ and $\sum_{i=1}^N \pi_{i}^{[1]}\gamma_{i}=0$.

Using these quantities, we let $\theta_n=\sum_{i,j=1}^N \pi_i^{[1]}\left(p^{[1]}\right)_{ij}^{(n)}\beta_{ij}$, which means the probability that both the starting and the ending nodes of an $n$-step random walk in layer one are cooperators, where the starting node $i$ is selected based on the reproductive value, $\pi_{i}^{\left[1\right]}$.
Analogously, for the inter-layer random walk defined previously, we let $\phi_{n,m}=\sum_{i,j=1}^N \pi_i^{[1]}\left(p^{[1,2]}\right)_{ij}^{(n,m)}\gamma_{ij}$. This quantity represents the probability that the beginning of the walk in layer one and the end of the walk in layer two both correspond to cooperators. Substituting $\theta_n$ and $\phi_{n,m}$ into equation~(\ref{eq:db_formula_main}) then gives the condition for selection to favor cooperation. In Supplementary Information section 2.2.2, we give examples illustrating how one can use network symmetry to obtain explicit expressions for these quantities in simple multilayer populations. For general multilayer networks, we also provide code for determining $\theta$, $\phi$, and evaluating equation~(\ref{eq:db_formula_main}).

\subsection{Rule for evolutionary dynamics in a two-layer ring network} \label{two_layer_circle}
We now consider an example on a two-layer ring network, where \emph{(i)} in each layer, a node is connected to two other nodes; and \emph{(ii)} node $i$ is connected to $j$ in layer one if and only if $i$'s associated node is connected to $j$'s associated node in layer two (see Fig.~\ref{fig:3}\textbf{a}). We study the initial strategy configuration of a single mutant cooperator in each layer. Let $d$ be the shortest distance between these two cooperator nodes. That is, if $i$ is a cooperator in layer one and $j$ is a cooperator in layer two, then $d$ is the length of the shortest path from $i$ to $j$ on the ring. When a node in layer one and its associated node in layer two are cooperators, $d=0$. The configuration shown in Fig.~\ref{fig:3}\textbf{a} is an example with $d=1$.

We find that cooperation is favored in the two-layer ring network only if equation~(\ref{eq:db_formula_main}) holds, where $\theta_1=-\left(N-1\right) /2$, $\theta_2=-\left(N-2\right) /2$, $\theta_3=-3\left(N-2\right) /4$, 
\begin{equation}
	\phi_{0,1}=-\sum_{\ell=1}^{N-1}\frac{\cos\frac{2\pi \ell d}{N}}{2N-1+\cos \frac{2\pi \ell}{N}},
\end{equation}
\begin{align}
	\phi_{2,0} &= 
	\begin{cases}
		-2\left(N-1\right)\phi_{0,1}-N+1 & d=0, \\
		& \\
		-2\left(N-1\right)\phi_{0,1}+1 & d\geqslant 1 ,
	\end{cases}
\end{align}
and 
\begin{align}
	\phi_{2,1} &= 
	\begin{cases}
		\left(4N^2-6N+3\right)\phi_{0,1}+2N^2-4N+3 & d=0 , 
		\\ & \\
		\left(4N^2-6N+3\right)\phi_{0,1}-\frac{5}{2}N+3 & d=1 , \\ & \\
		\left(4N^2-6N+3\right)\phi_{0,1}-2N+3 & d\geqslant 2.
	\end{cases}
\end{align}

\subsection{Small multilayer populations}\label{definition_non_isomorphic}
When mutant appearance is stochastic, the average fixation probability is used to measure which spatial structure facilitates cooperation. For example, many prior studies have relied on the assumption that a mutant cooperator appears in every node with the equal probability. By averaging over all initial locations with respect to a fixed mutant-appearance distribution, the remaining variables are population structure and the update rule. In addition to these two components, we also consider a more fine-grained approach that takes into account the mutants' initial positions within the population. In other words, we study the effects of spatial structure, update rule, and the initial strategy configuration on evolutionary dynamics \cite{2016-Chen-p39181-39181,McAvoy2021}.

We call the combination of a population structure and a mutant configuration a ``profile.'' In a single-layer network, two profiles $G$ and $H$ are isomorphic if there is a bijection $f:V\left(G\right)\rightarrow V\left(H\right)$ between the node sets of $G$ and $H$ such that \emph{(i)} any two nodes $i$ and $j$ of $G$ are adjacent if and only if $f\left(i\right)$ and $f\left(j\right)$ are adjacent in $H$; and \emph{(ii)} strategies of any node $u$ of $G$ and $f\left(u\right)$ of $H$ are identical. Otherwise, the two profiles are non-isomorphic (see examples in Supplementary Fig.~14).

Similarly, a pair of two-layer profiles $G$ and $H$ are isomorphic if there is a bijection $f:V\left(G\right)\rightarrow V\left(H\right)$ between the node sets of $G$ and $H$ such that \emph{(i)} in each layer, any two nodes $i$ and $j$ of $G$ are adjacent if and only if in the same layer $f\left(i\right)$ and $f\left(j\right)$ of $H$ are adjacent ; and \emph{(ii)} in each layer, the state of any node $u$ of $G$ and $f\left(u\right)$ of $H$ are identical. Otherwise, the two profiles are non-isomorphic. Supplementary Table 1 shows the number of non-isomorphic single-layer and non-isomorphic two-layer profiles for networks of size $N=3,4,5,6$. 
Note that the network in each layer is required to be connected.
The total number of non-isomorphic profiles is far greater for two-layer networks than single-layer ones. For example, for $N=3$ there are $26$ non-isomorphic two-layer profiles compared to $3$ such single-layer profiles; and for $N=6$ there are $36,394,472$ non-isomorphic two-layer profiles compared to 407 such single-layer profiles.

We analyze all non-isomorphic single-layer profiles for $N=3,4,5,6$ to obtain the proportion of profiles in which cooperation can be favored for some $b_1/c>0$ (or equivalently, the critical benefit-to-cost satisfies $0<(b_1/c)^*<\infty$) (see blue bars in Fig.~\ref{fig:6}).
When randomly choosing two single-layer profiles, for $N=6$, there are $407\times 407=165,649$ combinations. We take one as layer one and another as layer two. Since there are many ways for a node in layer one to correspond to a node in layer two (i.e. a multilayer ``superposition''), each combination can actually produce many two-layer non-isomorphic profiles. Assuming that such a combination generates $X$ two-layer non-isomorphic profiles, and of them $Y$ profiles make cooperation favored for some positive $b_1/c$ and $b_2/c$ (or equivalently, the region $(b_1/c,b_2/c)$ constrained by Eq.~(\ref{eq:db_formula_main}) partially overlaps with the first quadrant), we say coupling such two single-layer profiles makes cooperation favored with probability $Y/X$. Analyzing all such combinations, we obtain the proportion of couplings of a single-layer profile to a random single-layer profile that favor cooperation in both layers (see red bar in Fig.~\ref{fig:6} and Supplementary Table 2).

\subsection*{Data Availability}
All the network datasets used in this paper are freely and publicly available in the Colorado Index of Complex Networks (ICON) collection at https://icon.colorado.edu.

\subsection*{Code Availability}
Custom code is available at https://github.com/qisu1991/MultilayerPopulations.

\section*{Acknowledgements}
We thank Erol Ak\c{c}ay for helpful comments. 
This work is supported by the Simons Foundation (Math+X Grant to the University of Pennsylvania), the National Science Foundation (grants DMS-1907583, 2042144), and The David \& Lucile Packard Foundation (J.B.P.).

\newpage

\begin{center}
\Large\textbf{Supporting Information}
\end{center}

\setcounter{equation}{0}
\setcounter{figure}{0}
\setcounter{section}{0}
\setcounter{table}{0}
\renewcommand{\thesection}{SI.\arabic{section}}
\renewcommand{\thesubsection}{SI.\arabic{section}.\arabic{subsection}}
\renewcommand{\theequation}{SI.\arabic{equation}}
\renewcommand{\thetable}{SI.\arabic{table}}
\renewcommand{\figurename}{\footnotesize Supplementary~Figure}
\renewcommand{\tablename}{\footnotesize Supplementary~Table}

\section{Supplementary Methods} \label{sec:modelingFramework}

\subsection{Fixation probabilities under weak selection}
We begin with populations with two layers. In each time step, we choose a replacement event, $\left(\bm{R},\bm{\alpha}\right)$, which consists of a pair $\left(R^{\left[L\right]} ,\alpha^{\left[L\right]}\right)$ for each layer, $L$, where $\alpha^{\left[L\right]}:R^{\left[L\right]}\rightarrow\left\{1,\dots ,N\right\}$ is the offspring-to-parent map in layer $L$. We denote by $p_{\left(\bm{R},\bm{\alpha}\right)}\left(\vx\right)$ the probability of choosing $\left(\bm{R},\bm{\alpha}\right)$ in state $\vx\in\left\{0,1\right\}^{N}\times\left\{0,1\right\}^{N}$, where $x_{i}^{\left[L\right]}$ is $1$ if individual $i$ in layer $L$ has type $A$ and $0$ otherwise. We assume that $p_{\left(\bm{R},\bm{\alpha}\right)}\left(\vx\right)$ is a smooth function of $\delta$ in a small neighborhood of $\delta =0$ for every $\vx\in\left\{0,1\right\}^{N}\times\left\{0,1\right\}^{N}$ and every replacement event $\left(\bm{R},\bm{\alpha}\right)$.

In each layer, $L$, the map $\alpha^{\left[L\right]}:R^{\left[L\right]}\rightarrow\left\{1,\dots ,N\right\}$ extends to a map $\widetilde{\alpha}^{\left[L\right]}:\left\{1,\dots ,N\right\}\rightarrow\left\{1,\dots ,N\right\}$ defined by $\widetilde{\alpha}^{\left[L\right]}\left(i\right) =\alpha^{\left[L\right]}\left(i\right)$ if $i\in R^{\left[L\right]}$ and $\widetilde{\alpha}^{\left[L\right]}\left(i\right) =i$ if $i\not\in R^{\left[L\right]}$. For any state $\vx\in\left\{0,1\right\}^{N}\times\left\{0,1\right\}^{N}$, we write $\vx^{\left[L\right]}\in\left\{0,1\right\}^{N}$ for the state of the population in layer $L\in\left\{1,2\right\}$. This extension of $\bm{\alpha}$, denoted $\widetilde{\bm{\alpha}}$, gives an updated state $\vx_{\widetilde{\bm{\alpha}}}\in\left\{0,1\right\}^{N}\times\left\{0,1\right\}^{N}$ defined by $\left(\vx_{\widetilde{\bm{\alpha}}}\right)^{\left[L\right]}_{i}=x^{[L]}_{\widetilde{\alpha}^{\left[L\right]}\left(i\right)}$. 
For $\vx ,\vy\in\left\{0,1\right\}^{N}\times\left\{0,1\right\}^{N}$, we then have the transition probability
\begin{align}
	P_{\vx\rightarrow\vy} &= \sum_{\substack{\left(\bm{R},\bm{\alpha}\right) \\ \vx_{\widetilde{\bm{\alpha}}}=\vy}} p_{\left(\bm{R},\bm{\alpha}\right)}\left(\vx\right) . \label{eq:markovTransitions}
\end{align}

In addition to being smooth, we assume that the replacement rule satisfies the following:
\begin{fixation}
	There exists $\left(i_1,i_2\right)\in\left\{1,\dots ,N\right\}\times\left\{1,\dots ,N\right\}$, an integer $m\geqslant 1$, and a sequence of replacement events $\left\{\left(\bm{R}_{k},\bm{\alpha}_{k}\right)\right\}_{k=1}^{m}$ such that \emph{(i)} $p_{\left(\bm{R}_{k},\bm{\alpha}_{k}\right)}\left(\vx\right) >0$ for every $k\in\left\{1,\dots ,m\right\}$ and $\vx\in\left\{0,1\right\}^{N}\times\left\{0,1\right\}^{N}$; \emph{(ii)} for each $L\in\left\{1,2\right\}$, there exists $k_L$ such that $i_L\in R_{k_L}^{[L]}$; and \emph{(iii)} for $L\in\left\{1,2\right\}$, we have $\widetilde{\alpha}_{1}^{\left[L\right]}\circ\widetilde{\alpha}_{2}^{\left[L\right]}\circ\cdots\circ\widetilde{\alpha}_{m}^{\left[L\right]}\left(j\right) =i_{L}$ for every $j\in\left\{1,\dots ,N\right\}$.
\end{fixation}
\noindent As a consequence, at least a pair of locations (one in layer one and the other in layer two) can spread their genetic material throughout the rest of the population within the same layer. 
The Markov chain defined by \eq{markovTransitions} has four absorbing states, $\vA\vA$, $\vA\vB$, $\vB\vA$, and $\vB\vB$.
The first part represents the absorbing state in layer one and the second part in layer two, where  $\mathbf{A}$ ($\mathbf{B}$) corresponds to all-$A$ (all-$B$) in each layer.
As a result of the Fixation Axiom, all non-absorbing states are transient, and eventually the process must reach an absorbing state.

In our analysis, it will be convenient to work with a process having a unique stationary distribution. To do so, we consider a mutation-modified chain obtained by sending each absorbing state to a fixed, transient state $\bm{\xi}$ with probability $u>0$. This chain has transition probabilities
\begin{align}
	P_{\vx\rightarrow\vy}^{\MSS} &= 
	\begin{cases}
		u & \vx\in\left\{\vA\vA ,\vA\vB ,\vB\vA ,\vB\vB\right\} ,\ \vy = \bm{\xi} , \\
		& \\
		\left(1-u\right) P_{\vx\rightarrow\vy} & \vx\in\left\{\vA\vA ,\vA\vB ,\vB\vA ,\vB\vB\right\} ,\ \vy\neq\bm{\xi} , \\
		& \\
		P_{\vx\rightarrow\vy} & \vx\not\in\left\{\vA\vA ,\vA\vB ,\vB\vA ,\vB\vB\right\} ,
	\end{cases} \label{eq:MSS_chain}
\end{align}
where $\MSS$ indicates that $\bm{\xi}$ is regenerated. By the Fixation Axiom, this chain has a unique closed communication class and thus a unique stationary distribution, which we denote by $\pi_{\MSS}$.

The marginal probability that $i$ transmits its offspring to $j$ in layer $L$ and in state $\vx$ is
\begin{align}
	e_{ij}^{\left[L\right]}\left(\vx\right) &\coloneqq \sum_{\substack{\left(\bm{R},\bm{\alpha}\right) \\ \alpha^{\left[L\right]}\left(j\right) =i}} p_{\left(\bm{R},\bm{\alpha}\right)}\left(\vx\right) . \label{eq:marginalTransmission}
\end{align}
Let $e_{ij}^{\circ\left[L\right]}$ denote the marginal transmission probability from $i$ to $j$ (\eq{marginalTransmission}) under neutral drift ($\delta =0$), which we assume is independent of the state, $\vx$. The reproductive value (RV) of $i$ \cite{fisher-1930-clarendon,Maciejewski-2014-jtb,2019-Allen-p1147-1210} in layer $L$, denoted $\pi_{i}^{\left[L\right]}$, is (uniquely) defined by the system of equations
\begin{subequations}
	\begin{align}
		\sum_{j=1}^{N} e_{ij}^{\circ\left[L\right]} \pi_{j}^{\left[L\right]} &= \pi_{i}^{\left[L\right]} \sum_{j=1}^{N} e_{ji}^{\circ\left[L\right]} ; \\
		\sum_{i=1}^{N} \pi_{i}^{\left[L\right]} &= 1.
	\end{align}
\end{subequations}
Informally, this system says that the loss of ``value'' due to the death of $i$ is offset by the ``value'' $i$ propagates throughout the population due to reproduction. This distribution on $\left\{1,\dots ,N\right\}$ is convenient in what follows because the RV-weighted frequency of $A$ in layer $L$, denoted by $\widehat{\bm{\vx}}^{\left[L\right]}\coloneqq\sum_{i=1}^{N}\pi_{i}^{\left[L\right]}x_{i}^{\left[L\right]}$ for $\vx\in\left\{0,1\right\}^{N}\times\left\{0,1\right\}^{N}$, is a martingale under neutral drift \cite{2019-Allen-p1147-1210}.

In \eq{MSS_chain}, the change in RV-weighted frequency of $A$ in layer $L$ due to selection is
\begin{align}
	\delhatsel^{\left[L\right]}\left(\vx\right) &= \sum_{i,j=1}^{N} \pi_{i}^{\left[L\right]} \left( x_{j}^{\left[L\right]} - x_{i}^{\left[L\right]} \right) e_{ji}^{\left[L\right]}\left(\vx\right) .
\end{align}
Note that $\widehat{\triangle}_{\mathrm{sel}}^{[L]}(\vx) =0$ for $\vx\in\left\{\vA\vA ,\vA\vB ,\vB\vA ,\vB\vB\right\}$.
The overall change in RV-weighted frequency of $A$ in layer $L$ (due to selection or mutation) is
\begin{align}
	\delhat^{\left[L\right]}\left(\vx\right) &= 
	\begin{cases}
		-u\left(1-\widehat{\xi}^{\left[L\right]}\right) & \vx^{\left[L\right]} = \vA ,\ \vx^{\left[-L\right]}\in\left\{\vA ,\vB\right\} , \\
		& \\
		u\widehat{\xi}^{\left[L\right]} & \vx^{\left[L\right]} = \vB ,\ \vx^{\left[-L\right]}\in\left\{\vA ,\vB\right\} , \\
		& \\
		\delhatsel^{\left[L\right]}\left(\vx\right) & \vx\not\in\left\{\vA\vA ,\vA\vB ,\vB\vA ,\vB\vB\right\} ,
	\end{cases}
\end{align}
where $-L$ indicates the other layer (i.e. not layer $L$). Since $\E_{\MSS}\left[\delhat^{\left[L\right]}\right] =0$, taking $L=1$ for example,
we have
\begin{align}
	0 = \E_{\MSS}\left[\delhat^{\left[1\right]}\right] 
	&= \E_{\MSS}\left[\delhatsel^{\left[1\right]}\right] -u\pi_{\MSS}\left(\left\{\vA\right\}\times\left\{\vA ,\vB\right\}\right)\left(1-\widehat{\xi}^{\left[1\right]}\right) \nonumber \\
	&\quad + u\pi_{\MSS}\left(\left\{\vB\right\}\times\left\{\vA ,\vB\right\}\right)\widehat{\xi}^{\left[1\right]} ,
\end{align}
which gives
\begin{align}
	\E_{\MSS}\left[\delhatsel^{\left[L\right]}\right] &= 
	\begin{cases}
		\substack{u\pi_{\MSS}\left(\left\{\vA\right\}\times\left\{\vA ,\vB\right\}\right)\left(1-\widehat{\xi}^{\left[L\right]}\right) \\ - u\pi_{\MSS}\left(\left\{\vB\right\}\times\left\{\vA ,\vB\right\}\right)\widehat{\xi}^{\left[L\right]}} & L = 1 , \\
		& \\
		\substack{u\pi_{\MSS}\left(\left\{\vA ,\vB\right\}\times\left\{\vA\right\}\right)\left(1-\widehat{\xi}^{\left[L\right]}\right) \\ - u\pi_{\MSS}\left(\left\{\vA ,\vB\right\}\times\left\{\vB\right\}\right)\widehat{\xi}^{\left[L\right]}} & L = 2,
	\end{cases} \label{eq:delhatselMSS}
\end{align}
where $\left\{\vA\right\}\times\left\{\vA ,\vB\right\}$ indicates all cases that the state in layer one lies in $\left\{\vA\right\}$ and the state in layer two lies in $\left\{\vA ,\vB\right\}$ (analogous indications for other expressions). 
Let $ \rho_{A}^{\left[L\right]}\left(\bm{\xi}\right)$ denote the probability that the system becomes all-$A$ in layer $L$, starting from $\bm\xi$.
By standard results on rare-mutation evolutionary dynamics \cite{2006-Fudenberg-p251-262,McAvoy2021}, we have
\begin{subequations}\label{eq:lowMutationLimit}
	\begin{align}
		\lim_{u\rightarrow 0} \pi_{\MSS}\left(\left\{\vA\right\}\times\left\{\vA ,\vB\right\}\right) &= \rho_{A}^{\left[1\right]}\left(\bm{\xi}\right) ; \\
		\lim_{u\rightarrow 0} \pi_{\MSS}\left(\left\{\vB\right\}\times\left\{\vA ,\vB\right\}\right) &= 1-\rho_{A}^{\left[1\right]}\left(\bm{\xi}\right) ; \\
		\lim_{u\rightarrow 0} \pi_{\MSS}\left(\left\{\vA ,\vB\right\}\times\left\{\vA\right\}\right) &= \rho_{A}^{\left[2\right]}\left(\bm{\xi}\right) ; \\
		\lim_{u\rightarrow 0} \pi_{\MSS}\left(\left\{\vA ,\vB\right\}\times\left\{\vB\right\}\right) &= 1-\rho_{A}^{\left[2\right]}\left(\bm{\xi}\right) .
	\end{align}
\end{subequations}
Differentiating \eq{delhatselMSS} with respect to $u$ at $u=0$ (and using \eq{lowMutationLimit}) gives
\begin{align} \label{eq:rhoA_first}
	\rho_{A}^{\left[L\right]}\left(\bm{\xi}\right) &= \widehat{\xi}^{\left[L\right]} + \frac{d}{du}\Bigg\vert_{u=0}\E_{\MSS}\left[\delhatsel^{\left[L\right]}\right] .
\end{align}
Since the transition functions are smooth in $\delta$ and $u$, we have
\begin{align}
	\frac{d}{d\delta}\Bigg\vert_{\delta =0} \rho_{A}^{\left[L\right]}\left(\bm{\xi}\right) &= \frac{d}{du}\Bigg\vert_{u=0}\E_{\MSS}^{\circ}\left[\frac{d}{d\delta}\Bigg\vert_{\delta =0}\delhatsel^{\left[L\right]}\right] \label{eq:rhoAformula}
\end{align}
(see Proposition $4$ in paper \cite{2019-Allen-p1147-1210}). This result is a two-layer generalization of Theorem 1 in paper \cite{McAvoy2021}.

Finally, for $\vx\not\in\left\{\vA\vA ,\vA\vB ,\vB\vA ,\vB\vB\right\}$, consider the rare-mutation conditional distribution
\begin{align}
	\pi_{\RMC}\left(\vx\right) &\coloneqq K \frac{d}{du}\Big\vert_{u=0}\pi_{\MSS}\left(\vx\right) ,
\end{align}
where $K\coloneqq\left(\sum_{\vy\not\in\left\{\vA\vA ,\vA\vB ,\vB\vA ,\vB\vB\right\}}\frac{d}{du}\Big\vert_{u=0}\pi_{\MSS}\left(\vy\right)\right)^{-1}$. The normalizing factor ensures that $\pi_{\RMC}$ is a probability distribution on the non-absorbing states. This distribution satisfies:

\begin{lemma}\label{lem:stateFunction}
	For any state function $\varphi :\left\{0,1\right\}^{N}\times\left\{0,1\right\}^{N}\rightarrow\mathbb{R}$,
\begin{align}
	\mathbb{E}_{\RMC}^{\circ}\left[\varphi\right] 
	&= K^{\circ}\left(\substack{
		\varphi\left(\bm{\xi}\right) 
		- \widehat{\bm{\xi}}^{\left[1\right]}\widehat{\bm{\xi}}^{\left[2\right]}\varphi\left(\vA ,\vA\right) 
		-\widehat{\bm{\xi}}^{\left[1\right]}\left(1-\widehat{\bm{\xi}}^{\left[2\right]}\right)\varphi
		\left(\vA ,\vB\right) \\ 
		-\left(1-\widehat{\bm{\xi}}^{\left[1\right]}\right)\widehat{\bm{\xi}}^{\left[2\right]}
		\varphi\left(\vB ,\vA\right) -\left(1-\widehat{\bm{\xi}}^{\left[1\right]}\right)\left(1-\widehat{\bm{\xi}}^{\left[2\right]}\right)\varphi\left(\vB ,\vB\right)}\right) \nonumber \\
	&\qquad + \sum_{\left(\bm{R},\bm{\alpha}\right)} p_{\left(\bm{R},\bm{\alpha}\right)}^{\circ} \mathbb{E}_{\RMC}^{\circ}\left[\varphi_{\widetilde{\bm{\alpha}}}\right] .
\end{align}
\end{lemma}
\noindent The proof of this result is a straightforward adaptation of that of Lemma 1 in paper \cite{McAvoy2021} (details omitted).
Note that the state function $\varphi\left(\vx\right)$ there is replaced with
$\varphi\left(\bm{\xi}\right) 
- \widehat{\bm{\xi}}^{\left[1\right]}\widehat{\bm{\xi}}^{\left[2\right]}\varphi\left(\vA ,\vA\right) \\
-\widehat{\bm{\xi}}^{\left[1\right]}\left(1-\widehat{\bm{\xi}}^{\left[2\right]}\right)\varphi\left(\vA ,\vB\right)  
-\left(1-\widehat{\bm{\xi}}^{\left[1\right]}\right)\widehat{\bm{\xi}}^{\left[2\right]}\varphi\left(\vB ,\vA\right) -\left(1-\widehat{\bm{\xi}}^{\left[1\right]}\right)\left(1-\widehat{\bm{\xi}}^{\left[2\right]}\right)\varphi\left(\vB ,\vB\right)$.

\subsection{Selection condition for social goods}
Suppose that $B^{\left[L\right]}$ and $C^{\left[L\right]}$ are matrices, with $B_{ij}^{\left[L\right]}$ representing the benefit type $A$ at location $i$ provides to location $j$ (both in layer $L$). $C_{ij}^{\left[L\right]}$ is the corresponding cost (to $i$) for providing $B_{ij}^{\left[L\right]}$ to $j$. Type $B$ provides no benefits and incurs no costs. This formulation of payoffs allows for arbitrary ``social goods'' \citep{mcavoy-2020-nhb}, although for the sake of analysis we focus primarily on the case in which $B_{ij}^{\left[L\right]}=b_Lp_{ji}^{[L]}$ and $C_{ij}^{\left[L\right]}=c_Lp_{ij}^{[L]}$. In state $\vx\in\left\{0,1\right\}^{N}\times\left\{0,1\right\}^{N}$, the total payoff to individual at position $k$ due to layer $L$ is 
\begin{align}
u_{k}^{\left[L\right]}\left(\vx\right) &= \sum_{\ell =1}^{N} \left( -x_{k}^{\left[L\right]} C_{k\ell}^{\left[L\right]} + x_{\ell}^{\left[L\right]} B_{\ell k}^{\left[L\right]} \right) .
\end{align}

We assume that the probability that $i$ replaces $j$ in layer $L$, $e_{ij}^{\left[L\right]}$, depends on a vector $\mathbf{F}\in\left[0,\infty\right)^{N}$ that gives the fecundity values of the population. In traditional formulations, $F_{k}=\exp\left\{\delta u_{k}\left(\vx\right)\right\}$, where $\delta$ is the selection intensity and $u_{k}\left(\vx\right)$ is the payoff to individual $k$ \cite{2014-Maciejewski-p1003567-1003567}. For multilayer populations, the total payoff to $k$ is $u_{k}\left(\vx\right) =u_{k}^{\left[1\right]}\left(\vx\right) + u_{k}^{\left[2\right]}\left(\vx\right)$. Letting $m_{k;ij}^{\left[L\right]}$ be the marginal effect of individual $k$'s fitness on the probability that $i$ replaces $j$ in layer $L$ \cite{mcavoy-2020-nhb}, i.e.
\begin{align}
m_{k;ij}^{\left[L\right]} &\coloneqq \frac{\partial e_{ij}^{\left[L\right]}}{\partial F_{k}} \Bigg\vert_{\mathbf{F}=\mathbf{1}} , \label{eq:mkij}
\end{align}
we see that
\begin{align}
\frac{d}{d\delta}\Bigg\vert_{\delta =0} e_{ij}^{\left[L\right]}\left(\vx\right) &= \sum_{k=1}^{N} m_{k;ij}^{\left[L\right]} u_{k}\left(\vx\right) = \sum_{k=1}^{N} m_{k;ij}^{\left[L\right]} \left( u_{k}^{\left[1\right]}\left(\vx\right) + u_{k}^{\left[2\right]}\left(\vx\right) \right) .
\end{align}
Thus, it follows from the definition of $\delhatsel^{\left[L\right]}\left(\vx\right)$ that
\begin{align}
\frac{d}{d\delta}\Bigg\vert_{\delta =0} \delhatsel^{\left[L\right]}\left(\vx\right) &= \sum_{i,j,k=1}^{N} \pi_{i}^{\left[L\right]} m_{k;ji}^{\left[L\right]} \left( x_{j}^{\left[L\right]} - x_{i}^{\left[L\right]} \right) \left( u_{k}^{\left[1\right]}\left(\vx\right) + u_{k}^{\left[2\right]}\left(\vx\right) \right) \nonumber \\
&= \sum_{i,j,k=1}^{N} \pi_{i}^{\left[L\right]} m_{k;ji}^{\left[L\right]} \sum_{\ell =1}^{N} \left(\substack{-\left(x_{j}^{\left[L\right]}x_{k}^{\left[1\right]}-x_{i}^{\left[L\right]}x_{k}^{\left[1\right]}\right) C_{k\ell}^{\left[1\right]} + \left(x_{j}^{\left[L\right]}x_{\ell}^{\left[1\right]}-x_{i}^{\left[L\right]}x_{\ell}^{\left[1\right]}\right) B_{\ell k}^{\left[1\right]} \\ -\left(x_{j}^{\left[L\right]}x_{k}^{\left[2\right]}-x_{i}^{\left[L\right]}x_{k}^{\left[2\right]}\right) C_{k\ell}^{\left[2\right]} + \left(x_{j}^{\left[L\right]}x_{\ell}^{\left[2\right]}-x_{i}^{\left[L\right]}x_{\ell}^{\left[2\right]} \right) B_{\ell k}^{\left[2\right]}}\right) . \label{eq:delSelSocialGoods}
\end{align}
The probability that individual $i$ in layer $L$ and individual $j$ in layer $L'$ are both of type $A$ in the neutral RMC distribution is $x_{ij}^{\bm{\xi}\left[LL'\right]}\coloneqq\E_{\RMC}^{\circ}\left[ x_{i}^{\left[L\right]}x_{j}^{\left[L'\right]}\right]$. Taking the expectation of both sides of \eq{delSelSocialGoods}, combined with \ref{eq:rhoAformula} and the definition of the RMC distribution, gives
\begin{align}
\frac{d}{d\delta}\Bigg\vert_{\delta =0} \rho_{A}^{\left[L\right]}\left(\bm{\xi}\right) &= \frac{1}{K^{\circ}} \E_{\RMC}^{\circ}\left[\frac{d}{d\delta}\Bigg\vert_{\delta =0}\delhatsel^{\left[L\right]}\right] \nonumber \\
&= \frac{1}{K^{\circ}}\sum_{i,j,k=1}^{N} \pi_{i}^{\left[L\right]} m_{k;ji}^{\left[L\right]} \sum_{\ell =1}^{N} \left(\substack{-\left(x_{jk}^{\bm{\xi}\left[L1\right]}-x_{ik}^{\bm{\xi}\left[L1\right]}\right) C_{k\ell}^{\left[1\right]} + \left(x_{j\ell}^{\bm{\xi}\left[L1\right]}-x_{i\ell}^{\bm{\xi}\left[L1\right]}\right) B_{\ell k}^{\left[1\right]} \\ -\left(x_{jk}^{\bm{\xi}\left[L2\right]}-x_{ik}^{\bm{\xi}\left[L2\right]}\right) C_{k\ell}^{\left[2\right]} + \left(x_{j\ell}^{\bm{\xi}\left[L2\right]}-x_{i\ell}^{\bm{\xi}\left[L2\right]} \right) B_{\ell k}^{\left[2\right]}}\right) . \label{eq:fpDerivativeK}
\end{align}

Without a loss of generality, we assume now that $L=1$. To further simplify notation, let
\begin{align}
\beta_{ij}^{\bm{\xi}\left[1\right]} &\coloneqq \frac{x_{ij}^{\bm{\xi}\left[11\right]}}{K^{\circ}}
\end{align}
(and $\beta_{i}^{\bm{\xi}\left[1\right]}\coloneqq\beta_{ii}^{\bm{\xi}\left[1\right]}$). 
Using \lem{stateFunction}, these terms, which are associated to the first layer, satisfy the recurrence
\begin{align}
\beta_{ij}^{\bm{\xi}\left[1\right]} &= \xi_{i}^{\left[1\right]}\xi_{j}^{\left[1\right]} - \widehat{\bm{\xi}}^{\left[1\right]} + \sum_{\left(R^{\left[1\right]},\alpha^{\left[1\right]}\right)} p_{\left(R^{\left[1\right]},\alpha^{\left[1\right]}\right)}^{\circ} \beta_{\widetilde{\alpha}^{\left[1\right]}\left(i\right)\widetilde{\alpha}^{\left[1\right]}\left(j\right)}^{\bm{\xi}\left[1\right]} . \label{eq:etaRecurrence}
\end{align}
For the ``cross terms,'' which are associated to the two layers jointly, we let
\begin{align}
\gamma_{ij}^{\bm{\xi}\left[12\right]} &\coloneqq \frac{x_{ij}^{\bm{\xi}\left[12\right]}}{K^{\circ}} .
\end{align}
From \lem{stateFunction}, these terms satisfy the recurrence relation
\begin{align}
\gamma_{ij}^{\bm{\xi}\left[12\right]}  &= \xi_{i}^{\left[1\right]}\xi_{j}^{\left[2\right]} - \widehat{\bm{\xi}}^{\left[1\right]}\widehat{\bm{\xi}}^{\left[2\right]} + \sum_{\substack{\left(R^{\left[1\right]},\alpha^{\left[1\right]}\right) \\ \left(R^{\left[2\right]},\alpha^{\left[2\right]}\right)}} p_{\left(R^{\left[1\right]},\alpha^{\left[1\right]}\right)}^{\circ} p_{\left(R^{\left[2\right]},\alpha^{\left[2\right]}\right)}^{\circ} \gamma_{\widetilde{\alpha}^{\left[1\right]}\left(i\right)\widetilde{\alpha}^{\left[2\right]}\left(j\right)}^{\bm{\xi}\left[12\right]} . \label{eq:gammaRecurrence}
\end{align}
By \eq{fpDerivativeK}, we can then write
\begin{align}
\frac{d}{d\delta}\Bigg\vert_{\delta =0} \rho_{A}^{\left[1\right]}\left(\bm{\xi}\right) &= \sum_{i,j,k,\ell =1}^{N} \pi_{i}^{\left[1\right]} m_{k;ji}^{\left[1\right]} \left(\substack{-\left(\beta_{jk}^{\bm{\xi}\left[1\right]}-\beta_{ik}^{\bm{\xi}\left[1\right]}\right) C_{k\ell}^{\left[1\right]} + \left(\beta_{j\ell}^{\bm{\xi}\left[1\right]}-\beta_{i\ell}^{\bm{\xi}\left[1\right]}\right) B_{\ell k}^{\left[1\right]} \\ -\left(\gamma_{jk}^{\bm{\xi}\left[12\right]}-\gamma_{ik}^{\bm{\xi}\left[12\right]}\right) C_{k\ell}^{\left[2\right]} + \left(\gamma_{j\ell}^{\bm{\xi}\left[12\right]}-\gamma_{i\ell}^{\bm{\xi}\left[12\right]}\right) B_{\ell k}^{\left[2\right]}}\right) . \label{eq:fpDerivativeSG}
\end{align}

Note, however, that the recurrences of Equations~\ref{eq:etaRecurrence}~and~\ref{eq:gammaRecurrence} do not uniquely define $\beta_{ij}^{\bm\xi[1]}$ and $\gamma_{ij}^{\bm\xi[12]}$, respectively. If $\beta_{ij}^{\bm\xi[1]}$ \big(resp. $\gamma_{ij}^{\bm\xi[12]}$\big) is a solution to \eq{etaRecurrence} \big(resp. \eq{gammaRecurrence}\big), then so is $\beta_{ij}^{\bm\xi[1]} +C$ \big(resp. $\gamma_{ij}^{\bm\xi[12]} +C$\big) for any $C\in\mathbb{R}$. As a result of the Fixation Axiom, however, the space of solutions to each recurrence is exactly one-dimensional, which means that these are the only possible solutions. Moreover, since \eq{fpDerivativeSG} depends on differences of $\beta_{ij}^{\bm\xi[1]}$ and differences of $\gamma_{ij}^{\bm\xi[12]}$, it is irrelevant which solution to these recurrences is used. Therefore, we insist that
\begin{subequations}\label{eq:uniquenessConstraint}
\begin{align}
	\sum_{i=1}^{N} \pi_{i}^{\left[1\right]} \beta_{i}^{\bm{\xi}\left[1\right]} &= 0 ; \\
	\sum_{i=1}^{N} \pi_{i}^{\left[1\right]}\gamma_{ii}^{\bm{\xi}\left[12\right]} &= 0 .
\end{align}
\end{subequations}
These conditions are arbitrary but ensure that Equations~\ref{eq:etaRecurrence}~and~\ref{eq:gammaRecurrence} have unique solutions.

From Equations~\ref{eq:rhoA_first} and \ref{eq:fpDerivativeSG}, we have the following conclusion: 
compared with the neutral drift(i.e., $\delta=0$), under the initial strategy configuration $\bm{\xi}$, selection favors $A$-individuals in layer one if and only if 
\begin{equation}
\frac{d}{d\delta}\Bigg\vert_{\delta =0} \rho_{A}^{\left[1\right]}\left(\bm{\xi}\right)>0
\iff
\sum_{i,j,k,\ell =1}^{N} \pi_{i}^{\left[1\right]} m_{k;ji}^{\left[1\right]} \left(\substack{-\left(\beta_{jk}^{\bm{\xi}\left[1\right]}-\beta_{ik}^{\bm{\xi}\left[1\right]}\right) C_{k\ell}^{\left[1\right]} + \left(\beta_{j\ell}^{\bm{\xi}\left[1\right]}-\beta_{i\ell}^{\bm{\xi}\left[1\right]}\right) B_{\ell k}^{\left[1\right]} \\ -\left(\gamma_{jk}^{\bm{\xi}\left[12\right]}-\gamma_{ik}^{\bm{\xi}\left[12\right]}\right) C_{k\ell}^{\left[2\right]} + \left(\gamma_{j\ell}^{\bm{\xi}\left[12\right]}-\gamma_{i\ell}^{\bm{\xi}\left[12\right]}\right) B_{\ell k}^{\left[2\right]}}\right)>0. 
\end{equation}

~\\
Next, we compare the fixation probability of mutant $A$s in a population consisting of $B$s, i.e. $\rho_{A}^{[1]}\left(\bm{\xi}\right)$, with the fixation probability of mutant $B$s in a population consisting of $A$s, i.e. $\rho_{B}^{[1]}\left(\overline{\bm{\xi}}\right)$.
We assume that in $\bm\xi$ and $\overline{\bm\xi}$, states in both layers are symmetric \big($\overline{\xi}_{i}^{[1]}=1-\xi_i^{[1]}$ and $\overline{\xi}_{i}^{[2]}=1-\xi_i^{[2]}$ for every $i\in\{1,\cdots,N\}$\big).
For $\rho_{A}^{[1]}\left(\bm{\xi}\right)>\rho_{B}^{[1]}\left(\overline{\bm{\xi}}\right)$ in a two-layer population, selection favors $A$-individuals over $B$-individuals in layer one.
Using the fact that $\rho_{B}^{\left[1\right]}\left(\overline{\bm{\xi}}\right)=1-\rho_{A}^{\left[1\right]}\left(\overline{\bm{\xi}}\right)$, we see that
\begin{equation}
\begin{split}
	\rho_{A}^{[1]}\left(\bm{\xi}\right)>\rho_{B}^{[1]}\left(\overline{\bm{\xi}}\right) 
	\iff
	\sum_{i,j,k,\ell =1}^{N} \pi_{i}^{\left[1\right]} m_{k;ji}^{\left[1\right]} \left(\substack{-\left(\beta_{jk}^{\bm{\xi}\left[1\right]}-\beta_{ik}^{\bm{\xi}\left[1\right]}\right) C_{k\ell}^{\left[1\right]} + \left(\beta_{j\ell}^{\bm{\xi}\left[1\right]}-\beta_{i\ell}^{\bm{\xi}\left[1\right]}\right) B_{\ell k}^{\left[1\right]} \\ -\left(\gamma_{jk}^{\bm{\xi}\left[12\right]}-\gamma_{ik}^{\bm{\xi}\left[12\right]}\right) C_{k\ell}^{\left[2\right]} + \left(\gamma_{j\ell}^{\bm{\xi}\left[12\right]}-\gamma_{i\ell}^{\bm{\xi}\left[12\right]}\right) B_{\ell k}^{\left[2\right]}}\right)>0,
\end{split}
\end{equation}
where 
\begin{align}
\beta_{ij}^{\bm{\xi}\left[1\right]} &= 2\xi_{i}^{\left[1\right]}\xi_{j}^{\left[1\right]} - \xi_{i}^{\left[1\right]}-\xi_{j}^{\left[1\right]}+ \sum_{\left(R^{\left[1\right]},\alpha^{\left[1\right]}\right)} p_{\left(R^{\left[1\right]},\alpha^{\left[1\right]}\right)}^{\circ} \beta_{\widetilde{\alpha}^{\left[1\right]}\left(i\right)\widetilde{\alpha}^{\left[1\right]}\left(j\right)}^{\bm{\xi}\left[1\right]} 
\end{align}
and 
\begin{equation}
\begin{split}
	\gamma_{ij}^{\bm{\xi}\left[12\right]}  = 
	&2\xi_{i}^{\left[1\right]}\xi_{j}^{\left[2\right]}-\xi_{i}^{\left[1\right]}-\xi_{j}^{\left[2\right]} 
	-2\widehat{\bm{\xi}}^{\left[1\right]}\widehat{\bm{\xi}}^{\left[2\right]} 
	+\widehat{\bm{\xi}}^{\left[1\right]}+\widehat{\bm{\xi}}^{\left[2\right]} \\
	&+ \sum_{\substack{\left(R^{\left[1\right]},\alpha^{\left[1\right]}\right) \\ \left(R^{\left[2\right]},\alpha^{\left[2\right]}\right)}} p_{\left(R^{\left[1\right]},\alpha^{\left[1\right]}\right)}^{\circ} p_{\left(R^{\left[2\right]},\alpha^{\left[2\right]}\right)}^{\circ} \gamma_{\widetilde{\alpha}^{\left[1\right]}\left(i\right)\widetilde{\alpha}^{\left[2\right]}\left(j\right)}^{\bm{\xi}\left[12\right]} ,
\end{split}
\end{equation}
along with two additional constraints (Equation~\ref{eq:uniquenessConstraint}).

\subsection{Reduction to a population with one layer}
Let $|\xi^{[1]}|$ denote the number of $A$s in $\bm\xi^{[1]}$,  $\bm{\Xi}^{[1]}$ the set of all strategy configurations in layer one that has $|\xi^{[1]}|$ $A$s, and $|\bm{\Xi}^{[1]}|$ the number of components in $\bm{\Xi}^{[1]}$.
For example, for $|\xi^{[1]}|=1$, there are $N$ strategy configurations with an $A$ and thus $|\bm{\Xi}^{[1]}|=N$.
Suppose that the initial strategy configuration in layer one is stochastic rather than deterministic, namely being selected uniformly-at-random from $\bm{\Xi}^{[1]}$.
Averaging \eq{fpDerivativeSG} over $\bm{\Xi}^{[1]}$ gives
\begin{equation}
\begin{split}
	&\frac{d}{d\delta}\Bigg\vert_{\delta =0} \left[\frac{1}{|\bm{\Xi}^{[1]}|}\sum_{\bm\xi^{[1]}\in \bm{\Xi}^{[1]}}\rho_{A}^{\left[1\right]}\left(\bm{\xi}^{[1]},\bm{\xi}^{[2]}\right)\right] \\
	=& \sum_{i,j,k,\ell =1}^{N} \pi_{i}^{\left[1\right]} m_{k;ji}^{\left[1\right]} 
	\left(\substack{-\left(\sum_{\bm\xi^{[1]}\in \bm{\Xi}^{[1]}}\beta_{jk}^{\bm{\xi}\left[1\right]}/|\bm{\Xi}^{[1]}|-\sum_{\bm\xi^{[1]}\in \bm{\Xi}^{[1]}}\beta_{ik}^{\bm{\xi}\left[1\right]}/|\bm{\Xi}^{[1]}|\right) C_{k\ell}^{\left[1\right]} \\
		+ \left(\sum_{\bm\xi^{[1]}\in \bm{\Xi}^{[1]}}\beta_{j\ell}^{\bm{\xi}\left[1\right]}/|\bm{\Xi}^{[1]}|-\sum_{\bm\xi^{[1]}\in \bm{\Xi}^{[1]}}\beta_{i\ell}^{\bm{\xi}\left[1\right]}/|\bm{\Xi}^{[1]}|\right) B_{\ell k}^{\left[1\right]} \\ 
		-\left(\sum_{\bm\xi^{[1]}\in \bm{\Xi}^{[1]}}\gamma_{jk}^{\bm{\xi}\left[12\right]}/|\bm{\Xi}^{[1]}|-\sum_{\bm\xi^{[1]}\in \bm{\Xi}^{[1]}}\gamma_{ik}^{\bm{\xi}\left[12\right]}/|\bm{\Xi}^{[1]}|\right) C_{k\ell}^{\left[2\right]} \\
		+ \left(\sum_{\bm\xi^{[1]}\in \bm{\Xi}^{[1]}}\gamma_{j\ell}^{\bm{\xi}\left[12\right]}/|\bm{\Xi}^{[1]}|-\sum_{\bm\xi^{[1]}\in \bm{\Xi}^{[1]}}\gamma_{i\ell}^{\bm{\xi}\left[12\right]}/|\bm{\Xi}^{[1]}|\right) B_{\ell k}^{\left[2\right]}}\right) . \label{eq:fpDerivativeSG_uniform}
\end{split}
\end{equation}
Let $\gamma_{ij}^{[12]}: = \sum_{\bm\xi^{[1]}\in \bm{\Xi}^{[1]}}\gamma_{ij}^{\bm{\xi}\left[12\right]}/|\bm{\Xi}^{[1]}|$.
From \eq{gammaRecurrence}, we have the recurrence
\begin{align}
\gamma_{ij}^{\left[12\right]}  &= \frac{|\xi^{[1]}|}{N}\left(\xi_{j}^{\left[2\right]} - \widehat{\bm{\xi}}^{\left[2\right]}\right) + \sum_{\substack{\left(R^{\left[1\right]},\alpha^{\left[1\right]}\right) \\ \left(R^{\left[2\right]},\alpha^{\left[2\right]}\right)}} p_{\left(R^{\left[1\right]},\alpha^{\left[1\right]}\right)}^{\circ} p_{\left(R^{\left[2\right]},\alpha^{\left[2\right]}\right)}^{\circ} \gamma_{\widetilde{\alpha}^{\left[1\right]}\left(i\right)\widetilde{\alpha}^{\left[2\right]}\left(j\right)}^{\left[12\right]}  \label{eq:uniform}
\end{align}
with constraint $\sum_{i=1}^{N} \pi_{i}^{\left[1\right]}\gamma_{ii}^{\left[12\right]} = 0$.
We consider a recurrence
\begin{align}
\chi_{j}^{\left[2\right]}  &= \frac{|\xi^{[1]}|}{N}\left(\xi_{j}^{\left[2\right]} - \widehat{\bm{\xi}}^{\left[2\right]}\right) + \sum_{\left(R^{\left[2\right]},\alpha^{\left[2\right]}\right)} p_{\left(R^{\left[2\right]},\alpha^{\left[2\right]}\right)}^{\circ} \chi_{\widetilde{\alpha}^{\left[2\right]}\left(j\right)}^{\left[2\right]},  \label{eq:reduced_recurrence}
\end{align}
with constraint $\sum_{j=1}^{N} \pi_{j}^{\left[1\right]} \chi_{j}^{\left[2\right]}=0$,
which can further be simplified to be
\begin{align}
\chi_j^{[2]}=\frac{|\xi^{[1]}|}{N}\left(\xi^{\left[2\right]}_j-\widehat{\bm\xi}^{[2]}\right)
+\sum_{\ell=1}^N \left(e_{\ell j}^{[2]}\right)^{\circ}\chi_{\ell}^{[2]}+\left[1-\left(d_j^{[2]}\right)^{\circ}\right]\chi_j^{[2]},   \label{eq:reduced1}
\end{align}
where $\left(d_j^{[2]}\right)^{\circ}=\sum_{\ell=1}^N \left(e_{\ell j}^{[2]}\right)^{\circ}$.
Therefore, if $M_{ij}=\frac{\left(e_{ji}^{[2]}\right)^{\circ}}{\left(d_i^{[2]}\right)^{\circ}}$ is the transition matrix for the ancestral Markov chain in layer two, then
\begin{align}
\chi_j^{\bm\xi [2]}=\frac{|\xi^{[1]}|}{N\left(d_j^{[2]}\right)^{\circ}}\left[\xi_{j}^{\left[2\right]} - \widehat{\bm{\xi}}^{\left[2\right]}\right]
+\sum_{\ell =1}^N M_{j\ell }\chi_{\ell}^{\bm\xi [2]}.
\end{align}
Since $\left(\bm{D}^{[2]}\right)^{\circ}=\left[\pi_1^{[2]}\left(d_1^{[2]}\right)^{\circ},\pi_2^{[2]}\left(d_2^{[2]}\right)^{\circ},\cdots,\pi_N^{[2]}\left(d_N^{[2]}\right)^{\circ}\right]$ is stationary distribution for the ancestral Markov chain, and since
\begin{align}
\sum_{j\in V} \left(D^{[2]}\right)_j^{\circ} \frac{|\xi^{[1]}|}{N\left(d_j^{[2]}\right)^{\circ}}\left[\xi_{j}^{\left[2\right]} - \widehat{\bm{\xi}}^{\left[2\right]}\right]=0,
\end{align}
it follows that the reduced system for $\chi_j^{[2]}$, namely \eq{reduced1}, has a solution, which is also unique.
Since this solution also solves \eq{uniform} (which itself has a unique solution), it follows that the solution to \eq{uniform} must be independent of $i$. Thus, $\sum_{\bm\xi^{[1]}\in \bm{\Xi}^{[1]}}\gamma_{ik}^{\bm{\xi}\left[12\right]}/|\bm{\Xi}^{[1]}|=\sum_{\bm\xi^{[1]}\in \bm{\Xi}^{[1]}}\gamma_{jk}^{\bm{\xi}\left[12\right]}/|\bm{\Xi}^{[1]}|$ for every $i,j,k\in\left\{1,\dots ,N\right\}$. 
Besides, from Equations \ref{eq:mkij} and \ref{eq:etaRecurrence}, both $m_{k;ji}^{[1]}$ and $\sum_{\bm\xi^{[1]}\in \bm{\Xi}^{[1]}}\beta_{ik}^{\bm{\xi}\left[1\right]}/|\bm{\Xi}^{[1]}|$ in \eq{fpDerivativeSG_uniform} are independent of layer two.
Overall, when  in layer one $A$s are distributed randomly and with a uniform probability,  the effects of weak selection on the fixation probability of $A$ in layer one are independent of layer two, corresponding to the dynamics in a single-layer population.\\ 

Defining $\beta_{ij}^{\left[1\right]}:=\sum_{\bm\xi^{[1]}\in \bm{\Xi}^{[1]}}\beta_{ij}^{\bm{\xi}\left[1\right]}/|\bm{\Xi}^{[1]}|$, we arrive at 
\begin{equation}
\frac{d}{d\delta}\Bigg\vert_{\delta =0} \left[\frac{1}{|\bm{\Xi}^{[1]}|}\sum_{\bm\xi^{[1]}\in \bm{\Xi}^{[1]}}\rho_{A}^{\left[1\right]}\left(\bm{\xi}^{[1]},\bm{\xi}^{[2]}\right)\right]= 
\sum_{i,j,k,\ell =1}^{N} \pi_{i}^{\left[1\right]} m_{k;ji}^{\left[1\right]} 
\left(\substack{-\left(\beta_{jk}^{\left[1\right]}-\beta_{ik}^{\left[1\right]}\right) C_{k\ell}^{\left[1\right]} \\
	+ \left(\beta_{j\ell}^{\left[1\right]}-\beta_{i\ell}^{\left[1\right]}\right) B_{\ell k}^{\left[1\right]} }\right),
\end{equation}
where $\beta_{i}^{\left[1\right]}$ and $\beta_{ij}^{\left[1\right]}$ can be obtained by solving
\begin{align}
\begin{cases}
	\beta_{i}^{[1]}=\sum_{\left(R^{\left[1\right]},\alpha^{\left[1\right]}\right)} p_{\left(R^{\left[1\right]},\alpha^{\left[1\right]}\right)}^{\circ} \beta_{\widetilde{\alpha}^{\left[1\right]}\left(i\right)}^{\left[1\right]} \\
	& \\
	\beta_{ij}^{[1]}=\frac{|\xi^{[1]}|(|\xi^{[1]}|-N)}{N(N-1)}+  \sum_{\left(R^{[1]},\alpha^{[1]}\right)}p_{\left(R^{[1]},\alpha^{[1]}\right)}^{\circ}
	\beta^{[1]}_{\widetilde{\alpha}^{\left[1\right]}\left(i\right)\widetilde{\alpha}^{\left[1\right]}\left(j\right)}
\end{cases}
\end{align}
and constraint $\sum_{i=1}^{N} \pi_{i}^{\left[1\right]} \beta_{i}^{\left[1\right]} = 0 $. 

\clearpage
\section{Supplementary Results}
\subsection{Applications to specific update rules}
Here, we consider applications of \S\ref{sec:modelingFramework} to death-birth (DB), pairwise-comparison (PC), and birth-death (BD) updating. 
Again, we focus on layer one without the loss of generality. In each case below, we assume that the population structure is an undirected, unweighted graph with adjacency matrix $\left(w_{ij}^{\left[L\right]}\right)_{i,j=1}^{N}$. Let $w_{i}^{\left[L\right]}\coloneqq\sum_{j=1}^{N}w_{ij}^{\left[L\right]}$ denote node $i$'s degree in layer $L$.
We consider the random walk in a two-layer network: $p_{ij}^{\left[L\right]}\coloneqq w_{ij}^{\left[L\right]}/w_{i}^{\left[L\right]}$ is the probability of moving from node $i$ to $j$ in a one-step random walk in the network of layer $L$, $\left(p^{\left[L\right]}\right)_{ij}^{(n)}$ the probability of moving from node $i$ to $j$ in a $n$-step walk in the network of layer $L$, 
and $\left(p^{[1,2]}\right)_{ij}^{(n,m)}$ the probability that a walker starting at node $i$ in layer one terminates at node $j$ in layer two after a $n$-step walk in layer one and a following $m$-step walk in layer two (the beginning of the second random walk corresponds to the end of the first).

\subsubsection{DB updating in both layers}\label{sec:DBboth}
Under DB updating, the marginal effect of $k$ on $j$ replacing $i$ is
\begin{align}
m_{k;ji}^{\left[1\right]} &= \frac{1}{N}p_{ij}^{\left[1\right]}\left(\delta_{j,k}-p_{ik}^{\left[1\right]}\right) .
\end{align}
Let $f_{ijk}=-\beta_{ij}^{\bm{\xi}\left[1\right]}C_{jk}^{\left[1\right]}+\beta_{ik}^{\bm{\xi}\left[1\right]}B_{kj}^{\left[1\right]}-\gamma_{ij}^{\bm{\xi}\left[12\right]}C_{jk}^{\left[2\right]}+\gamma_{ik}^{\bm{\xi}\left[12\right]}B_{kj}^{\left[2\right]}$. The reproductive value for DB updating is $\pi_{i}^{\left[1\right]}=w_{i}^{\left[1\right]}/\sum_{k=1}^{N}w_{k}^{\left[1\right]}$ \citep{2019-Allen-p1147-1210,mcavoy-2020-nhb}. Since $\pi_{i}^{\left[1\right]} p_{ij}^{\left[1\right]}=\pi_{j}^{\left[1\right]} p_{ji}^{\left[1\right]}$ for every $i$ and $j$, we have
\begin{align}
\frac{d}{d\delta}\Bigg\vert_{\delta =0} \rho_{A}^{\left[1\right]}\left(\bm{\xi}\right) &= \sum_{i,j,k,\ell=1}^{N} \pi_{i}^{\left[1\right]} m_{k;ji}^{\left[1\right]} \left( f_{jk\ell} - f_{ik\ell} \right) \nonumber \\
&= -\frac{1}{N} \sum_{i,k,\ell=1}^{N} \pi_{i}^{\left[1\right]} \left(p^{\left[1\right]}\right)_{ik}^{\left(2\right)} f_{ik\ell} + \frac{1}{N} \sum_{i,j,\ell=1}^{N} \pi_{j}^{\left[1\right]} p_{ji}^{\left[1\right]} f_{jj\ell} \nonumber \\
&\quad + \frac{1}{N} \sum_{i,k,\ell=1}^{N} \pi_{i}^{\left[1\right]} p_{ik}^{\left[1\right]} f_{ik\ell} - \frac{1}{N} \sum_{i,j,\ell=1}^{N} \pi_{i}^{\left[1\right]} p_{ij}^{\left[1\right]} f_{ij\ell} \nonumber \\
&= \frac{1}{N} \sum_{i,\ell=1}^{N} \pi_{i}^{\left[1\right]} f_{ii\ell} - \frac{1}{N} \sum_{i,j,\ell=1}^{N} \pi_{i}^{\left[1\right]} \left(p^{\left[1\right]}\right)_{ij}^{\left(2\right)} f_{ij\ell} \nonumber \\
&= \frac{1}{N} \sum_{i=1}^{N} \pi_{i}^{\left[1\right]} \sum_{\ell =1}^{N} \left( \substack{-\beta_{ii}^{\bm\xi\left[1\right]} C_{i\ell}^{\left[1\right]} + \beta_{i\ell}^{\bm\xi\left[1\right]} B_{\ell i}^{\left[1\right]} \\ -\gamma_{ii}^{\bm\xi\left[12\right]} C_{i\ell}^{\left[2\right]} + \gamma_{i\ell}^{\bm\xi\left[12\right]} B_{\ell i}^{\left[2\right]}}\right) \nonumber \\
&\quad - \frac{1}{N} \sum_{i,j=1}^{N} \pi_{i}^{\left[1\right]} \left(p^{\left[1\right]}\right)_{ij}^{\left(2\right)} \sum_{\ell =1}^{N} \left(\substack{-\beta_{ij}^{\bm\xi\left[1\right]} C_{j\ell}^{\left[1\right]} + \beta_{i\ell}^{\bm\xi\left[1\right]} B_{\ell j}^{\left[1\right]} \\ -\gamma_{ij}^{\bm\xi\left[12\right]} C_{j\ell}^{\left[2\right]} + \gamma_{i\ell}^{\bm\xi\left[12\right]} B_{\ell j}^{\left[2\right]}}\right) . \label{eq:dFormulaDB}
\end{align}
Therefore, $\frac{d}{d\delta}\Big\vert_{\delta =0} \rho_{A}^{\left[1\right]}\left(\bm{\xi}\right) >0$ is satisfied if and only if
\begin{align} 
\sum_{i=1}^{N} \pi_{i}^{\left[1\right]} \sum_{\ell =1}^{N} \left( \substack{-\beta_{ii}^{\bm\xi\left[1\right]} C_{i\ell}^{\left[1\right]} + \beta_{i\ell}^{\bm\xi\left[1\right]} B_{\ell i}^{\left[1\right]} \\ -\gamma_{ii}^{\bm\xi\left[12\right]} C_{i\ell}^{\left[2\right]} + \gamma_{i\ell}^{\bm\xi\left[12\right]} B_{\ell i}^{\left[2\right]}}\right) &> \sum_{i,j=1}^{N} \pi_{i}^{\left[1\right]} \left(p^{\left[1\right]}\right)_{ij}^{\left(2\right)} \sum_{\ell =1}^{N} \left(\substack{-\beta_{ij}^{\bm\xi\left[1\right]} C_{j\ell}^{\left[1\right]} + \beta_{i\ell}^{\bm\xi\left[1\right]} B_{\ell j}^{\left[1\right]} \\ -\gamma_{ij}^{\bm\xi\left[12\right]} C_{j\ell}^{\left[2\right]} + \gamma_{i\ell}^{\bm\xi\left[12\right]} B_{\ell j}^{\left[2\right]}}\right).
\label{eq:dFormulaDB2}
\end{align}
In particular, for $B_{ij}^{\left[L\right]}=b_Lp_{ji}^{[L]}$ and $C_{ij}^{\left[L\right]}=cp_{ij}^{[L]}$, the above condition is
\begin{align}
\sum_{i=1}^{N} \pi_{i}^{\left[1\right]} \sum_{\ell =1}^{N} \left( \substack{-\beta_{ii}^{\bm\xi\left[1\right]} cp_{i\ell}^{\left[1\right]} + \beta_{i\ell}^{\bm\xi\left[1\right]} b_1p_{i\ell}^{\left[1\right]} \\ -\gamma_{ii}^{\bm\xi\left[12\right]} cp_{i\ell}^{\left[2\right]} + \gamma_{i\ell}^{\bm\xi\left[12\right]} b_2p_{i\ell}^{\left[2\right]}}\right) &> \sum_{i,j=1}^{N} \pi_{i}^{\left[1\right]} \left(p^{\left[1\right]}\right)_{ij}^{\left(2\right)} \sum_{\ell =1}^{N} \left(\substack{-\beta_{ij}^{\bm\xi\left[1\right]} cp_{j\ell}^{\left[1\right]} + \beta_{i\ell}^{\bm\xi\left[1\right]} b_1p_{j\ell}^{\left[1\right]} \\ -\gamma_{ij}^{\bm\xi\left[12\right]} cp_{j\ell}^{\left[2\right]} + \gamma_{i\ell}^{\bm\xi\left[12\right]} b_2p_{j\ell}^{\left[2\right]}}\right) . \label{eq:db_formula}
\end{align}
Defining $\theta_n^{\bm\xi[1]}=\sum_{i,j=1}^N \pi_i^{[1]}\left(p^{[1]}\right)_{ij}^{(n)}\beta_{ij}^{\bm{\xi}[1]}$ 
and $\phi_{n,m}^{\bm\xi[12]}=\sum_{i,j=1}^N \pi_i^{[1]}\left(p^{[1,2]}\right)_{ij}^{(n,m)}\gamma^{\bm{\xi}[12]}_{ij}$, we can write \eq{db_formula} to be 
\begin{equation}
\left(\theta_1^{\bm\xi[1]}-\theta_3^{\bm\xi[1]}\right)b_1+\left(\phi_{0,1}^{\bm\xi[12]}-\phi_{2,1}^{\bm\xi[12]}\right)b_2+\left(\theta_2^{\bm\xi[1]}+\phi_{2,0}^{\bm\xi[12]}\right)c>0.
\label{eq:db_formula}
\end{equation}\\

We now turn to Equations~\ref{eq:etaRecurrence}~and~\ref{eq:gammaRecurrence}. For $i=j$,
\begin{align}
\beta_{i}^{\bm{\xi}\left[1\right]} &= \xi_{i}^{\left[1\right]} - \widehat{\bm{\xi}}^{\left[1\right]} + \frac{1}{N}\sum_{k=1}^{N} p_{ik}^{\left[1\right]} \beta_{k}^{\bm{\xi}\left[1\right]} + \left(1-\frac{1}{N}\right)\beta_{i}^{\bm{\xi}\left[1\right]} ,
\end{align}
which gives
\begin{align}
\beta_{i}^{\bm{\xi}\left[1\right]} &= N\left(\xi_{i}^{\left[1\right]} - \widehat{\bm{\xi}}^{\left[1\right]}\right) + \sum_{k=1}^{N} p_{ik}^{\left[1\right]} \beta_{k}^{\bm{\xi}\left[1\right]} .  \label{eq:eta_db1}
\end{align}
For $i\neq j$,
\begin{align}
\beta_{ij}^{\bm{\xi}\left[1\right]} &= \xi_{i}^{\left[1\right]}\xi_{j}^{\left[1\right]} - \widehat{\bm{\xi}}^{\left[1\right]} + \frac{1}{N} \sum_{k=1}^{N} p_{ik}^{\left[1\right]} \beta_{kj}^{\bm{\xi}\left[1\right]} + \frac{1}{N} \sum_{k=1}^{N} p_{jk}^{\left[1\right]} \beta_{ik}^{\bm{\xi}\left[1\right]} + \left(1-\frac{2}{N}\right)\beta_{ij}^{\bm{\xi}\left[1\right]} ,
\end{align}
which gives
\begin{align}
\beta_{ij}^{\bm{\xi}\left[1\right]} &= \frac{N}{2}\left(\xi_{i}^{\left[1\right]}\xi_{j}^{\left[1\right]} - \widehat{\bm{\xi}}^{\left[1\right]}\right) + \frac{1}{2} \sum_{k=1}^{N} p_{ik}^{\left[1\right]} \beta_{kj}^{\bm{\xi}\left[1\right]} + \frac{1}{2} \sum_{k=1}^{N} p_{jk}^{\left[1\right]} \beta_{ik}^{\bm{\xi}\left[1\right]} .  \label{eq:eta_db2}
\end{align}
For $i,j\in\left\{1,\dots ,N\right\}$, the cross-terms satisfy
\begin{align}
\gamma_{ij}^{\bm{\xi}\left[12\right]} &= \xi_{i}^{\left[1\right]}\xi_{j}^{\left[2\right]} - \widehat{\bm{\xi}}^{\left[1\right]}\widehat{\bm{\xi}}^{\left[2\right]} + \frac{1}{N^{2}}\sum_{k_{1},k_{2}=1}^{N} p_{ik_{1}}^{\left[1\right]}p_{jk_{2}}^{\left[2\right]}\gamma_{k_{1}k_{2}}^{\bm{\xi}\left[12\right]} +\frac{1}{N}\left(1-\frac{1}{N}\right) \sum_{k_{1}=1}^{N} p_{ik_{1}}^{\left[1\right]} \gamma_{k_{1}j}^{\bm{\xi}\left[12\right]} \nonumber \\
&\quad +\frac{1}{N}\left(1-\frac{1}{N}\right) \sum_{k_{2}=1}^{N} p_{jk_{2}}^{\left[2\right]} \gamma_{ik_{2}}^{\bm{\xi}\left[12\right]} +\left(1-\frac{1}{N}\right)^{2}\gamma_{ij}^{\bm{\xi}\left[12\right]} ,
\end{align}
which gives
\begin{align}
\gamma_{ij}^{\bm{\xi}\left[12\right]} &= \frac{N^{2}}{2N-1}\left(\xi_{i}^{\left[1\right]}\xi_{j}^{\left[2\right]} - \widehat{\bm{\xi}}^{\left[1\right]}\widehat{\bm{\xi}}^{\left[2\right]}\right) + \frac{1}{2N-1}\sum_{k_{1},k_{2}=1}^{N} p_{ik_{1}}^{\left[1\right]}p_{jk_{2}}^{\left[2\right]}\gamma_{k_{1}k_{2}}^{\bm{\xi}\left[12\right]} \nonumber \\
&\quad +\frac{N-1}{2N-1}\sum_{k_{1}=1}^{N} p_{ik_{1}}^{\left[1\right]} \gamma_{k_{1}j}^{\bm{\xi}\left[12\right]} +\frac{N-1}{2N-1} \sum_{k_{2}=1}^{N} p_{jk_{2}}^{\left[2\right]} \gamma_{ik_{2}}^{\bm{\xi}\left[12\right]} .   \label{eq:eta_gamma_db}
\end{align}
With the constraints of \eq{uniquenessConstraint}, these recurrences have unique solutions.

\subsubsection{PC updating in both layers}\label{sec:PCboth}
Under PC updating, the marginal effect of $k$ on $j$ replacing $i$ is
\begin{align}
m_{k;ji}^{\left[1\right]} &= \frac{1}{4N} p_{ij}^{\left[1\right]} \left(\delta_{j,k}-\delta_{i,k}\right) .
\end{align}
As before, let $f_{ijk}=-\beta_{ij}^{\bm{\xi}\left[1\right]}C_{jk}^{\left[1\right]}+\beta_{ik}^{\bm{\xi}\left[1\right]}B_{kj}^{\left[1\right]}-\gamma_{ij}^{\bm{\xi}\left[12\right]}C_{jk}^{\left[2\right]}+\gamma_{ik}^{\bm{\xi}\left[12\right]}B_{kj}^{\left[2\right]}$. 
The reproductive value for PC updating is again $\pi_{i}^{\left[1\right]}=w_{i}^{\left[1\right]}/\sum_{k=1}^{N}w_{k}^{\left[1\right]}$ \citep{mcavoy-2020-nhb}. 
The derivative of $\rho_{A}^{\left[1\right]}\left(\bm{\xi}\right)$ is then
\begin{align}
\frac{d}{d\delta}\Bigg\vert_{\delta =0} \rho_{A}^{\left[1\right]}\left(\bm{\xi}\right) &= \sum_{i,j,k,\ell=1}^{N} \pi_{i}^{\left[1\right]} m_{k;ji}^{\left[1\right]} \left( f_{jk\ell} - f_{ik\ell} \right) \nonumber \\
&= \frac{1}{4N} \sum_{i,j,\ell=1}^{N} \pi_{i}^{\left[1\right]} p_{ij}^{\left[1\right]} f_{jj\ell} - \frac{1}{4N} \sum_{i,j,\ell=1}^{N} \pi_{i}^{\left[1\right]} p_{ij}^{\left[1\right]} f_{ji\ell} \nonumber \\
&\quad - \frac{1}{4N} \sum_{i,j,\ell=1}^{N} \pi_{i}^{\left[1\right]} p_{ij}^{\left[1\right]} f_{ij\ell} + \frac{1}{4N} \sum_{i,j,\ell=1}^{N} \pi_{i}^{\left[1\right]} p_{ij}^{\left[1\right]} f_{ii\ell} \nonumber \\
&= \frac{1}{2N} \sum_{i,\ell=1}^{N} \pi_{i}^{\left[1\right]} f_{ii\ell} - \frac{1}{2N} \sum_{i,j,\ell=1}^{N} \pi_{i}^{\left[1\right]} p_{ij}^{\left[1\right]} f_{ij\ell} \nonumber \\
&= \frac{1}{2N} \sum_{i=1}^{N} \pi_{i}^{\left[1\right]} \sum_{\ell =1}^{N} \left( \substack{-\beta_{ii}^{\bm\xi\left[1\right]} C_{i\ell}^{\left[1\right]} + \beta_{i\ell}^{\bm\xi\left[1\right]} B_{\ell i}^{\left[1\right]} \\ -\gamma_{ii}^{\bm\xi\left[12\right]} C_{i\ell}^{\left[2\right]} + \gamma_{i\ell}^{\bm\xi\left[12\right]} B_{\ell i}^{\left[2\right]}}\right) \nonumber \\
&\quad - \frac{1}{2N} \sum_{i,j=1}^{N} \pi_{i}^{\left[1\right]} p_{ij}^{\left[1\right]} \sum_{\ell =1}^{N} \left(\substack{-\beta_{ij}^{\bm\xi\left[1\right]} C_{j\ell}^{\left[1\right]} + \beta_{i\ell}^{\bm\xi\left[1\right]} B_{\ell j}^{\left[1\right]} \\ -\gamma_{ij}^{\bm\xi\left[12\right]} C_{j\ell}^{\left[2\right]} + \gamma_{i\ell}^{\bm\xi\left[12\right]} B_{\ell j}^{\left[2\right]}}\right) . \label{eq:dFormulaPC}
\end{align}
Thus, $\frac{d}{d\delta}\Big\vert_{\delta =0} \rho_{A}^{\left[1\right]}\left(\bm{\xi}\right) >0$ if and only if
\begin{align}
\sum_{i=1}^{N} \pi_{i}^{\left[1\right]} \sum_{\ell =1}^{N} \left( \substack{-\beta_{ii}^{\bm\xi\left[1\right]} C_{i\ell}^{\left[1\right]} + \beta_{i\ell}^{\bm\xi\left[1\right]} B_{\ell i}^{\left[1\right]} \\ -\gamma_{ii}^{\bm\xi\left[12\right]} C_{i\ell}^{\left[2\right]} + \gamma_{i\ell}^{\bm\xi\left[12\right]} B_{\ell i}^{\left[2\right]}}\right) &> \sum_{i,j=1}^{N} \pi_{i}^{\left[1\right]} p_{ij}^{\left[1\right]} \sum_{\ell =1}^{N} \left(\substack{-\beta_{ij}^{\bm\xi\left[1\right]} C_{j\ell}^{\left[1\right]} + \beta_{i\ell}^{\bm\xi\left[1\right]} B_{\ell j}^{\left[1\right]} \\ -\gamma_{ij}^{\bm\xi\left[12\right]} C_{j\ell}^{\left[2\right]} + \gamma_{i\ell}^{\bm\xi\left[12\right]} B_{\ell j}^{\left[2\right]}}\right) .
\end{align}
In particular, for $B_{ij}^{\left[L\right]}=b_Lp_{ji}^{[L]}$ and $C_{ij}^{\left[L\right]}=cp_{ij}^{[L]}$, this above  condition is 
\begin{align}
\sum_{i=1}^{N} \pi_{i}^{\left[1\right]} \sum_{\ell =1}^{N} \left( \substack{-\beta_{ii}^{\bm\xi\left[1\right]} cp_{i\ell}^{\left[1\right]} + \beta_{i\ell}^{\bm\xi\left[1\right]} b_1p_{i\ell}^{\left[1\right]} \\ -\gamma_{ii}^{\bm\xi\left[12\right]} cp_{i\ell}^{\left[2\right]} + \gamma_{i\ell}^{\bm\xi\left[12\right]} b_2p_{i\ell}^{\left[2\right]}}\right)  &> \sum_{i,j=1}^{N} \pi_{i}^{\left[1\right]} p_{ij}^{\left[1\right]} \sum_{\ell =1}^{N} \left(\substack{-\beta_{ij}^{\bm\xi\left[1\right]} cp_{j\ell}^{\left[1\right]} + \beta_{i\ell}^{\bm\xi\left[1\right]} b_1p_{j\ell}^{\left[1\right]} \\ -\gamma_{ij}^{\bm\xi\left[12\right]} cp_{j\ell}^{\left[2\right]} + \gamma_{i\ell}^{\bm\xi\left[12\right]} b_2p_{j\ell}^{\left[2\right]}}\right),  \label{eq:pc_formula}
\end{align}
which can be further written to be
\begin{equation}
\left(\theta_1^{\bm\xi[1]}-\theta_2^{\bm\xi[1]}\right)b_1+\left(\phi_{0,1}^{\bm\xi[12]}-\phi_{1,1}^{\bm\xi[12]}\right)b_2+\left(\theta_1^{\bm\xi[1]}+\phi_{1,0}^{\bm\xi[12]}\right)c>0.
\label{eq:pc_formula_main}
\end{equation}

Turning to Equations~\ref{eq:etaRecurrence}~and~\ref{eq:gammaRecurrence}, we see that for $i=j$,
\begin{align}
\beta_{i}^{\bm{\xi}\left[1\right]} &= \xi_{i}^{\left[1\right]} - \widehat{\bm{\xi}}^{\left[1\right]} + \frac{1}{2N} \sum_{k=1}^{N} p_{ik}^{\left[1\right]} \beta_{k}^{\bm{\xi}\left[1\right]} + \left(1-\frac{1}{2N}\right)\beta_{i}^{\bm{\xi}\left[1\right]} ,
\end{align}
which gives
\begin{align}
\beta_{i}^{\bm{\xi}\left[1\right]} &= 2N\left(\xi_{i}^{\left[1\right]} - \widehat{\bm{\xi}}^{\left[1\right]}\right) + \sum_{k=1}^{N} p_{ik}^{\left[1\right]} \beta_{k}^{\bm{\xi}\left[1\right]} .
\end{align}
For $i\neq j$,
\begin{align}
\beta_{ij}^{\bm{\xi}\left[1\right]} &= \xi_{i}^{\left[1\right]}\xi_{j}^{\left[1\right]} - \widehat{\bm{\xi}}^{\left[1\right]} + \frac{1}{2N} \sum_{k=1}^{N} p_{ik}^{\left[1\right]} \beta_{kj}^{\bm{\xi}\left[1\right]} \nonumber \\
&\quad + \frac{1}{2N} \sum_{k=1}^{N} p_{jk}^{\left[1\right]} \beta_{ik}^{\bm{\xi}\left[1\right]} + \left(1-\frac{1}{N}\right) \beta_{ij}^{\bm{\xi}\left[1\right]} ,
\end{align}
which gives
\begin{align}
\beta_{ij}^{\bm{\xi}\left[1\right]} &= N\left(\xi_{i}^{\left[1\right]}\xi_{j}^{\left[1\right]} - \widehat{\bm{\xi}}^{\left[1\right]}\right) + \frac{1}{2} \sum_{k=1}^{N} p_{ik}^{\left[1\right]} \beta_{kj}^{\bm{\xi}\left[1\right]} + \frac{1}{2} \sum_{k=1}^{N} p_{jk}^{\left[1\right]} \beta_{ik}^{\bm{\xi}\left[1\right]} .
\end{align}
Finally, for $i,j\in\left\{1,\dots ,N\right\}$, the cross-terms satisfy
\begin{align}
\gamma_{ij}^{\bm{\xi}\left[12\right]} &= \xi_{i}^{\left[1\right]}\xi_{j}^{\left[2\right]} - \widehat{\bm{\xi}}^{\left[1\right]}\widehat{\bm{\xi}}^{\left[2\right]} + \frac{1}{4N^{2}}\sum_{k_{1},k_{2}=1}^{N} p_{ik_{1}}^{\left[1\right]}p_{jk_{2}}^{\left[2\right]}\gamma_{k_{1}k_{2}}^{\bm{\xi}\left[12\right]} +\frac{1}{2N}\left(1-\frac{1}{2N}\right) \sum_{k_{1}=1}^{N} p_{ik_{1}}^{\left[1\right]} \gamma_{k_{1}j}^{\bm{\xi}\left[12\right]} \nonumber \\
&\quad +\frac{1}{2N}\left(1-\frac{1}{2N}\right) \sum_{k_{2}=1}^{N} p_{jk_{2}}^{\left[2\right]} \gamma_{ik_{2}}^{\bm{\xi}\left[12\right]} +\left(1-\frac{1}{2N}\right)^{2}\gamma_{ij}^{\bm{\xi}\left[12\right]} ,
\end{align}
which gives
\begin{align}
\gamma_{ij}^{\bm{\xi}\left[12\right]} &= \frac{4N^{2}}{4N-1}\left(\xi_{i}^{\left[1\right]}\xi_{j}^{\left[2\right]} - \widehat{\bm{\xi}}^{\left[1\right]}\widehat{\bm{\xi}}^{\left[2\right]}\right) + \frac{1}{4N-1}\sum_{k_{1},k_{2}=1}^{N} p_{ik_{1}}^{\left[1\right]}p_{jk_{2}}^{\left[2\right]}\gamma_{k_{1}k_{2}}^{\bm{\xi}\left[12\right]} \nonumber \\
&\quad +\frac{2N-1}{4N-1} \sum_{k_{1}=1}^{N} p_{ik_{1}}^{\left[1\right]} \gamma_{k_{1}j}^{\bm{\xi}\left[12\right]} +\frac{2N-1}{4N-1} \sum_{k_{2}=1}^{N} p_{jk_{2}}^{\left[2\right]} \gamma_{ik_{2}}^{\bm{\xi}\left[12\right]}
\end{align}
Once again, these recurrences have unique solutions with the constraints of \eq{uniquenessConstraint}.

\subsubsection{BD updating in both layers}\label{sec:BDboth}
Under BD updating, the marginal effect of $k$ on $j$ replacing $i$ is
\begin{align}
m_{k;ji}^{\left[1\right]} &= \frac{1}{N}\left(\delta_{j,k}-\frac{1}{N}\right) p_{ji}^{\left[1\right]} .
\end{align}
As for the earlier cases, let $f_{ijk}=-\beta_{ij}^{\bm{\xi}\left[1\right]}C_{jk}^{\left[1\right]}+\beta_{ik}^{\bm{\xi}\left[1\right]}B_{kj}^{\left[1\right]}-\gamma_{ij}^{\bm{\xi}\left[12\right]}C_{jk}^{\left[2\right]}+\gamma_{ik}^{\bm{\xi}\left[12\right]}B_{kj}^{\left[2\right]}$. The reproductive value for BD updating is $\pi_{i}^{\left[1\right]}=\left(w_{i}^{\left[1\right]}\right)^{-1}/\sum_{k=1}^{N}\left(w_{k}^{\left[1\right]}\right)^{-1}$ \cite{mcavoy-2020-nhb}. Therefore,
\begin{align}
\frac{d}{d\delta}\Bigg\vert_{\delta =0} \rho_{A}^{\left[1\right]}\left(\bm{\xi}\right) &= \sum_{i,j,k,\ell=1}^{N} \pi_{i}^{\left[1\right]} m_{k;ji}^{\left[1\right]} \left( f_{jk\ell} - f_{ik\ell} \right) \nonumber \\
&= -\frac{1}{N^{2}}\sum_{i,j,k,\ell=1}^{N} \pi_{i}^{\left[1\right]} p_{ji}^{\left[1\right]} f_{jk\ell} + \frac{1}{N}\sum_{i,j,\ell=1}^{N} \pi_{i}^{\left[1\right]} p_{ji}^{\left[1\right]} f_{jj\ell} \nonumber \\
&\quad +\frac{1}{N^{2}}\sum_{i,j,k,\ell=1}^{N} \pi_{i}^{\left[1\right]} p_{ji}^{\left[1\right]} f_{ik\ell} - \frac{1}{N}\sum_{i,j,\ell=1}^{N} \pi_{i}^{\left[1\right]} p_{ji}^{\left[1\right]} f_{ij\ell} \nonumber \\
&= \frac{1}{N}\sum_{i,j,\ell=1}^{N} \pi_{i}^{\left[1\right]} p_{ji}^{\left[1\right]} f_{ii\ell} - \frac{1}{N}\sum_{i,j,\ell=1}^{N} \pi_{i}^{\left[1\right]} p_{ji}^{\left[1\right]} f_{ij\ell} \nonumber \\
&= \frac{1}{N}\sum_{i,j=1}^{N} \pi_{i}^{\left[1\right]} p_{ji}^{\left[1\right]}\sum_{\ell = 1}^N \left(\substack{-\beta_{ii}^{\bm{\xi}\left[1\right]}C_{i\ell}^{\left[1\right]}+\beta_{i\ell}^{\bm{\xi}\left[1\right]}B_{\ell i}^{\left[1\right]} \\ -\gamma_{ii}^{\bm{\xi}\left[12\right]}C_{i\ell}^{\left[2\right]}+\gamma_{i\ell}^{\bm{\xi}\left[12\right]}B_{\ell i}^{\left[2\right]}}\right) \nonumber \\
&\quad - \frac{1}{N}\sum_{i,j=1}^{N} \pi_{i}^{\left[1\right]} p_{ji}^{\left[1\right]}\sum_{\ell=1}^N \left(\substack{-\beta_{ij}^{\bm{\xi}\left[1\right]}C_{j\ell}^{\left[1\right]}+\beta_{i\ell}^{\bm{\xi}\left[1\right]}B_{\ell j}^{\left[1\right]} \\ -\gamma_{ij}^{\bm{\xi}\left[12\right]}C_{j\ell}^{\left[2\right]}+\gamma_{i\ell}^{\bm{\xi}\left[12\right]}B_{\ell j}^{\left[2\right]}}\right) .
\end{align}
Thus, $\frac{d}{d\delta}\Big\vert_{\delta =0} \rho_{A}^{\left[1\right]}\left(\bm{\xi}\right) >0$ if and only if
\begin{align}
\sum_{i,j=1}^{N} \pi_{i}^{\left[1\right]} p_{ji}^{\left[1\right]} \sum_{\ell=1}^N\left(\substack{-\beta_{ii}^{\bm{\xi}\left[1\right]}C_{i\ell}^{\left[1\right]}+\beta_{i\ell}^{\bm{\xi}\left[1\right]}B_{\ell i}^{\left[1\right]} \\ -\gamma_{ii}^{\bm{\xi}\left[12\right]}C_{i\ell}^{\left[2\right]}+\gamma_{i\ell}^{\bm{\xi}\left[12\right]}B_{\ell i}^{\left[2\right]}}\right) &> \sum_{i,j=1}^{N} \pi_{i}^{\left[1\right]} p_{ji}^{\left[1\right]} \sum_{\ell=1}^N\left(\substack{-\beta_{ij}^{\bm{\xi}\left[1\right]}C_{j\ell}^{\left[1\right]}+\beta_{i\ell}^{\bm{\xi}\left[1\right]}B_{\ell j}^{\left[1\right]} \\ -\gamma_{ij}^{\bm{\xi}\left[12\right]}C_{j\ell}^{\left[2\right]}+\gamma_{i\ell}^{\bm{\xi}\left[12\right]}B_{\ell j}^{\left[2\right]}}\right) .
\end{align}
In particular, for $B_{ij}^{\left[L\right]}=b_Lp_{ji}^{[L]}$ and $C_{ij}^{\left[L\right]}=cp_{ij}^{[L]}$, the above  condition is
\begin{align}
\sum_{i,j=1}^{N} \pi_{i}^{\left[1\right]} p_{ji}^{\left[1\right]}
\sum_{\ell=1}^N
\left(\substack{-\beta_{ii}^{\bm{\xi}\left[1\right]}cp_{i\ell}^{\left[1\right]}+\beta_{i\ell}^{\bm{\xi}\left[1\right]}b_{1}p_{i\ell}^{\left[1\right]} \\ -\gamma_{ii}^{\bm{\xi}\left[12\right]}cp_{i\ell}^{\left[2\right]}+\gamma_{i\ell}^{\bm{\xi}\left[12\right]}b_{2}p_{i\ell}^{\left[2\right]}}\right) &> \sum_{i,j=1}^{N} \pi_{i}^{\left[1\right]} p_{ji}^{\left[1\right]} \sum_{\ell=1}^N\left(\substack{-\beta_{ij}^{\bm{\xi}\left[1\right]}cp_{j\ell}^{\left[1\right]}+\beta_{i\ell}^{\bm{\xi}\left[1\right]}b_{1}p_{j\ell}^{\left[1\right]} \\ -\gamma_{ij}^{\bm{\xi}\left[12\right]}cp_{j\ell}^{\left[2\right]}+\gamma_{i\ell}^{\bm{\xi}\left[12\right]}b_{2}p_{j\ell}^{\left[2\right]}}\right) . \label{eq:bd_formula}
\end{align}

For $i=j$, we have
\begin{align}
\beta_{i}^{\bm{\xi}\left[1\right]} &= \xi_{i}^{\left[1\right]} - \widehat{\bm{\xi}}^{\left[1\right]} + \sum_{\left(R^{\left[1\right]},\alpha^{\left[1\right]}\right)} p_{\left(R^{\left[1\right]},\alpha^{\left[1\right]}\right)}^{\circ} \beta_{\widetilde{\alpha}^{\left[1\right]}\left(i\right)}^{\bm{\xi}\left[1\right]} \nonumber \\
&= \xi_{i}^{\left[1\right]} - \widehat{\bm{\xi}}^{\left[1\right]} + \frac{1}{N} \sum_{k=1}^{N} p_{ki}^{\left[1\right]} \beta_{k}^{\bm{\xi}\left[1\right]} + \left(1-\frac{1}{N} \sum_{k=1}^{N} p_{ki}^{\left[1\right]}\right) \beta_{i}^{\bm{\xi}\left[1\right]} ,
\end{align}
which gives
\begin{align}
\beta_{i}^{\bm{\xi}\left[1\right]} &= \frac{N\left(\xi_{i}^{\left[1\right]} - \widehat{\bm{\xi}}^{\left[1\right]}\right) + \sum_{k=1}^{N} p_{ki}^{\left[1\right]} \beta_{k}^{\bm{\xi}\left[1\right]}}{\sum_{k=1}^{N} p_{ki}^{\left[1\right]}} .
\end{align}
For $i\neq j$,
\begin{align}
\beta_{ij}^{\bm{\xi}\left[1\right]} &= \xi_{i}^{\left[1\right]}\xi_{j}^{\left[1\right]} - \widehat{\bm{\xi}}^{\left[1\right]} + \sum_{\left(R^{\left[1\right]},\alpha^{\left[1\right]}\right)} p_{\left(R^{\left[1\right]},\alpha^{\left[1\right]}\right)}^{\circ} \beta_{\widetilde{\alpha}^{\left[1\right]}\left(i\right)\widetilde{\alpha}^{\left[1\right]}\left(j\right)}^{\bm{\xi}\left[1\right]} \nonumber \\
&= \xi_{i}^{\left[1\right]}\xi_{j}^{\left[1\right]} - \widehat{\bm{\xi}}^{\left[1\right]} + \frac{1}{N} \sum_{k=1}^{N} p_{ki}^{\left[1\right]} \beta_{kj}^{\bm{\xi}\left[1\right]} + \frac{1}{N} \sum_{k=1}^{N} p_{kj}^{\left[1\right]} \beta_{ik}^{\bm{\xi}\left[1\right]} \nonumber \\
&\quad + \left(1-\frac{1}{N} \sum_{k=1}^{N} p_{ki}^{\left[1\right]}-\frac{1}{N} \sum_{k=1}^{N} p_{kj}^{\left[1\right]}\right)\beta_{ij}^{\bm{\xi}\left[1\right]} ,
\end{align}
which gives
\begin{align}
\beta_{ij}^{\bm{\xi}\left[1\right]} &= \frac{N\left(\xi_{i}^{\left[1\right]}\xi_{j}^{\left[1\right]} - \widehat{\bm{\xi}}^{\left[1\right]}\right) + \sum_{k=1}^{N} p_{ki}^{\left[1\right]} \beta_{kj}^{\bm{\xi}\left[1\right]} + \sum_{k=1}^{N} p_{kj}^{\left[1\right]} \beta_{ik}^{\bm{\xi}\left[1\right]}}{\sum_{k=1}^{N} p_{ki}^{\left[1\right]}+\sum_{k=1}^{N} p_{kj}^{\left[1\right]}} .
\end{align}
Finally, we have
\begin{align}
\gamma_{ij}^{\bm{\xi}\left[12\right]} &= \xi_{i}^{\left[1\right]}\xi_{j}^{\left[2\right]} - \widehat{\bm{\xi}}^{\left[1\right]}\widehat{\bm{\xi}}^{\left[2\right]} + \sum_{\substack{\left(R^{\left[1\right]},\alpha^{\left[1\right]}\right) \\ \left(R^{\left[2\right]},\alpha^{\left[2\right]}\right)}} p_{\left(R^{\left[1\right]},\alpha^{\left[1\right]}\right)}^{\circ} p_{\left(R^{\left[2\right]},\alpha^{\left[2\right]}\right)}^{\circ} \gamma_{\widetilde{\alpha}^{\left[1\right]}\left(i\right)\widetilde{\alpha}^{\left[2\right]}\left(j\right)}^{\bm{\xi}\left[12\right]} \nonumber \\
&= \xi_{i}^{\left[1\right]}\xi_{j}^{\left[2\right]} - \widehat{\bm{\xi}}^{\left[1\right]}\widehat{\bm{\xi}}^{\left[2\right]} + \frac{1}{N^{2}}\sum_{k_{1},k_{2}=1}^{N} p_{k_{1}i}^{\left[1\right]} p_{k_{2}j}^{\left[2\right]} \gamma_{k_{1}k_{2}}^{\bm{\xi}\left[12\right]} \nonumber \\
&\quad + \frac{1}{N}\left(1-\frac{1}{N}\sum_{k_{2}=1}^{N}p_{k_{2}j}^{\left[2\right]}\right)\sum_{k_{1}=1}^{N} p_{k_{1}i}^{\left[1\right]} \gamma_{k_{1}j}^{\bm{\xi}\left[12\right]} + \frac{1}{N}\left(1-\frac{1}{N}\sum_{k_{1}=1}^{N}p_{k_{1}i}^{\left[1\right]}\right)\sum_{k_{2}=1}^{N} p_{k_{2}j}^{\left[2\right]} \gamma_{ik_{2}}^{\bm{\xi}\left[12\right]} \nonumber \\
&\quad + \left(1-\frac{1}{N}\sum_{k_{1}=1}^{N}p_{k_{1}i}^{\left[1\right]}\right)\left(1-\frac{1}{N}\sum_{k_{2}=1}^{N}p_{k_{2}j}^{\left[2\right]}\right) \gamma_{ij}^{\bm{\xi}\left[12\right]} ,
\end{align}
which gives
\begin{align}
\gamma_{ij}^{\bm{\xi}\left[12\right]} &= \frac{\substack{N^{2}\left(\xi_{i}^{\left[1\right]}\xi_{j}^{\left[2\right]} - \widehat{\bm{\xi}}^{\left[1\right]}\widehat{\bm{\xi}}^{\left[2\right]}\right) + \sum_{k_{1},k_{2}=1}^{N} p_{k_{1}i}^{\left[1\right]} p_{k_{2}j}^{\left[2\right]} \gamma_{k_{1}k_{2}}^{\bm{\xi}\left[12\right]} \\ + \left(N-\sum_{k_{2}=1}^{N}p_{k_{2}j}^{\left[2\right]}\right)\sum_{k_{1}=1}^{N} p_{k_{1}i}^{\left[1\right]} \gamma_{k_{1}j}^{\bm{\xi}\left[12\right]} + \left(N-\sum_{k_{1}=1}^{N}p_{k_{1}i}^{\left[1\right]}\right)\sum_{k_{2}=1}^{N} p_{k_{2}j}^{\left[2\right]} \gamma_{ik_{2}}^{\bm{\xi}\left[12\right]}}}{N\sum_{k_{1}=1}^{N}p_{k_{1}i}^{\left[1\right]} + N\sum_{k_{2}=1}^{N}p_{k_{2}j}^{\left[2\right]} - \left(\sum_{k_{1}=1}^{N}p_{k_{1}i}^{\left[1\right]}\right)\left(\sum_{k_{2}=1}^{N}p_{k_{2}j}^{\left[2\right]}\right)} .
\end{align}
These recurrences give unique solutions with the constraints of \eq{uniquenessConstraint}.

\subsubsection{Mixed DB and PC updating}
Consider now the case in which the two layers are updated using different rules. We study two cases here, with layer one always the layer of interest. In the first case, layers one and two undergo DB and PC updating, respectively. \eq{dFormulaDB} remains the same, as does the recurrence for $\beta_{ij}^{\bm\xi[1]}$ derived in \S\ref{sec:DBboth}. The only modification necessary is to the cross-terms, $\gamma_{ij}^{\bm\xi[12]}$. In the second case, layers one and two undergo PC and DB updating, respectively. Similarly, both \eq{dFormulaPC} and $\beta_{ij}^{\bm\xi[1]}$ still hold from \S\ref{sec:PCboth}, but we must make changes to the cross-terms, $\gamma_{ij}^{\bm\xi[12]}$.

\subsubsection{DB updating in layer one, PC updating in layer two}
For $i,j\in\left\{1,\dots ,N\right\}$, with DB updating in layer one and PC updating in layer two,
\begin{align}
\gamma_{ij}^{\bm{\xi}\left[12\right]} &= \xi_{i}^{\left[1\right]}\xi_{j}^{\left[2\right]} - \widehat{\bm{\xi}}^{\left[1\right]}\widehat{\bm{\xi}}^{\left[2\right]} + \frac{1}{2N^{2}}\sum_{k_{1},k_{2}=1}^{N} p_{ik_{1}}^{\left[1\right]}p_{jk_{2}}^{\left[2\right]}\gamma_{k_{1}k_{2}}^{\bm{\xi}\left[12\right]} +\frac{1}{N}\left(1-\frac{1}{2N}\right) \sum_{k_{1}=1}^{N} p_{ik_{1}}^{\left[1\right]} \gamma_{k_{1}j}^{\bm{\xi}\left[12\right]} \nonumber \\
&\quad +\frac{1}{2N}\left(1-\frac{1}{N}\right) \sum_{k_{2}=1}^{N} p_{jk_{2}}^{\left[2\right]} \gamma_{ik_{2}}^{\bm{\xi}\left[12\right]} +\left(1-\frac{1}{N}\right)\left(1-\frac{1}{2N}\right)\gamma_{ij}^{\bm{\xi}\left[12\right]} ,
\end{align}
giving
\begin{align}
\gamma_{ij}^{\bm{\xi}\left[12\right]} &= \frac{2N^{2}}{3N-1}\left(\xi_{i}^{\left[1\right]}\xi_{j}^{\left[2\right]} - \widehat{\bm{\xi}}^{\left[1\right]}\widehat{\bm{\xi}}^{\left[2\right]}\right) + \frac{1}{3N-1}\sum_{k_{1},k_{2}=1}^{N} p_{ik_{1}}^{\left[1\right]}p_{jk_{2}}^{\left[2\right]}\gamma_{k_{1}k_{2}}^{\bm{\xi}\left[12\right]} \nonumber \\
&\quad + \frac{2N-1}{3N-1} \sum_{k_{1}=1}^{N} p_{ik_{1}}^{\left[1\right]} \gamma_{k_{1}j}^{\bm{\xi}\left[12\right]} +\frac{N-1}{3N-1} \sum_{k_{2}=1}^{N} p_{jk_{2}}^{\left[2\right]} \gamma_{ik_{2}}^{\bm{\xi}\left[12\right]} .
\end{align}
These terms are the only modifications to \S\ref{sec:DBboth} needed to evaluate $\frac{d}{d\delta}\Big\vert_{\delta =0} \rho_{A}^{\left[1\right]}\left(\bm{\xi}\right)$.

\subsubsection{PC updating in layer one, DB updating in layer two}
For $i,j\in\left\{1,\dots ,N\right\}$, with PC updating in layer one and DB updating in layer two,
\begin{align}
\gamma_{ij}^{\bm{\xi}\left[12\right]} &= \xi_{i}^{\left[1\right]}\xi_{j}^{\left[2\right]} - \widehat{\bm{\xi}}^{\left[1\right]}\widehat{\bm{\xi}}^{\left[2\right]} + \frac{1}{2N^{2}}\sum_{k_{1},k_{2}=1}^{N} p_{ik_{1}}^{\left[1\right]}p_{jk_{2}}^{\left[2\right]}\gamma_{k_{1}k_{2}}^{\bm{\xi}\left[12\right]} +\frac{1}{2N}\left(1-\frac{1}{N}\right) \sum_{k_{1}=1}^{N} p_{ik_{1}}^{\left[1\right]} \gamma_{k_{1}j}^{\bm{\xi}\left[12\right]} \nonumber \\
&\quad +\frac{1}{N}\left(1-\frac{1}{2N}\right) \sum_{k_{2}=1}^{N} p_{jk_{2}}^{\left[2\right]} \gamma_{ik_{2}}^{\bm{\xi}\left[12\right]} +\left(1-\frac{1}{N}\right)\left(1-\frac{1}{2N}\right)\gamma_{ij}^{\bm{\xi}\left[12\right]} ,
\end{align}
giving
\begin{align}
\gamma_{ij}^{\bm{\xi}\left[12\right]} &= \frac{2N^{2}}{3N-1}\left(\xi_{i}^{\left[1\right]}\xi_{j}^{\left[2\right]} - \widehat{\bm{\xi}}^{\left[1\right]}\widehat{\bm{\xi}}^{\left[2\right]}\right) + \frac{1}{3N-1}\sum_{k_{1},k_{2}=1}^{N} p_{ik_{1}}^{\left[1\right]}p_{jk_{2}}^{\left[2\right]}\gamma_{k_{1}k_{2}}^{\bm{\xi}\left[12\right]} \nonumber \\
&\quad +\frac{N-1}{3N-1} \sum_{k_{1}=1}^{N} p_{ik_{1}}^{\left[1\right]} \gamma_{k_{1}j}^{\bm{\xi}\left[12\right]} +\frac{2N-1}{3N-1} \sum_{k_{2}=1}^{N} p_{jk_{2}}^{\left[2\right]} \gamma_{ik_{2}}^{\bm{\xi}\left[12\right]} .
\end{align}
These terms are the only modifications to \S\ref{sec:PCboth} needed to evaluate $\frac{d}{d\delta}\Big\vert_{\delta =0} \rho_{A}^{\left[1\right]}\left(\bm{\xi}\right)$.

\subsubsection{DB updating in both layers with accumulated payoffs across interactions}
In this part, accumulated payoffs rather than edge-weighted payoffs are used to update strategies, i.e. 
$B_{ij}^{\left[L\right]}=b^{[L]}w_{ji}^{[L]}$ and $C_{ij}^{\left[L\right]}=cw_{ij}^{[L]}$.
\eq{dFormulaDB2} still predicts the replacement of $A$-players to $B$-players.
We introduce $\widetilde{\theta}_{n,m}^{\bm{\xi}[1]}=\sum_{i,j,k=1}^N \pi_i^{[1]}\left(p^{[1]}\right)_{ik}^{(n)}w_k^{[1]}\left(p^{[1]}\right)_{kj}^{(m)}\beta_{ij}^{\bm{\xi}[1]}$ and 
$\widetilde{\phi}_{n,m}^{\bm{\xi}[12]}=\sum_{i,j,k=1}^N \pi_i^{[1]}\left(p^{[1]}\right)_{ik}^{(n)}w_k^{[2]}\left(p^{[2]}\right)_{kj}^{(m)}\gamma_{ij}^{\bm{\xi}[12]}$.
Inserting $B_{ij}^{\left[L\right]}$ and $C_{ij}^{\left[L\right]}$ into \eq{dFormulaDB2}, we have 
\begin{equation}
\begin{split}
	\frac{d}{d\delta}\Bigg\vert_{\delta =0} \rho_{A}^{[1]}\left(\bm{\xi}\right) >0 \iff 
	&\left(\widetilde\theta_{0,1}^{\bm{\xi}[1]}-\widetilde\theta_{2,1}^{\bm{\xi}[1]}\right)b_1+\left(\widetilde\phi_{0,1}^{[12]}-\widetilde\phi_{2,1}^{\bm{\xi}[12]}\right)b_2 \\
	&-\left(\widetilde\theta_{0,0}^{\bm{\xi}[1]}-\widetilde\theta_{2,0}^{\bm{\xi}[1]}+\widetilde\phi_{0,0}^{\bm{\xi}[12]}-\widetilde\phi_{2,0}^{\bm{\xi}[12]}\right)c>0 ,
\end{split}
\end{equation}
where $\beta_{ij}^{\bm{\xi}[1]}$ and $\gamma_{ij}^{\bm{\xi}[12]}$ can be obtained by solving \eq{eta_db2} and \eq{eta_gamma_db} with two additional constraints, $\sum_{i=1}^N\pi_i^{[1]}\beta_i^{\bm{\xi}[1]}=0$ and $\sum_{i=1}^N \pi_i^{[1]}\gamma_{ii}^{\bm{\xi}[12]}=0$.

\subsection{Applications to specific networks}
\subsubsection{Two-layer ring network}
In this part, we show the application of \eq{db_formula_main} in a two-layer ring network of size $N$.
Rings in two layers are symmetric, as shown in Supplementary Fig. \ref{fig:S15}.
Initially, in each layer, there are an $A$-individual and $N-1$ $B$-individuals.
Here we take $\xi_1^{[1]}=1$ and $\xi_j^{[1]}=0$ for $j\ne 1$, $\xi_i^{[2]}=1$ and $\xi_j^{[2]}=0$ for $j\ne i$.
Let $d$ denote the distance between positions of $A$-players in layer one and two, namely the shortest distance between nodes $1$ and $i$.
For example, $d$ is $1$ in the configuration illustrated in Supplementary Fig.~\ref{fig:S15}.

We begin with death-birth updating used in both layers.
Substituting \eq{eta_gamma_db} into $\phi_{n,m}^{\bm\xi[12]}$ gives
\begin{equation} 
\begin{split}
	\phi_{n,m}^{\bm\xi[12]}=&\frac{1}{2N-1}\phi_{n+1,m+1}^{\bm\xi[12]}+\frac{N-1}{2N-1}\phi_{n+1,m}^{\bm\xi[12]}+\frac{N-1}{2N-1}\phi_{n,m+1}^{\bm\xi[12]}\\
	&+\frac{N^2}{2N-1}\sum_{i,j=1}^N \pi_{i}^{[1]}\left(p^{[1,2]}\right)_{ij}^{(n,m)}\xi_{i}^{[1]}\xi_{j}^{[2]}-\frac{N^2}{2N-1}\widehat{\bm\xi}^{[1]}\widehat{\bm\xi}^{[2]}.
	\label{eq:ring_ring_eq1}
\end{split}
\end{equation}
Since the structures in both layers are symmetric, $\left(p^{[1,2]}\right)_{ij}^{(n,m)}=\left(p_{ij}^{[1]}\right)^{(n+m)}$ and 
$\phi_{n+1,m}^{\bm\xi[12]}=\phi_{n,m+1}^{\bm\xi[12]}$.
Defining $\phi_{n,m}^{\bm\xi[12]}:=\phi_{n+m}^{\bm\xi[12]}$, we can rewrite \eq{ring_ring_eq1} as
\begin{equation} \label{recurrence_ring}
\begin{split}
	\phi_{n+m+2}^{\bm\xi[12]}-\phi_{n+m+1}^{\bm\xi[12]}=&(1-2N)\left(\phi_{n+m+1}^{\bm\xi[12]}-\phi_{n+m}^{\bm\xi[12]}\right)\\
	&-N^2\sum_{i,j=1}^N \pi_{i}^{[1]}\left(p_{ij}^{[1]}\right)^{(n+m)}\xi_{i}^{[1]}\xi_{j}^{[2]}
	+N^2\widehat{\bm\xi}^{[1]}\widehat{\bm\xi}^{[2]} \\
	=&\left(1-2N\right)^{n+m+1}(\phi_1^{\bm\xi[12]}-\phi_0^{\bm\xi[12]}) \\
	&-N^2\sum_{k=0}^{n+m}(1-2N)^k\left[\sum_{i,j=1}^N \pi_{i}^{[1]}\left(p_{ij}^{[1]}\right)^{(n+m-k)}\xi_{i}^{[1]}\xi_{j}^{[2]}
	-\widehat{\bm\xi}^{[1]}\widehat{\bm\xi}^{[2]}\right].  
\end{split}
\end{equation}
Using $\phi_0^{\bm\xi[12]}=0$, we arrive at 
\begin{equation} \label{eq:ring_ring_psi1}
\begin{split}
	\phi_1^{\bm\xi[12]}= \frac{\phi_{n+m+2}^{\bm\xi[12]}-\phi_{n+m+1}^{\bm\xi[12]}}{(1-2N)^{n+m+1}}
	+N^2\sum_{k=0}^{n+m}\frac{1}{(1-2N)^{k+1}}\left[\pi_{1}^{[1]}\left(p^{[1]}\right)_{1i}^{(k)}
	-\pi_{1}^{[1]}\pi_{i}^{[1]}\right].
\end{split}
\end{equation}
In the following, we calculate the quantity $\left(p^{[1]}\right)_{1i}^{(k)}$, the probability of moving from node $1$ to $i$ in a ring network with $N$ nodes in a $k$-step random walk.
The Markov transition matrix for such a symmetric random walk is given by a $N\times N$ matrix:
\begin{align}
\bm M = \frac{1}{2}
\begin{pmatrix}
	0 & 1 & 0  & \cdots  & 0 & 1   \\
	1 & 0 & 1  & \cdots  & 0 & 0   \\
	0 & 1 & 0  & \cdots  & 0 & 0   \\
	\vdots & \vdots  & \vdots & \ddots   & \vdots & \vdots \\
	0 & 0 & 0  & \cdots  & 0 & 1   \\
	1 & 0 & 0  & \cdots  & 1 & 0   \\
\end{pmatrix}.
\end{align}
Let $\bm p_{k}$ be the vector of probabilities at the $k-$th step, so that the $i$-th component of $\bm p_{k}$ is the probability that the random walker is found at node $i$ at step $k$. 
Then we have
\begin{equation}
\bm p_{k+1}=\bm M\bm p_k, \quad
\bm p_0 = 
\begin{pmatrix}
	1 & 0 & 0  & \cdots  & 0 & 0   \\
\end{pmatrix}^{\text{T}},
\end{equation}
which gives
\begin{equation}
\bm p_{k}=\bm M^k\bm p_{0}.
\end{equation}
A further analysis to $\bm M$ gives
\begin{equation}
\bm M=\frac{1}{2}\left(\bm Q+\bm Q^{\text{T}}\right),  \label{eq:M_decomposition}
\end{equation}
where
\begin{align}
\bm Q =
\begin{pmatrix}
	0 & 1 & 0  & \cdots  & 0 & 0   \\
	0 & 0 & 1  & \cdots  & 0 & 0   \\
	0 & 0 & 0  & \cdots  & 0 & 0   \\
	\vdots & \vdots  & \vdots & \ddots   & \vdots & \vdots \\
	0 & 0 & 0  & \cdots  & 0 & 1   \\
	1 & 0 & 0  & \cdots  & 0 & 0   \\
\end{pmatrix}.
\end{align}
Since the column vectors of $\bm Q$ are orthonormal, it is an orthogonal matrix, i.e. $\bm Q^{\text{T}}\bm Q=\bm I$.
The eigenvalues of $\bm Q$, $\lambda$,  satisfy
\begin{equation}
\text{det}\left(\lambda \bm I-\bm Q\right)=\lambda^N-1=0.
\end{equation}
Then we have $\bm Q$'s eigenvalues, given by the roots of unity
\begin{equation}
\lambda_\ell = \text{exp}(\text{i}\omega_\ell), \quad \omega_\ell=\frac{2\pi\ell}{N}, \quad \ell=0,\dots ,N-1.
\end{equation}
Letting $\bm v_\ell=\left(1,\lambda_\ell,\lambda_\ell^2,\cdots,\lambda_\ell^{N-1}\right)^{\text{T}}$, it is easily seen that $\bm Q\bm v_\ell =\lambda_\ell \bm v_\ell$.
Combining this equation with $\bm Q^{\text{T}}\bm Q=\bm I$, we have $\bm Q^{\text{T}}\bm v_\ell=\bm Q^{-1}\bm v_\ell=\lambda^{-1}_\ell \bm v_\ell$. 
The eigenvalues $\lambda_\ell$ are distinct, and the vectors $\bm v_\ell$ form  an orthogonal basis of $\mathbb{C}^N$ (with respect to the standard sesquilinear inner product).
Using \eq{M_decomposition}, we have
\begin{equation}
\bm M\bm v_\ell = \frac{1}{2}\left(\lambda_\ell+\lambda^{-1}_\ell\right)\bm v_\ell
= \mathrm{cos}(\omega_\ell)\bm v_\ell.
\end{equation}
Since $\bm v_\ell$ form an orthogonal basis of $\mathbb{C}^N$, $\mathrm{cos}(\omega_\ell)$, $\ell=0,\cdots,N-1$, form the complete set of eigenvalues of $\bm M$.  
Based on the orthogonal basis $\bm v_\ell$, we have
\begin{equation}
\bm p_0=\frac{1}{N}\sum_{\ell=0}^{N-1}\bm v_\ell, 
\quad
\bm p_k=\bm M^k \bm p_0=\frac{1}{N}\sum_{\ell=0}^N \left(\mathrm{cos}(\omega_\ell)\right)^k \bm v_\ell.
\end{equation}
We obtain
\begin{equation} \label{eq:p1ih}
\left(p^{[1]}\right)_{1i}^{(k)}=\frac{1}{N}\sum_{\ell=0}^{N-1}\left(\cos\left(\frac{2\pi \ell}{N}\right)\right)^{k}\cos\left(\frac{2\pi \ell(i-1)}{N}\right).
\end{equation}
Furthermore, we have $\lim\limits_{n+m \to \infty }\left(p^{[1]}\right)_{ij}^{(n+m)}=\pi_j^{[1]}$ and $\lim\limits_{n+m \to \infty }{\phi_{n+m}^{\bm\xi[12]}}=\sum_{i,j=1}^N \pi_i^{[1]}\pi_j^{[2]}\gamma_{i,j}^{\bm\xi[12]}$.
In \eq{ring_ring_psi1}, let $n+m\rightarrow \infty$, with \eq{p1ih}, we have
\begin{equation}
\phi_1^{\bm\xi[12]}=-\sum_{\ell=1}^{N-1}\frac{\cos\frac{2\pi \ell(i-1)}{N}}{2N-1+\cos \frac{2\pi \ell}{N}}=-\sum_{\ell=1}^{N-1}\frac{\cos\frac{2\pi \ell d}{N}}{2N-1+\cos \frac{2\pi \ell}{N}}.
\end{equation}
Applying the recurrence relation in \eq{ring_ring_psi1}, we obtain
\begin{subequations} \label{eq:db_circle_phi}
\begin{align}
	\phi_2^{\bm\xi[12]} &= -2(N-1)\phi_1^{\bm\xi [12]}-N\delta_{d,0}+1, \\
	\phi_3^{\bm\xi[12]} &= (4N^2-6N+3)\phi_1^{\bm\xi [12]}-\frac{N}{2}\delta_{d,1}+2N(N-1)\delta_{d,0}-2N+3,
\end{align}
\end{subequations}
where $\delta_{i,j}=1$ for $i=j$ and $0$ otherwise.
Moreover, referring to Equation 70 in \cite{McAvoy2021}, we have $\theta_1^{\bm\xi[1]}=-(N-1)/2$, $\theta_2^{\bm\xi[1]}=-(N-2)/2$, and $\theta_3^{\bm\xi[1]}=-3(N-2)/4$.
Inserting $\theta_1^{\bm\xi[1]}$, $\theta_2^{\bm\xi[1]}$, $\theta_3^{\bm\xi[1]}$, and $ \phi_{0,1}^{\bm\xi[12]}= \phi_1^{\bm\xi[12]}$, $ \phi_{2,0}^{\bm\xi[12]}= \phi_2^{\bm\xi[12]}$, $ \phi_{2,1}^{\bm\xi[12]}= \phi_3^{\bm\xi[12]}$ into \eq{db_formula_main}, we have the rule for $A$-individuals replacing $B$-individuals in the two-layer ring network.
~\\

Next, we assume that pairwise-comparison (PC) updating is used in both layers. Using an analysis analogous to that of Equations \ref{eq:ring_ring_eq1}-\ref{eq:db_circle_phi}, we see that
\begin{subequations} \label{eq:pc_circle_phi}
\begin{align}
	\theta_1^{\bm\xi[1]} &= 1-N ; \\
	\theta_2^{\bm\xi[1]} &= 2-N ; \\
	\phi_{1,0}^{\bm\xi[12]} &= -\sum_{\ell=1}^{N-1}\frac{4\cos\frac{2\pi \ell d}{N}}{4N-1+\cos \frac{2\pi \ell}{N}} ; \\
	\phi_{0,1}^{\bm\xi[12]} &= -\sum_{\ell=1}^{N-1}\frac{4\cos\frac{2\pi \ell d}{N}}{4N-1+\cos \frac{2\pi \ell}{N}} ; \\
	\phi_{1,1}^{\bm\xi[12]} &= (2-4N)\phi_{0,1}^{\bm\xi[12]}-4N\delta_{d,0}+4 .
\end{align}
\end{subequations}
By substituting \eq{pc_circle_phi} into \eq{pc_formula_main}, we arrive at the rule for the evolution of cooperation under PC updating.
In particular, when the distance between two mutants is $1$, i.e. $d=1$, we have $\phi_{0,1}^{\bm\xi[12]}-\phi_{1,1}^{\bm\xi[12]}>0$.
Therefore, even though cooperation can never evolve in layer one alone for any $b_1/c$ under PC updating, coupling the two layers can favor cooperation in layer one provided
\begin{equation}
\frac{b_2}{c}>-\frac{\left(\theta_1^{\bm\xi[1]}-\theta_2^{\bm\xi[1]}\right)b_1/c+\theta_1^{\bm\xi[1]}+\phi_{1,0}^{\bm\xi[12]}}{\phi_{0,1}^{\bm\xi[12]}-\phi_{1,1}^{\bm\xi[12]}}.
\end{equation}

\subsubsection{Two-layer star network}
Here, we turn to the application of \eq{db_formula_main} to a two-layer star network of size $N$, as shown in Supplementary Fig.~\ref{fig:S16}.
We begin with death-birth updating in both layers.
In this two-layer star network, nodes $2, \cdots, N-1$ are symmetric in terms of both structure and configuration, which gives  $\beta_2^{\bm\xi[1]}=\cdots=\beta_{N-1}^{\bm\xi[1]}$.
For simplicity, in the following, we denote $\beta_1^{\bm\xi[1]}$ by $\beta_1$, $\beta_N^{\bm\xi[1]}$ by $\beta_N$, and $\beta_l^{\bm\xi[1]}$ by $\beta_{\bullet}$ for  $2\leqslant l\leqslant N-2$.
Using this symmetry property in \eq{eta_db1}, we have 
\begin{subequations} \label{eq:db_star_eta}
\begin{align}
	\beta_1 &= N\left(1-\frac{1}{2(N-1)}\right)+\beta_N ; \\
	\beta_{\bullet} &= -\frac{N}{2(N-1)}+\beta_N ; \\
	\beta_N &= -\frac{N}{2(N-1)}+\frac{1}{N-1}\beta_1+\frac{N-2}{N-1}\beta_{\bullet}.
\end{align}
\end{subequations}
Combining with $\sum_{i=1}^N \pi_i^{[1]}\beta_{i}^{\bm\xi[1]}=0$, we get $\beta_1=-\frac{7N-4N^2}{4(N-1)}$, $\beta_{\bullet}=-\frac{3N}{4(N-1)}$, and $\beta_N=-\frac{N}{4(N-1)}$.
The number of variables $\beta_{ij}^{\bm\xi[1]}$ for $i\ne j$ is up to $N(N-1)/2$.
According to the symmetry of nodes $2,\cdots,N-1$,
we can describe $\beta_{ij}^{\bm\xi[1]}$ by four variables, i.e., 
$\beta_{1l}^{\bm\xi[1]}$ by $\beta_{1\circ}$, 
$\beta_{1N}^{\bm\xi[1]}$ by $\beta_{1N}$, 
$\beta_{ls}^{\bm\xi[1]}$ by $\beta_{\bullet\circ}$, 
and  $\beta_{lN}^{[1]}$ by $\beta_{\bullet N}(\bm\xi)$ for $2\leqslant l\ne s\leqslant N-2$.
The second equation of \eq{eta_db2} can be written as
\begin{subequations}
\begin{align}
	\beta_{1\circ} &= -\frac{N}{4(N-1)}+\frac{1}{2}\beta_{\bullet N}+\frac{1}{2}\beta_{1N} ; \\
	\beta_{1N} &= -\frac{N}{4(N-1)}+\frac{N-2}{2(N-1)}\beta_{1\circ}+\frac{1}{2(N-1)}\beta_{1}+\frac{1}{2}\beta_{N} ; \\
	\beta_{\bullet \circ} &= -\frac{N}{4(N-1)}+\beta_{\bullet N} ; \\
	\beta_{\bullet N} &= -\frac{N}{4(N-1)}+\frac{1}{2(N-1)}\beta_{1\circ}+\frac{N-3}{2(N-1)}\beta_{\bullet \circ}+\frac{1}{2(N-1)}\beta_{\bullet}+\frac{1}{2}\beta_{N}.
\end{align}
\end{subequations}
Combining with $\beta_1$, $\beta_{\bullet}$, $\beta_N$, we get
\begin{subequations} \label{eq:star_star_eta_db}
\begin{align}
	\beta_{1\circ}&=-\frac{11N^2-11N+6}{4(3N-1)(N-1)} ; \\
	\beta_{1N}&=-\frac{2N^2-3N+3}{2(3N-1)(N-1)} ; \\
	\beta_{\bullet \circ}&=-\frac{15N^2-15N+6}{4(3N-1)(N-1)} ; \\
	\beta_{\bullet N}&=-\frac{6N^2-7N+3}{2(3N-1)(N-1)}.
\end{align}
\end{subequations}
Using these values in $\theta_n^{\bm\xi[L]}$, we arrive at
\begin{subequations} \label{eq:star_star_theta}
\begin{align}
	\theta_1^{\bm\xi[1]} &= -\frac{2N-3}{2(N-1)} ; \\
	\theta_2^{\bm\xi[1]} &= -\frac{3(N-2)}{4(N-1)} ; \\
	\theta_3^{\bm\xi[1]} &= -\frac{2N-3}{2(N-1)}.
\end{align}
\end{subequations}
The number of variables $\gamma_{ij}^{\bm\xi[12]}$ is of order $N^2$.
Using the symmetry property, we can describe $\gamma_{ij}^{\bm\xi[12]}$ by nine variables, i.e. 
$\gamma_{11}^{\bm\xi[12]}$ by $\gamma_{11}$, 
$\gamma_{1s}^{\bm\xi[12]}$ by $\gamma_{1\circ}$,
$\gamma_{1N}^{\bm\xi[12]}$ by $\gamma_{1N}$,
$\gamma_{\ell 1}^{\bm\xi[12]}$ by $\gamma_{\bullet 1}$,
$\gamma_{\ell s}^{\bm\xi[12]}$ by $\gamma_{\bullet \circ}$, 
$\gamma_{\ell N}^{\bm\xi[12]}$ by $\gamma_{\bullet N}$,
$\gamma_{N1}^{\bm\xi[12]}$ by $\gamma_{N1}$,
$\gamma_{Ns}^{\bm\xi[12]}$ by $\gamma_{N\circ}$,
$\gamma_{NN}^{\bm\xi[12]}$ by $\gamma_{NN}$ for $1\leqslant l,s\leqslant N$.
Then, the third equation of \eq{eta_gamma_db} can be written to be
\begin{subequations}  \label{eq:star_star_gamma_db}
\begin{align}
	\gamma_{11} =& \frac{N-2}{2N-1}\gamma_{1\circ}+\frac{1}{2N-1}\gamma_{1N}+\frac{N-1}{2N-1}\gamma_{N1}+\frac{N-2}{(2N-1)(N-1)}\gamma_{N\circ} \nonumber \\
	&+\frac{1}{(2N-1)(N-1)}\gamma_{NN}-\frac{N^2}{4(2N-1)(N-1)^2}; \\
	\gamma_{1\circ} =& \frac{N-1}{2N-1}\gamma_{11}+\frac{1}{2N-1}\gamma_{N1}+\frac{N-1}{2N-1}\gamma_{N\circ}-\frac{N^2}{4(2N-1)(N-1)^2}; \\
	\gamma_{1N} =& \frac{N-1}{2N-1}\gamma_{11}+\frac{1}{2N-1}\gamma_{N1}+\frac{N-1}{2N-1}\gamma_{NN}-\frac{N^2}{4(2N-1)(N-1)^2}+\frac{N^2}{(2N-1)}; \\
	\gamma_{\bullet 1} =& \frac{N-2}{2N-1}\gamma_{\bullet \circ}+\frac{1}{2N-1}\gamma_{\bullet N}+\frac{N-1}{2N-1}\gamma_{N1}+\frac{N-2}{(2N-1)(N-1)}\gamma_{N\circ} \nonumber \\
	&+\frac{1}{(2N-1)(N-1)}\gamma_{NN}-\frac{N^2}{4(2N-1)(N-1)^2}; \\
	\gamma_{\bullet \circ} =& \frac{N-1}{2N-1}\gamma_{\bullet 1}+\frac{1}{2N-1}\gamma_{N1}+\frac{N-1}{2N-1}\gamma_{N\circ}-\frac{N^2}{4(2N-1)(N-1)^2}; \\
	\gamma_{\bullet N} =& \frac{N-1}{2N-1}\gamma_{\bullet 1}+\frac{1}{2N-1}\gamma_{N1}+\frac{N-1}{2N-1}\gamma_{NN}-\frac{N^2}{4(2N-1)(N-1)^2}; \\
	\gamma_{N1} =& \frac{1}{2N-1}\gamma_{11}+\frac{N-2}{(2N-1)(N-1)^2}\gamma_{1\circ}+\frac{1}{(2N-1)(N-1)^2}\gamma_{1N}
	+\frac{N-2}{2N-1}\gamma_{\bullet 1} \nonumber \\
	&+\frac{(N-2)^2}{(2N-1)(N-1)^2}\gamma_{\bullet \circ}+\frac{N-2}{(2N-1)(N-1)^2}\gamma_{\bullet N}+
	\frac{N-2}{2N-1}\gamma_{N\circ}+\frac{1}{2N-1}\gamma_{NN} \nonumber \\
	&-\frac{N^2}{4(2N-1)(N-1)^2}; \\
	\gamma_{N\circ} =& \frac{1}{(2N-1)(N-1)}\gamma_{11}+\frac{1}{2N-1}\gamma_{1\circ}+\frac{N-2}{(2N-1)(N-1)}\gamma_{\bullet 1} \nonumber \\
	&+\frac{N-2}{2N-1}\gamma_{\bullet \circ}+\frac{N-1}{2N-1}\gamma_{N1}-\frac{N^2}{4(2N-1)(N-1)^2}; \\
	\gamma_{NN} =& \frac{1}{(2N-1)(N-1)}\gamma_{11}+\frac{1}{2N-1}\gamma_{1N}+\frac{N-2}{(2N-1)(N-1)}\gamma_{\bullet 1} \nonumber \\
	&+\frac{N-2}{2N-1}\gamma_{\bullet N}+\frac{N-1}{2N-1}\gamma_{N1}-\frac{N^2}{4(2N-1)(N-1)^2} .
\end{align}
\end{subequations}
Combining with $\sum_{i=1}^N \pi_i^{[1]}\gamma_{ii}^{\bm\xi[12]}=0$, we obtain
\begin{subequations} \label{eq:star_star_eta_db2}
\begin{align}
	\gamma_{11} &= \frac{N}{\sigma_1}\left(16N^5-82N^4+157N^3-142N^2+60N-8\right); \\
	\gamma_{1\circ} &= -\sigma_2; \\
	\gamma_{1N} &= \frac{1}{\sigma_1}\left(48N^7-208N^6+306N^5-121N^4-116N^3+124N^2-32N\right); \\
	\gamma_{\bullet 1} &= -\frac{\sigma_3}{\sigma_1}; \\
	\gamma_{\bullet \circ} &= -\frac{N^2}{\sigma_1}\left(16N^4-50N^3+57N^2-28N+4\right); \\
	\gamma_{\bullet N} &= -\sigma_2; \\
	\gamma_{N1} &= -\frac{1}{\sigma_1}\left(16N^5-82N^4+161N^3-156N^2+76N-16\right); \\
	\gamma_{N\circ} &= -\frac{\sigma_3}{16(N-1)^4(6N^2-7N+2)}; \\
	\gamma_{NN} &= \frac{N}{\sigma_1}\left(16N^5-82N^4+157N^3-142N^2+60N-8\right),
\end{align}
\end{subequations}
where 
\begin{subequations}
\begin{align}
	\sigma_1 &= 16(2N-1)(3N-2)(N-1)^4; \\
	\sigma_2 &= \frac{N(15N^3-44N^2+44N-16)}{16(3N-2)(N-1)^4}; \\
	\sigma_3 &= N(16N^5-62N^4+99N^3-82N^2+36N-8).
\end{align}
\end{subequations}
Using these values in $\phi_{n,m}^{\bm\xi[12]}$, we arrive at
\begin{subequations}\label{eq:star_star_psi}
\begin{align}
	\phi_{0,1}^{\bm\xi[12]} =& -\frac{N(8N^5-52N^4+112N^3-107N^2+46N-8)}{8(2N-1)(3N-2)(N-1)^4}; \\
	\phi_{2,0}^{\bm\xi[12]} =& -\frac{N^2(N-2)}{2(2N-1)(3N-2)(N-1)^2}; \\
	\phi_{2,1}^{\bm\xi[12]} =& -\frac{N(8N^5-44N^4+84N^3-79N^2+38N-8)}{8(2N-1)(3N-2)(N-1)^4}.
\end{align}
\end{subequations}

In layer one alone (no coupling between layer one and layer two), selection favors $A$ replacing $B$ whenever
$\left(\theta_3^{\bm\xi[1]}-\theta_1^{\bm\xi[1]}\right)b_1-\theta_2^{\bm\xi[1]}c>0$.
However, \eq{star_star_theta} shows that $\theta_3^{\bm\xi[1]}-\theta_1^{\bm\xi[1]}=0$ and $\theta_2^{\bm\xi[1]}>0$.
$A$-individuals therefore are disfavored to replace $B$-individuals in a single layer for any $b_1/c$.
The situation is exactly the same in a separate layer two. 

When the two layers are coupled, inserting \eq{star_star_theta} and \eq{star_star_psi} into \eq{db_formula_main}, we have the condition for selection to favor $A$ relative to $B$,
\begin{equation}
\frac{b_1}{c}>\frac{18N^4-55N^3+64N^2-33N+6}{2N^2(2N-1)}.
\end{equation}
Analogously, selection favors $A$-individuals replacing $B$-individuals in layer two if
\begin{equation}
\frac{b_2}{c}>\frac{18N^4-55N^3+64N^2-33N+6}{2N^2(2N-1)}.
\end{equation}
Overall, if both $b_1/c$ and $b_2/c$ exceed these thresholds, coupling the two layers can favor $A$ replacing $B$ in both layers, which could never happen without the coupling.   
~\\

Next, assuming pairwise-comparison updating used in both layers, by an analogous analysis to Equations \ref{eq:db_star_eta}-\ref{eq:star_star_psi}, we have
\begin{subequations} \label{eq:pc_star_phi}
\begin{align}
	\theta_1^{\bm\xi[1]} &= -\frac{2N-3}{(N-1)} ; \\
	\theta_2^{\bm\xi[1]} &= -\frac{3(N-2)}{2(N-1)} ; \\
	\phi_{1,0}^{\bm\xi[12]} &= -\frac{N(16N^5-84N^4+130N^3-79N^3+20N-1)}{2(N-1)^3(24N^3-26N^2+9N-2)} ; \\
	\phi_{0,1}^{\bm\xi[12]} &= \frac{N^2(16N^4-52N^3+62N^2-29N+5)}{2(N-1)^3(24N^3-26N^2+9N-1)} ; \\
	\phi_{1,1}^{\bm\xi[12]} &= -\frac{N(8N^4-30N^3+34N^2-13N+2)}{2(N-1)^3(12N^2-7N+1)} .
\end{align}
\end{subequations}
By substituting \eq{pc_star_phi} into \eq{pc_formula_main}, we have the rule to predict the evolution of cooperation under PC updating.
Note that $\phi_{0,1}^{\bm\xi[12]}-\phi_{1,1}^{\bm\xi[12]}>0$.
Therefore, even if cooperation can never evolve under PC updating in layer one alone for any $b_1/c$, coupling two layers can favor cooperation for 
\begin{equation}
\frac{b_2}{c}>-\frac{\left(\theta_1^{\bm\xi[1]}-\theta_2^{\bm\xi[1]}\right)b_1/c+\theta_1^{\bm\xi[1]}+\phi_{1,0}^{\bm\xi[12]}}{\phi_{0,1}^{\bm\xi[12]}-\phi_{1,1}^{\bm\xi[12]}}.
\end{equation} 

Finally, if birth-death updating is used in both layers, we can simplify \eq{bd_formula} to be 
\begin{equation}
\begin{split}
	&b_1\sum_{i,j,\ell=1}^N \pi_i^{[1]}p_{ji}^{[1]}p_{i\ell}^{[1]}
	\left(\beta_{i\ell}^{\bm\xi[1]}-\beta_{j\ell}^{\bm\xi[1]}\right)
	+b_2\sum_{i,j,\ell=1}^N \pi_i^{[1]}p_{ji}^{[1]}p_{i\ell}^{[2]}
	\left(\gamma_{i\ell}^{\bm\xi[12]}-\gamma_{j\ell}^{\bm\xi[12]}\right) \\
	&-c\left[\sum_{i,j=1}^N \pi_i^{[1]}p_{ji}^{[1]}
	\left(\beta_{ii}^{\bm\xi[1]}-\beta_{ji}^{\bm\xi[1]}\right)+
	\sum_{i,j=1}^N \pi_i^{[1]}p_{ji}^{[1]}
	\left(\gamma_{ii}^{\bm\xi[12]}-\gamma_{ji}^{\bm\xi[12]}\right)\right]>0 \\
	&\quad \coloneqq b_1\theta_b^{\bm\xi[1]}+b_2\phi_b^{\bm\xi[12]}-c\left(\theta_c^{\bm\xi[1]}+\phi_c^{\bm\xi[12]}\right)>0.
\end{split}
\end{equation}
By an analysis analogous to Equations \ref{eq:db_star_eta}-\ref{eq:star_star_eta_db} and \ref{eq:star_star_gamma_db}-\ref{eq:star_star_eta_db2}, we have
\begin{subequations} \label{eq:bd_star_phi}
\begin{align}
	\theta_b^{\bm\xi[1]} &= -\frac{N(N^2-4N+5)}{2(N^2-2N+2)}; \\
	\theta_c^{\bm\xi[1]} &= -\frac{N^3-4N^2+6N-3}{N^2-2N+2}; \\
	\phi_b^{\bm\xi[12]} &= \frac{N^2(N^5-3N^4+N^3+8N^2-13N+6)}{(N^2-2N+2)^2(N^6-3N^5+2N^4+5N^3-10N^2+2N+4)};\\
	\phi_c^{\bm\xi[12]} &= \frac{N^2(N-1)^2(N^7-7N^6+13N^5+N^4-28N^3+26N^2+3N-10)}
	{(N^2-2N+2)^2(2N^8-8N^7+9N^6+9N^5-32N^4+19N^3+14N^2-10N-4)}.
\end{align}
\end{subequations}
\eq{bd_star_phi} says that $\phi_b^{\bm\xi[12]}>0$. 
Therefore, under BD updating, even if cooperation cannot evolve in a separate layer one for any $b_1/c$, coupling two layers can favor cooperation in layer one provided
\begin{equation}
\frac{b_2}{c}>-\frac{\theta_b^{\bm\xi[1]}}{\phi_b^{\bm\xi[12]}}b_1/c+\theta_c^{\bm\xi[1]}+\phi_c^{\bm\xi[12]}.
\end{equation} 

\subsection{Extensions}
\subsubsection{Different network sizes in different layers}
In more general cases, a node appearing in one layer does not necessarily exist in the other layer, and vice versa.
Even when a node exists in both of the layers, it could be isolated or disconnected from the majority of other individuals within the same layer, which could imply that this individual has negligible effects on the population dynamics. To some degree, such a node could be considered non-existent in the corresponding layer.

Here we investigate the case where node sets in different layers overlap to some degree but are not necessarily identical.
Let $V^{[L]}$ denote the set of nodes in layer $L$ and $N^{[L]}$ the number of nodes accordingly.
Under death-birth updating in both layers, with $B_{ij}^{\left[L\right]}=b_Lp_{ji}^{[L]}$ and $C_{ij}^{\left[L\right]}=cp_{ij}^{[L]}$, $\frac{d}{d\delta}\Big\vert_{\delta =0} \rho_{A}^{\left[1\right]}\left(\bm{\xi}\right) >0$ holds if and only if
\begin{equation}
\left(\theta_1^{\bm\xi[1]}-\theta_3^{\bm\xi[1]}\right)b_1+\left(\phi_{0,1}^{\bm\xi[12]}-\phi_{2,1}^{\bm\xi[12]}\right)b_2+\left(\theta_2^{\bm\xi[1]}+\phi_{2,0}^{\bm\xi[12]}\right)c>0.
\label{eq:db_formula_difsize}
\end{equation}
Here, $\theta_n^{\bm\xi[1]}=\sum_{i,j\in V^{[1]}} \pi_i^{[1]}\left(p^{[1]}\right)_{ij}^{(n)}\beta_{ij}^{\bm{\xi}[1]}$ 
and $\phi_{n,m}^{\bm\xi[12]}=\sum_{i\in V^{[1]},j\in V^{[2]}} \pi_i^{[1]}\left(p^{[1,2]}\right)_{ij}^{(n,m)}\gamma^{\bm{\xi}[12]}_{ij}$.
$\beta_{ij}^{\bm{\xi}[1]}$ and $\gamma^{\bm{\xi}[12]}_{ij}$ can be obtained by solving
\begin{subequations}
\begin{align}
	\beta_{i}^{\bm{\xi}[1]}=& N^{[1]}\left(\xi_{i}^{[1]}-\widehat{\bm\xi}^{[1]}\right)+\sum_{k\in V^{[1]}} p_{ik}^{[1]}\beta_{k}^{\bm{\xi}[1]};  \\
	\beta_{ij}^{\bm{\xi}[1]}=& \frac{N^{[1]}}{2}\left(\xi_{i}^{[1]}\xi_{j}^{[1]}-\widehat{\bm\xi}^{[1]}\right)
	+\frac{1}{2}\sum_{k\in V^{[1]}} p_{ik}^{[1]}\beta_{kj}^{\bm{\xi}[1]}+\frac{1}{2}\sum_{k\in V^{[1]}} p_{jk}^{[1]}\beta_{ki}^{\bm{\xi}[1]}; \\
	\gamma^{\bm{\xi}[12]}_{ij}
	=&\frac{N^{[1]}N^{[2]}}{N^{[1]}+N^{[2]}-1}\left[\xi_{i}^{[1]}\xi_{j}^{[2]}-\widehat{\bm\xi}^{[1]}\widehat{\bm\xi}^{[2]}\right]
	+\frac{1}{N^{[1]}+N^{[2]}-1}\sum_{k_1\in V^{[1]},k_2\in V^{[2]}} p_{ik_1}^{[1]}p_{jk_2}^{[2]}\gamma^{\bm{\xi}[12]}_{k_1k_2} \nonumber \\
	&+\frac{N^{[2]}-1}{N^{[1]}+N^{[2]}-1}\sum_{k_1\in V^{[1]}} p_{ik_1}^{[1]}\gamma^{\bm{\xi}[12]}_{k_1j}  
	+\frac{N^{[1]}-1}{N^{[1]}+N^{[2]}-1}\sum_{k_2\in V^{[2]}} p_{jk_2}^{[2]}\gamma^{\bm{\xi}[12]}_{ik_2} .
\end{align}
\end{subequations}
and two additional constraints $\sum_{i\in V^{[1]}}\pi_i^{[1]}\beta_{i}^{\bm{\xi}[1]}=0$ and $\sum_{i\in V^{[1]}\cap V^{[2]}} \pi_i^{[1]}\gamma^{\bm{\xi}[12]}_{ii}=0$. 

\subsubsection{An arbitrary number of layers}
The number of layers may differ in multilayer systems.
Here we consider a multilayer population with an arbitrary number of layers, denoted by $M$.
Under death-birth updating in both layers, with $B_{ij}^{\left[L\right]}=b_Lp_{ji}^{[L]}$ and $C_{ij}^{\left[L\right]}=cp_{ij}^{[L]}$, $\frac{d}{d\delta}
\Big\vert_{\delta =0} \rho_{A}^{\left[1\right]}\left(\bm{\xi}\right) >0$ holds if and only if
\begin{equation}
\left(\theta_1^{\bm\xi[1]}-\theta_3^{\bm\xi[1]}\right)b_1+\theta_2^{\bm\xi[1]}c+\sum_{L=2}^M \left[\left(\phi_{0,1}^{\bm\xi[1L]}-\phi_{2,1}^{\bm\xi[1L]}\right)b_L+\phi_{2,0}^{\bm\xi[1L]}c\right]>0,
\label{eq:db_formula_main_multilayer}
\end{equation}
where $\phi_{n,m}^{\bm\xi[1L]}=\sum_{i\in V^{[1]},j\in V^{[L]}} \pi_i^{[1]}\left(p^{[1,L]}\right)_{ij}^{(n,m)}\gamma^{\bm{\xi}[1L]}_{ij}$.
$\beta_{ij}^{\bm{\xi}[1]}$ and $\gamma^{\bm{\xi}[1L]}_{ij}$ can be obtained by solving
\begin{subequations}
\begin{align}
	\beta_{i}^{\bm{\xi}[1]}=& N\left(\xi_{i}^{[1]}-\widehat{\bm\xi}^{[1]}\right)+\sum_{k=1}^N p_{ik}^{[1]}\beta_{k}^{\bm{\xi}[1]} ; \\
	\beta_{ij}^{\bm{\xi}[1]}=& \frac{N}{2}\left(\xi_{i}^{[1]}\xi_{j}^{[1]}-\widehat{\bm\xi}^{[1]}\right)
	+\frac{1}{2}\sum_{k=1}^N p_{ik}^{[1]}\beta_{kj}^{\bm{\xi}[1]}+\frac{1}{2}\sum_{k=1}^N p_{jk}^{[1]}\beta_{ki}^{\bm{\xi}[1]} ; \\
	\gamma^{\bm{\xi}[1L]}_{ij}
	=&\frac{N^2}{2N-1}\left[\xi_{i}^{[1]}\xi_{j}^{[L]}-\widehat{\bm\xi}^{[1]}\widehat{\bm\xi}^{[L]}\right]
	+\frac{1}{2N-1}\sum_{k_1,k_L=1}^N p_{ik_1}^{[1]}p_{jk_L}^{[L]}\gamma^{\bm{\xi}[1L]}_{k_1k_L} \nonumber \\
	&+\frac{N-1}{2N-1}\sum_{k_1=1}^N p_{ik_1}^{[1]}\gamma^{\bm{\xi}[1L]}_{k_1j}  
	+\frac{N-1}{2N-1}\sum_{k_L=1}^N p_{jk_L}^{[L]}\gamma^{\bm{\xi}[1L]}_{ik_L} .
\end{align}
\end{subequations}
and $M$ additional constraints $\sum_{i=1}^N \pi_i^{[1]}\beta_{i}^{\bm{\xi}[1]}=0$ and $\sum_{i=1}^N \pi_i^{[1]}\gamma^{\bm{\xi}[1L]}_{ii}=0$  for $L\ne 1$.

\subsection{Empirical social networks} \label{empirical_populations}
We investigate six empirical social networks (see datasets in \text{https://comunelab.fbk.eu/data.php}).
Of these networks, some have more than two layers.
Although our method can tackle with an arbitrary number of layers, for simplicity, here we only study the evolutionary dynamics in a two-layer network.
We form a two-layer network by using only two layers of the original network or sort all layers into two categories.
The following are details of the six datasets: \\
1. CS-AARHUS (CA):  the multiplex social network consists of five kinds of online and offline relationships (Facebook, Leisure, Work, Co-authorship, Lunch) between the employees of Computer Science department at Aarhus. 
In this work, we form a two-layer network based on online relationship (facebook, coauthor) and offline relationship (lunch, leisure, work);\\
2. KAPFERER TAILOR SHOP (KTS): layers represent two different types of interaction, recorded at two different times (seven months apart) over a period of one month. 
One consists of work- and assistance-related interactions, and the other consists of friendship and socioemotional interactions.
We form a two-layer network based on friendship over two months and work relationship over two months;\\
3. KRACKHARDT HIGH TECH (KHT): the multiplex social network consists of 3 kinds of relationships (Advice, Friendship and ``Reports to") between managers of a high-tech company. 
We form a two-layer network based on friendship and work relationship (``Report to");\\
4. LAZEGA LAW FIRM (LLF): the multiplex social network consists of 3 kinds of relationship (Co-work, Friendship and Advice) between partners and associates of a corporate law partnership. 
We form a two-layer network based on friendship and work relationship (Co-work);\\
5. PEDGETT FLORENTINE FAMILIES (PFF): the multiplex social network consists of 2 layers (marriage alliances and business relationships) describing Florentine families in the Renaissance.
We form a two-layer network based on marriage alliances and business relationships;\\
6. VICKERS-CHAN-7THGRADERS MULTIPLEX NETWORK (VC7): the data were collected by Vickers from 29 seventh grade students in a school in Victoria, Australia. Students were asked to nominate their classmates on a number of relations including the following three (layers): Who do you get on with in the class? Who are your best friends in the class? Who would you prefer to work with?
We form a two-layer network by friendship (the second question) and work relationship (the third question).

\clearpage
\section{Supplementary Discussion}
Our results have been derived in a completely general mathematical framework, whose power we have illustrated through systematic analysis of all small networks, as well as extensive sampling of larger random networks with different degree distributions. We have also analyzed six empirical two-layer networks across diverse real-world communities, where we find that coupling promotes the spread of prosocial behavior, especially by strategic design of incentives in one layer. 
Our analysis of these six empirical networks has been confined to the simple donation game, which is not a perfectly accurate description of all the real-world social interactions that occur in these empirical settings. 
But this simple model hopefully captures the key, qualitative tension between prosocial, selfish, and even antisocial behavior \cite{2006-Ohtsuki-p502-505,2009-Tarnita-p570-581,2009-Tarnita-p8601-8604,2014-Allen-p113-151,Fotouhi2018,2017-Allen-p227-230}. 
The qualitative conclusions we draw from it are driven by the empirical network structures and the behavioral dynamics that arise when individuals garner influence across domains.

The literature on evolutionary game theory commonly assumes that new types (innovations or mutations) appear uniformly at random within a population. This assumption simplifies the mathematical analysis of population dynamics, and it is also scientifically reasonable when death rates are uniform and mutants are initially rare \cite{2006-Ohtsuki-p502-505,2009-Tarnita-p8601-8604,2013-Chen-p637-664,2014-Debarre-p3409-3409,2014-Allen-p113-151,2017-Allen-p227-230,2019-Qi-p20190041-20190041}. However, non-uniform arrangements of mutants can lead to completely different outcomes. For example, one arrangement might favor cooperation while another suppresses it \cite{2016-Chen-p39181-39181}, as we have seen in the multilayer context as well. More generally, we have proven that when the mutant in a layer is introduced randomly and uniformly, then the threshold required for cooperation to be favored is independent of the other layers (see Supplementary Information section 1.3 and Supplementary Fig.~17). In other words, averaging the dynamics over a uniform initial mutant distribution obscures the effects of one layer on another. And so we conclude that the common assumption used in the field turns out to be pathological special case that is not representative of the effects of mutation in general.

Our study of multilayer games has used single-layer networks as the primary reference point for comparison. However, there are substantial similarities between the process we study here and evolutionary set theory \cite{2009-Tarnita-p8601-8604}, a framework in which different sets represent different social categories (or different types of social relationships), and each individual falls into one or more of these sets. The crucial difference between evolutionary set theory and multilayer networks is that, in the former, individuals adopt a single strategy and apply it against all other members of his or her set(s); and set membership can change in time. In our setting, on the other hand, domain membership is fixed but we allow for separate behaviors in different domains of interactions. In this sense, the framework of multilayer networks is orthogonal to evolutionary set theory. Moreover, in the context of multilayer games it is not the strategy one uses in a layer that determines your influence in that layer; rather, all your strategies matter.

The last two decades have seen extensive investigation into the effects of spatial structure on evolutionary games \cite{2005-Lieberman-p312-316}. Most of these studies are based on a single (one-layer) population structure, limited to one of a few different update rules. While the use of multilayer networks in evolutionary dynamics is not new \cite{2012-Wang-p48001-48001,2012-Gomez-Gardenes-p620-620,Santos2014,2015-Wang-p124-124,Kleineberg2018}, to our knowledge our work provides the rigorous mathematical results on evolution in multilayer populations. These results are applicable to an arbitrary number of layers and any connectivity structure within each layer, and so they allow for efficient exploration of diverse multilayer structures. They also apply to a broad class of evolutionary update rules, including mixtures across layers. Many questions remain for future work in this area, including the effects of different interaction and replacement structures in each layer; the dynamics of producers of other kinds of social goods; the implications of strategy ``spillover'' from one layer to another; the structure correlations among layers; and dynamic social categories that can change over time. As modeling techniques grow more sophisticated to reflect the complexity of human and non-human societies, a better empirical understanding of interdependence of social domains (such as behavioral experiments with human subjects) will be crucial for predicting the dynamics of prosocial behaviors.

\clearpage

\makeatletter
\@fpsep\textheight
\makeatother

\begin{table}[!h]
	\begin{adjustwidth}{0in}{0in}
		\caption{\label{Tab1} The number of all non-isomorphic single-layer and two-layer profiles on networks of size $N=3$, 4, 5, and 6. Note that the network in each layer is required to be connected.}
		\begin{tabular}{p{3cm}<{\centering}| p{5cm}<{\centering} p{5cm}<{\centering}}
			\toprule
			$N$ &  number of non-isomorphic single-layer profiles 
			& number of non-isomorphic two-layer profiles\\
			\midrule
			3 & 3 & 26\\
			4&  11& 1,028\\
			5& 58& 114,992\\
			6& 407& 36,394,472\\
			\bottomrule
		\end{tabular}
	\end{adjustwidth}
\end{table}

\newpage

\begin{table}
	\centering
	\caption{\label{Table 2} We consider single-layer profiles of networks of size $N=6$. 
		Concretely, we randomly choose a pair of profiles from all 407 non-isomorphic single-layer profiles (see Supplementary Table~\ref{Tab1}).
		Note that the two chosen single-layer profiles can be identical, and so there are $407^2$ such pairs.
		Let $(b_1/c)^*$ and $(b_2/c)^*$ respectively denote the (single-layer) thresholds of the two profiles.
		(A) Frequency of choosing two single-layer profiles that one has $(b_1/c)^*$ corresponding to the row and the other $(b_2/c)^*$ corresponding to the column.
		(B) Frequency of choosing two single-layer profiles: [1] one has $(b_1/c)^*$ corresponding to the row and the other $(b_2/c)^*$ corresponding to the column; [2] with coupling the two layers, cooperation can be favored in layer one---there exist positive values of $b_1/c$ and $b_2/c$ favoring cooperation in layer one. 
		(C) Frequency of choosing two single-layer profiles: [1] one has $(b_1/c)^*$ corresponding to the row and the other $(b_2/c)^*$ corresponding to the column; [2] with coupling the two layers, cooperation can be favored in both layers---there exist positive values of $b_1/c$ and $b_2/c$ favoring cooperation in both layers. 
	}
	\begin{tabular}{|c|c|ccc|} 
		\hline
		\multicolumn{1}{|c}{A} &  & \multicolumn{3}{c|}{$(b_2/c)^*$}  \\ 
		\cline{3-5}
		\multicolumn{1}{|c}{} &  & $(-\infty,0)$ & $(0,\infty)$ &  $\infty$                \\ 
		\hline
		\multirow{3}{*}{$(b_1/c)^*$}     & $(-\infty,0)$ & 27.91\% & 24.14\% & 0.78\% \\
		& $(0,\infty)$ & 24.14\% & 20.89\% & 0.67\%   \\
		& $\infty$ & 0.78\% & 0.67\% & 0.02\%\quad\\
		\hline
	\end{tabular}
	\begin{tabular}{|c|c|ccc|} 
		\hline
		\multicolumn{1}{|c}{B} &  & \multicolumn{3}{c|}{$(b_2/c)^*$}  \\ 
		\cline{3-5}
		\multicolumn{1}{|c}{} &  & $(-\infty,0)$ & $(0,\infty)$ &  $\infty$                \\ 
		\hline
		\multirow{3}{*}{$(b_1/c)^*$}     & $(-\infty,0)$ & 13.66\% & 9.58\% & 0.37\% \\
		& $(0,\infty)$ & 24.14\% & 20.89\% & 0.67\%   \\
		& $\infty$ & 0.30\% & 0.24\% & 0.007\%\\
		\hline
	\end{tabular}
	\begin{tabular}{|c|c|ccc|} 
		\hline
		\multicolumn{1}{|c}{C} &  & \multicolumn{3}{c|}{$(b_2/c)^*$}  \\ 
		\cline{3-5}
		\multicolumn{1}{|c}{} &  & $(-\infty,0)$ & $(0,\infty)$ &  $\infty$                \\ 
		\hline
		\multirow{3}{*}{$(b_1/c)^*$}     & $(-\infty,0)$ & 1.12\% & 9.58\% & 0.14\% \\
		& $(0,\infty)$ & 9.58\% & 19.99\% & 0.24\%   \\
		& $\infty$ & 0.14\% & 0.24\% & 0.003\%\\
		\hline
	\end{tabular}
\end{table}

\begin{figure}
	\centering
	\includegraphics[width=0.6\textwidth]{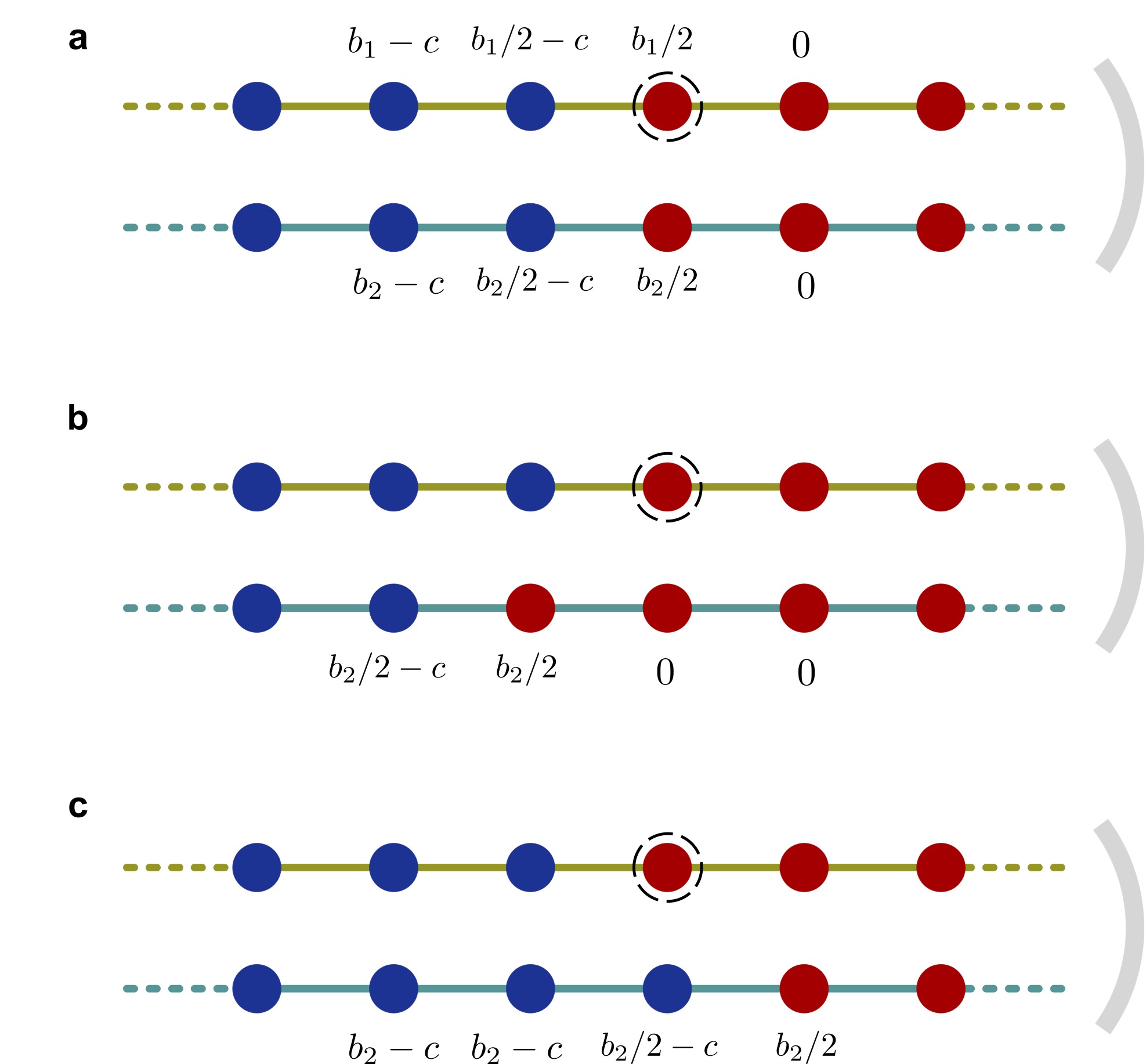}
	\caption{\label{fig:S1} \textbf{Intuition about cooperation-promoting effects by coupling layers}.
		We consider a two-layer ring network of infinite size and three initial strategy configurations, as shown in panels \textbf{a}-\textbf{c}.
		Values next to nodes are players' payoffs derived in corresponding layers.
		We study the expansion of cooperator clusters in layer one.
		In layer one, cooperator cluster expands only when the defector at the boundary (dashed circle) dies and the neighboring cooperator succeeds in dispersing its offspring to the vacant site.
		This happens with probability positively related to the difference between the neighboring cooperator's and the neighboring defector's payoff, i.e. $u_C-u_D$.
		In the absence of layer two, $u_C-u_D=b_1/2-c$.
		With layer two, payoffs obtained in layer two matter and $u_C-u_D=(b_1+b_2)/2-2c$ (\textbf{a}), $u_C-u_D=(b_1+b_2)/2-c$ (\textbf{b}), $u_C-u_D=(b_1+b_2)/2-2c$ (\textbf{c}).
		For $b_2>2c$, layer two provides more advantages to the cooperator in layer one than to the defector.
		The introduction of layer two thus promotes the expansion of cooperator cluster in layer one.
	}
\end{figure}

\newpage

\begin{figure}
	\centering
	\includegraphics[width=0.7\textwidth]{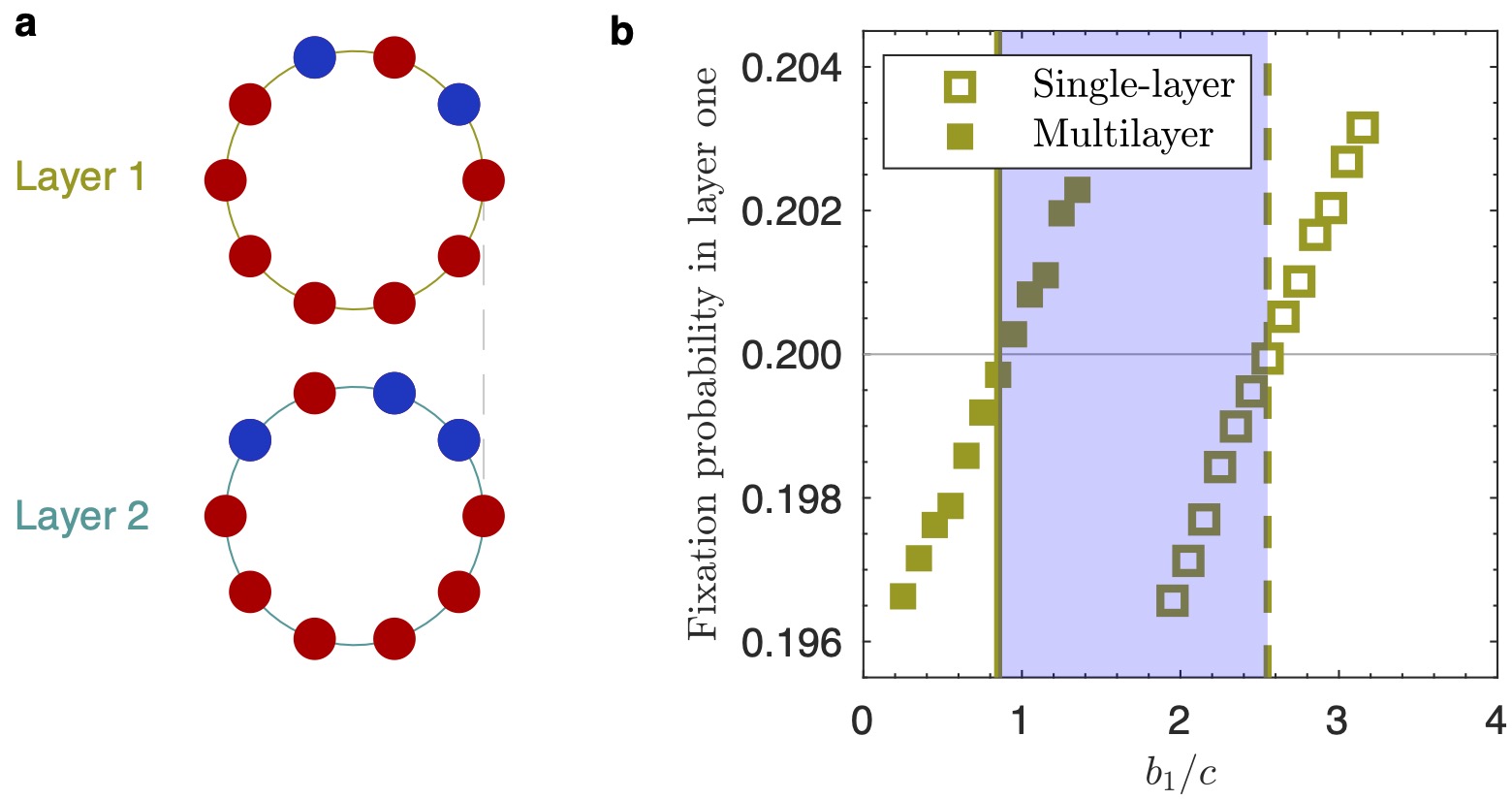}
	\caption{\label{fig:S2} \textbf{Multilayer games can promote cooperation}. We present an example in which cooperation in layer one is facilitated by coupling with games in layer two. 
		The networks consist of a ``ring" in each layer, with each node connected to two neighbors. 
		Nodes occupying the same position of both layers represent the same player, as illustrated by the dotted line. 
		\textbf{a}, An initial strategy configuration that contains two cooperators in layer one and three cooperators in layer two.
		\textbf{b}, The probability that cooperation eventually fixes in layer one, starting from the initial configuration shown in panel \textbf{a}. 
		We compare two scenarios: when the two layers evolve independently (open squares) versus when the two layers are coupled (solid squares). Cooperation is favored by selection if it is more likely to fix than under neutrality ($\delta=0$, indicated by the horizontal line). 
		Cooperation is favored by selection when the benefit-to-cost ratio exceeds a critical value, $(b_1/c)^*$ (vertical line). 
		For the benefit-to-cost ratio indicated in light blue, coupling to layer two promotes cooperation in layer one even though it would be disfavored by selection in layer one alone. 
		Dots indicate results from $10^7$ replicate Monte Carlo simulations, and lines indicate analytical predictions.
		Parameters: $b_2=10$, $c=1$, and $\delta=0.02$.
	}
\end{figure}

\newpage

\begin{figure}
	\centering
	\includegraphics[width=0.5\textwidth]{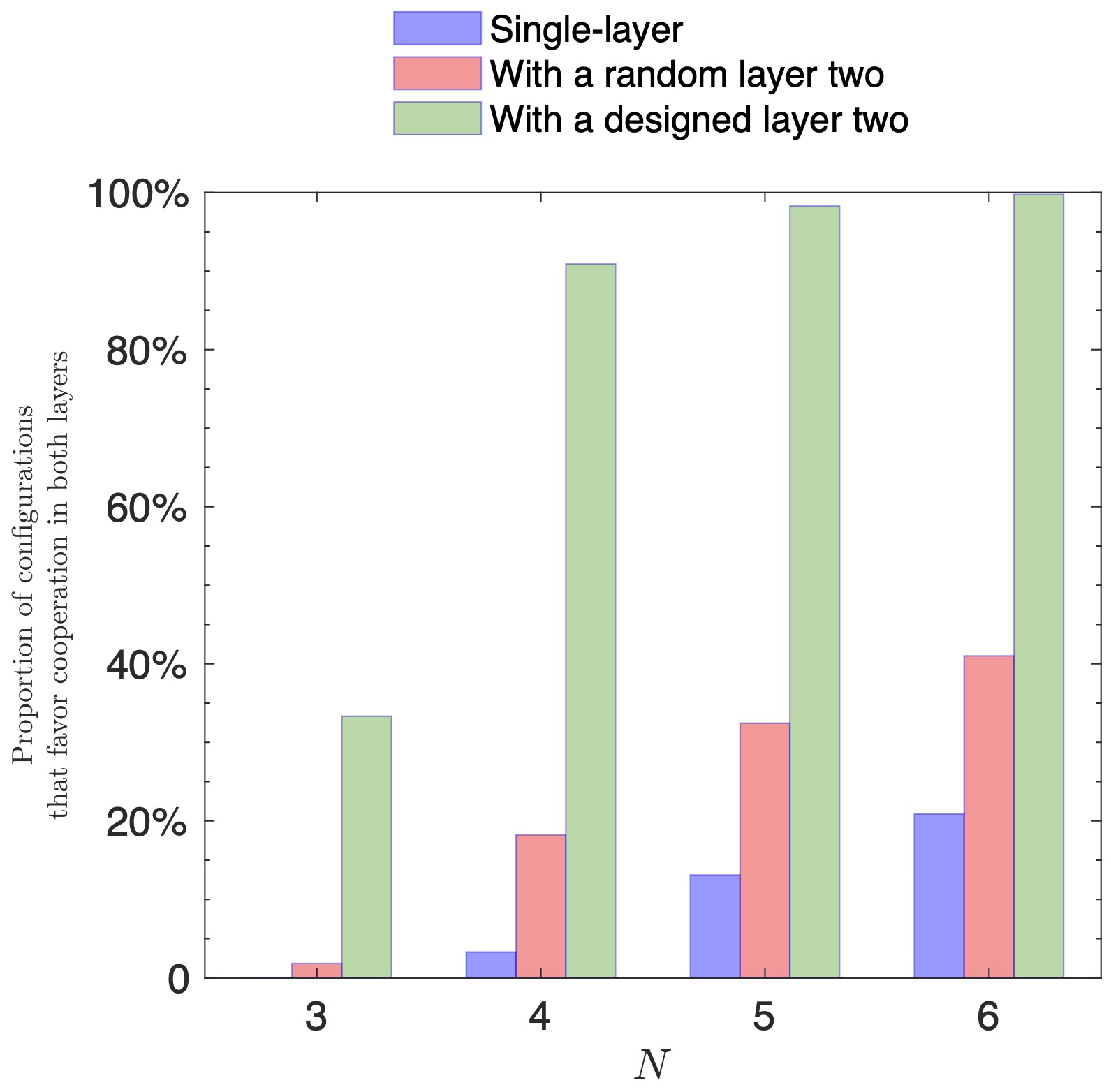}
	\caption{\label{fig:S3} \textbf{Proportion of networks that permit the evolution of cooperation in both layers}. 
		We systematically analyze all networks of $N=3$, 4, 5, or 6 individuals, including all mutant configurations of a single cooperator in each layer. 
		Blue bars indicate the proportion of two random networks and mutant configurations, in which selection can favor cooperation in both of them simultaneously for some $b_1/c>0$ and $b_2/c>0$ when each of them evolve separately. 
		Coupling the two networks increases frequency of cases in which selection can favor cooperation in both of them simultaneously (i.e.~selection favors cooperation in both layers for some choice of $b_1/c>0$ and $b_2/c>0$, red).  
		Given a network in layer one, coupling to a deliberately chosen network in layer two further increases the chance that selection can favor cooperation in both layers (green).
	}
\end{figure}

\newpage

\begin{figure}
	\centering
	\includegraphics[width=0.7\textwidth]{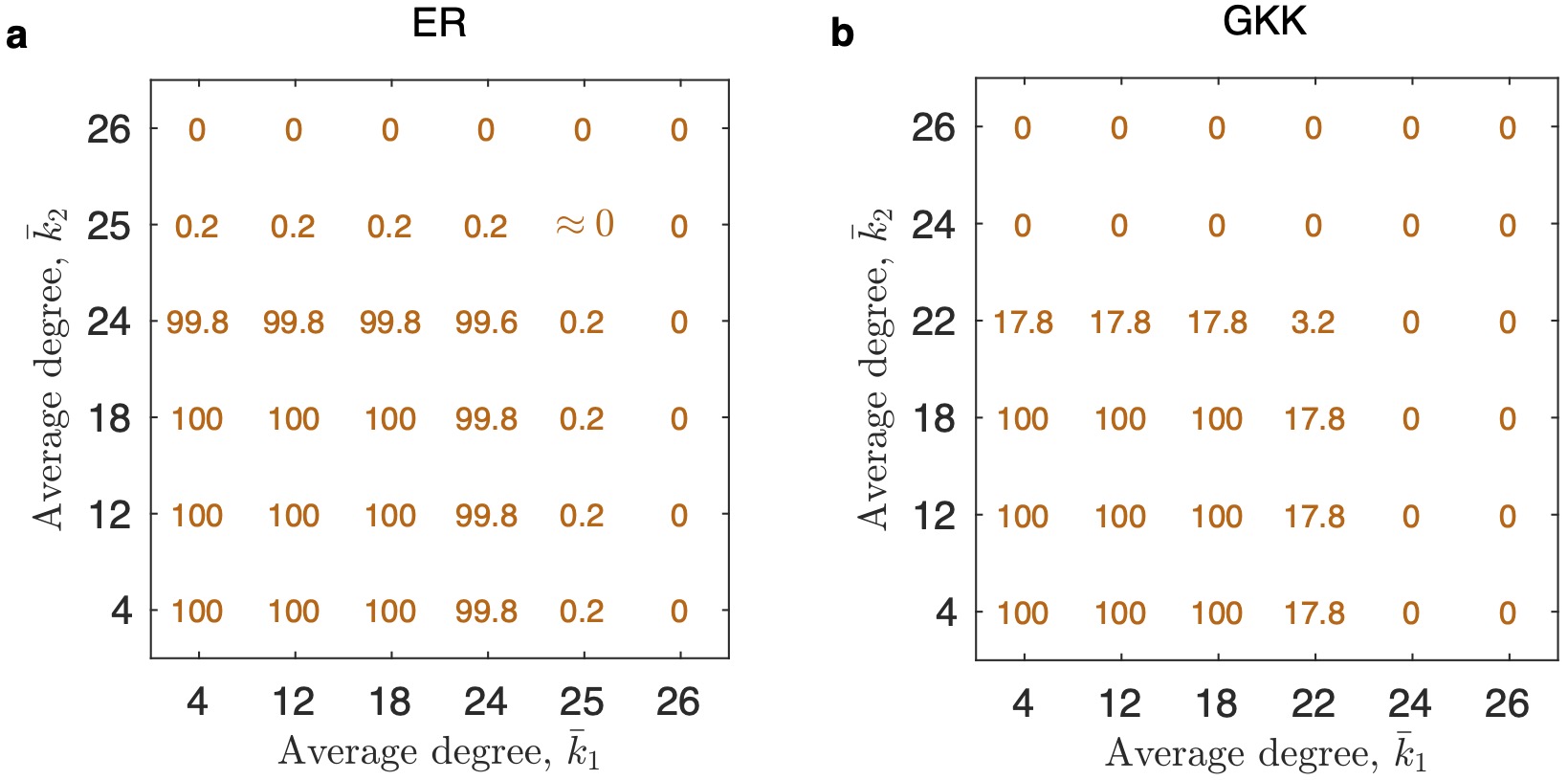}
	\caption{\label{fig:S4} \textbf{Proportion of random networks making cooperation favored possible when layers evolve independently}. 
		We sampled 100 two-layer ER networks of size $N=50$, and 100 two-layer GKK networks of size $N=50$, for each pair of average node degrees, $\bar{k}_1$ and $\bar{k}_2$,  in layer one and layer two, as indicated.
		For each two-layer network we analyzed all 2,500 initial configurations of a single mutant cooperator in each layer. 
		Two layers evolve separately. 
		\textbf{a}, Proportion (percentage) of sampled two-layer ER networks and configurations in which selection can favor cooperation in both layers, for some positive values of $b_1/c$ and $b_2/c$. 
		\textbf{b}, Proportion (percentage) of sampled two-layer GKK networks and configurations in which selection can favor cooperation in both layers, for some positive values of $b_1/c$ and $b_2/c$.
	}
\end{figure}

\newpage

\begin{figure}
	\centering
	\includegraphics[width=0.7\textwidth]{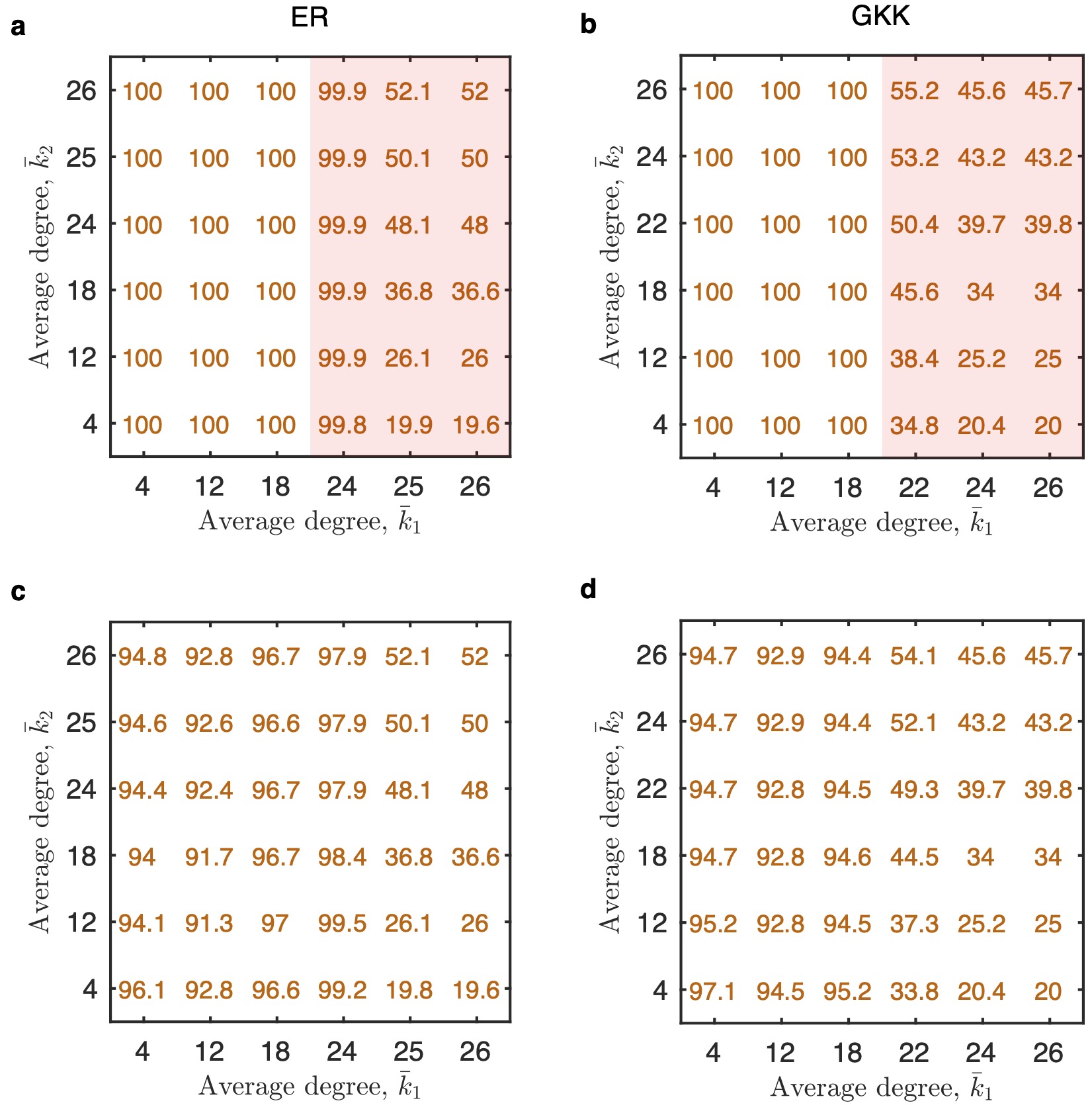}
	\caption{\label{fig:S5} \textbf{Multilayer coupling increases the proportion of selection favoring cooperation in layer one or reduces the threshold required in layer one}. 100 two-layer ER networks of size $N=50$, and 100 two-layer GKK networks of size $N=50$, for each pair of average node degrees, $\bar{k}_1$ and $\bar{k}_2$,  in layer one and layer two, as indicated.
		For each two-layer network we analyzed all 2,500 initial configurations of a single mutant cooperator in each layer.
		\textbf{a}, Proportion (percentage) of the sampled two-layer ER networks and configurations in which selection can favor cooperation in layer one, for some positive values of $b_1/c$ and $b_2/c$. 
		Highlighted entries indicate regimes when coupling increases the frequency of selection for cooperation in layer one compared to independent evolution in each layer.
		\textbf{b}, Proportion (percentage) of the sampled two-layer GKK networks and configurations in which selection can favor cooperation in layer one, for some positive values of $b_1/c$ and $b_2/c$.
		\textbf{c}, Proportion (percentage) of the sampled two-layer ER networks and configurations in which coupling promotes selection favoring cooperation in layer one: [1] for $(b_1/c)^*>0$, $b_1/c$ required by selection favoring cooperation in layer one is reduced below $(b_1/c)^*$; 
		[2] for $(b_1/c)^*<0$, there exist positive values of $(b_1/c)$ and $(b_1/c)$ making cooperation favored in layer one.
		\textbf{d}, Proportion (percentage) of the sampled two-layer GKK networks and configurations in which coupling promotes selection favoring cooperation in layer one.}
\end{figure}

\newpage

\begin{figure}
	\centering
	\includegraphics[width=0.7\textwidth]{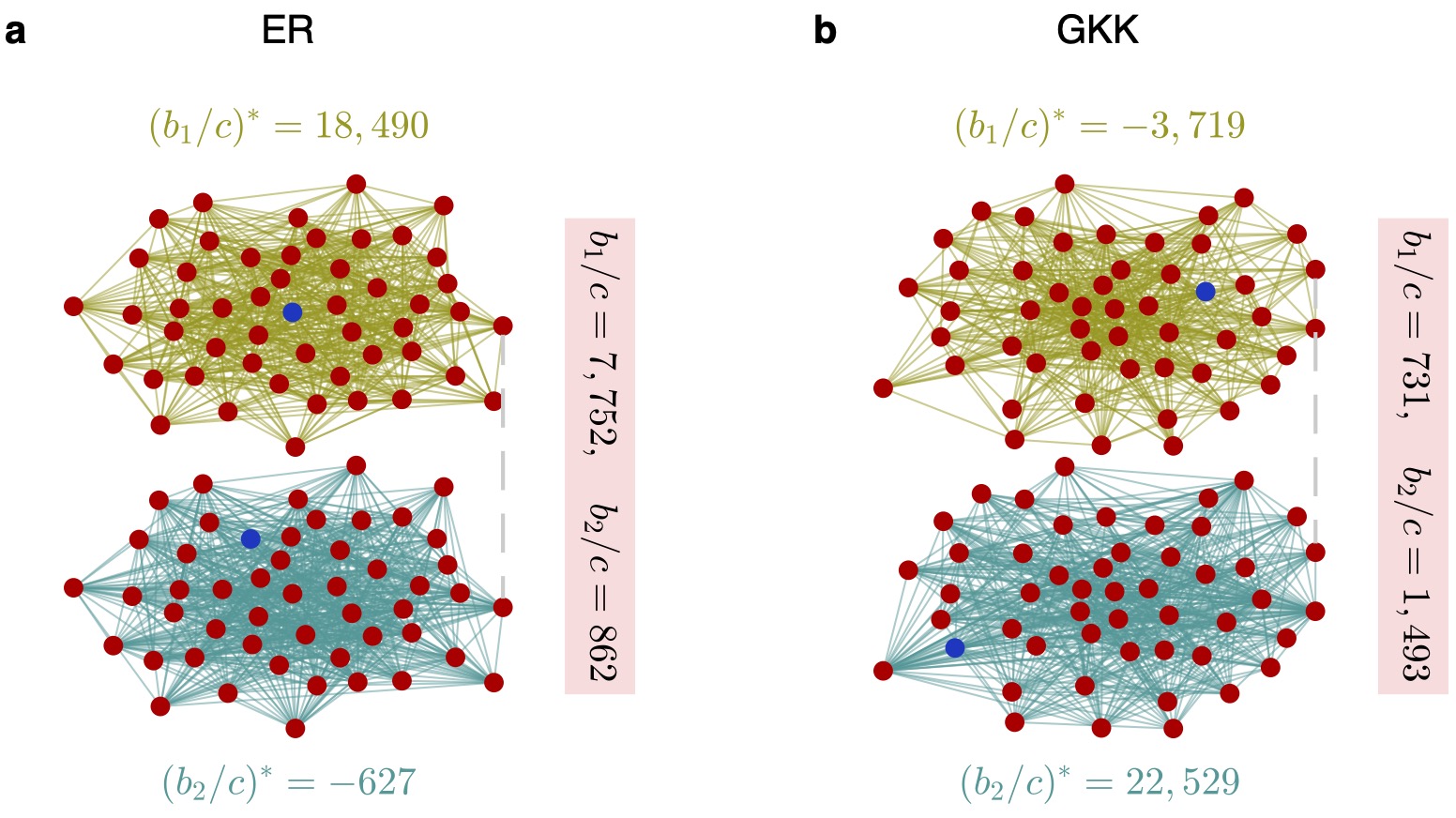}
	\caption{\label{fig:S6} \textbf{Examples of coupling two layers reducing the threshold in one layer and making cooperation favored possible in the other layer}.
		Let $(b_1/c)^*$ and $(b_2/c)^*$ respectively denote the thresholds in layer one and two when layers evolve separately.
		In \textbf{a},  $(b_1/c)^*>0$ and $(b_2/c)^*<0$.
		Coupling two layers reduces the value of $b_1/c$ required to favor cooperation in layer one below $(b_1/c)^*$, and meanwhile makes cooperation favored for some positive $b_1/c$ and $b_2/c$.
		Panel \textbf{b} shows an example of two-layer GKK networks.
	}
\end{figure}

\newpage

\begin{figure}
	\centering
	\includegraphics[width=0.7\textwidth]{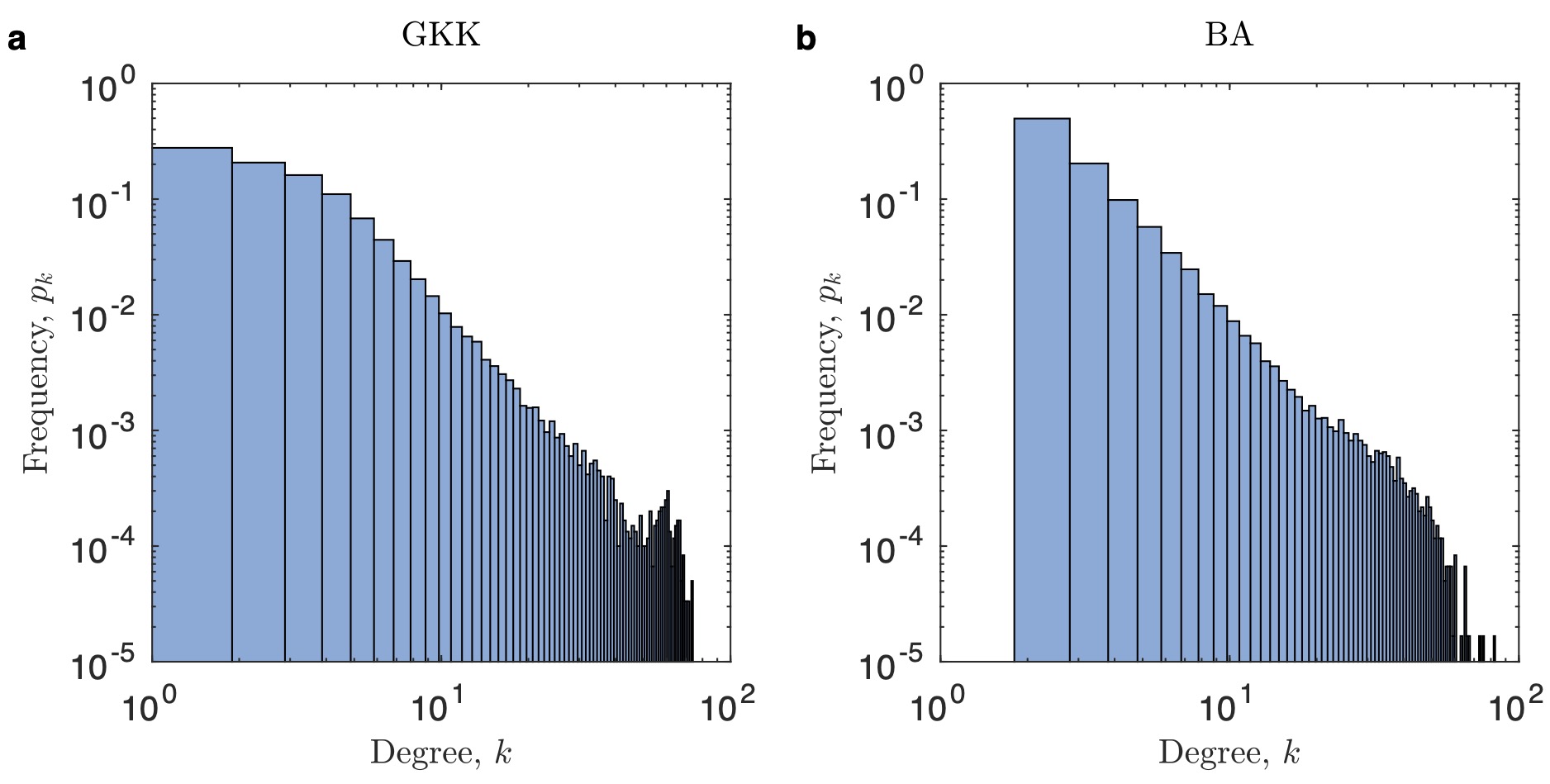}
	\caption{\label{fig:S7} \textbf{Node degree distribution of approximate scale-free networks with size  $N=300$ and average degree $\bar{k}=4$}.
		\textbf{a}, Networks generated by Goh-Kahng-Kim (GKK) algorithm with exponent $\gamma=2.5$.
		\textbf{b}, Networks generated by Barab\'{a}si-Albert (BA) algorithm.
		$p_k$ represents the proportion of nodes with degree $k$.}
\end{figure}

\newpage

\begin{figure}
	\centering
	\includegraphics[width=0.7\textwidth]{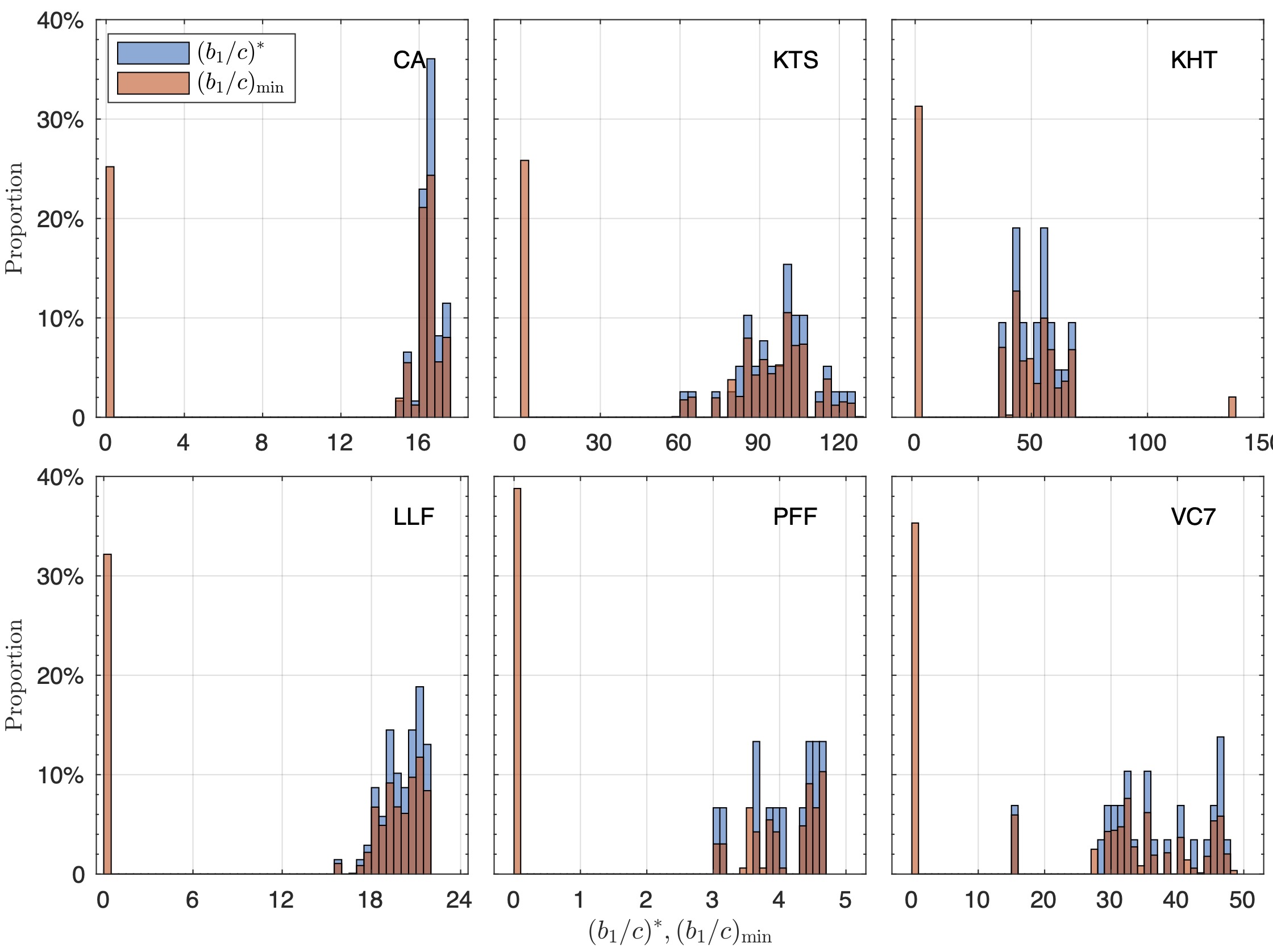}
	\caption{\label{fig:S8} \textbf{Distribution of minimal benefit-to-cost ratios required for cooperation to be favored in six real-world two-layer networks.}
		We analyze all initial configurations with a mutant cooperator in each layer.
		Let $(b_1/c)^*$ denote the critical benefit-to-cost ratio required for cooperation to be favored in layer one when layers evolve separately.
		The blue histogram shows the distribution of $(b_1/c)^*$ for all initial configurations.
		When layers are coupled, let $(b_1/c)_\text{min}$ denote the minimum ratio required to favor cooperation in layer one, for all $b_2/c>0$ in layer two.
		The red histogram presents the distribution of $(b_1/c)_\text{min}$ for all initial configurations.
		Coupling layers tends to permit cooperation to fix selectively for smaller benefit-to-cost ratio in layer one, including many cases where cooperation is favored despite providing no immediate benefit in layer one at all, i.e. $(b_1/c)_\text{min}=0$.
	}
\end{figure}

\newpage

\begin{figure}
	\centering
	\includegraphics[width=0.7\textwidth]{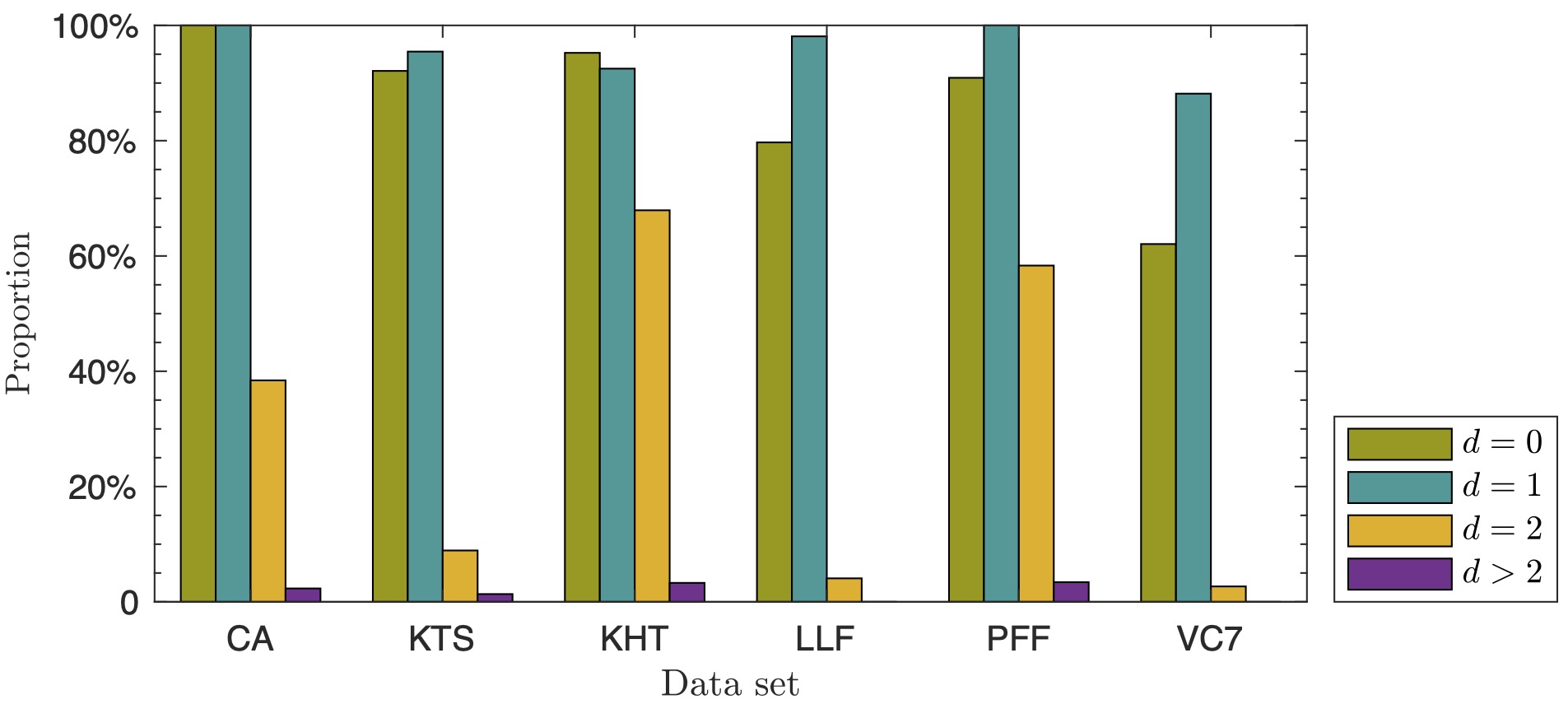}
	\caption{\label{fig:S9} \textbf{Mutants' positions in both layers decides if coupling layers can reduce $(b_1/c)_\text{min}$ to $0$}.
		When layers are coupled, let $(b_1/c)_\text{min}$ denote the minimum ratio required to favor cooperation in layer one, for all $b_2/c>0$ in layer two.
		Assuming the mutant lies in node $i$ in layer one and node $j$ in layer two, mutants' distance $d$ is the distance between $i$'s associated node and $j$ in layer two.
		For example, if nodes $i$ and $j$ refer to the same player, mutants' distance is $d=0$.
		If node $j$ is connected to $i$'s associated node in layer two, mutants' distance is $d=1$.
		$d=2$ means that node $j$ is next nearest to $i$'s associated node and $d>2$ means that $j$ is at least three-step away from $i$'s associated node in layer two.
		We analyze all initial configurations with a single mutant cooperator in each layer.
		For all initial configurations with distance $d$, we report the proportion of the initial configuration giving $(b_1/c)_\text{min}=0$.
		The closer mutants are in layer two, the more likely coupling layers can reduce the benefit-to-cost ratio required for cooperation to be favored in layer one to zero.
	}
\end{figure}

\newpage

\begin{figure}
	\centering
	\includegraphics[width=0.7\textwidth]{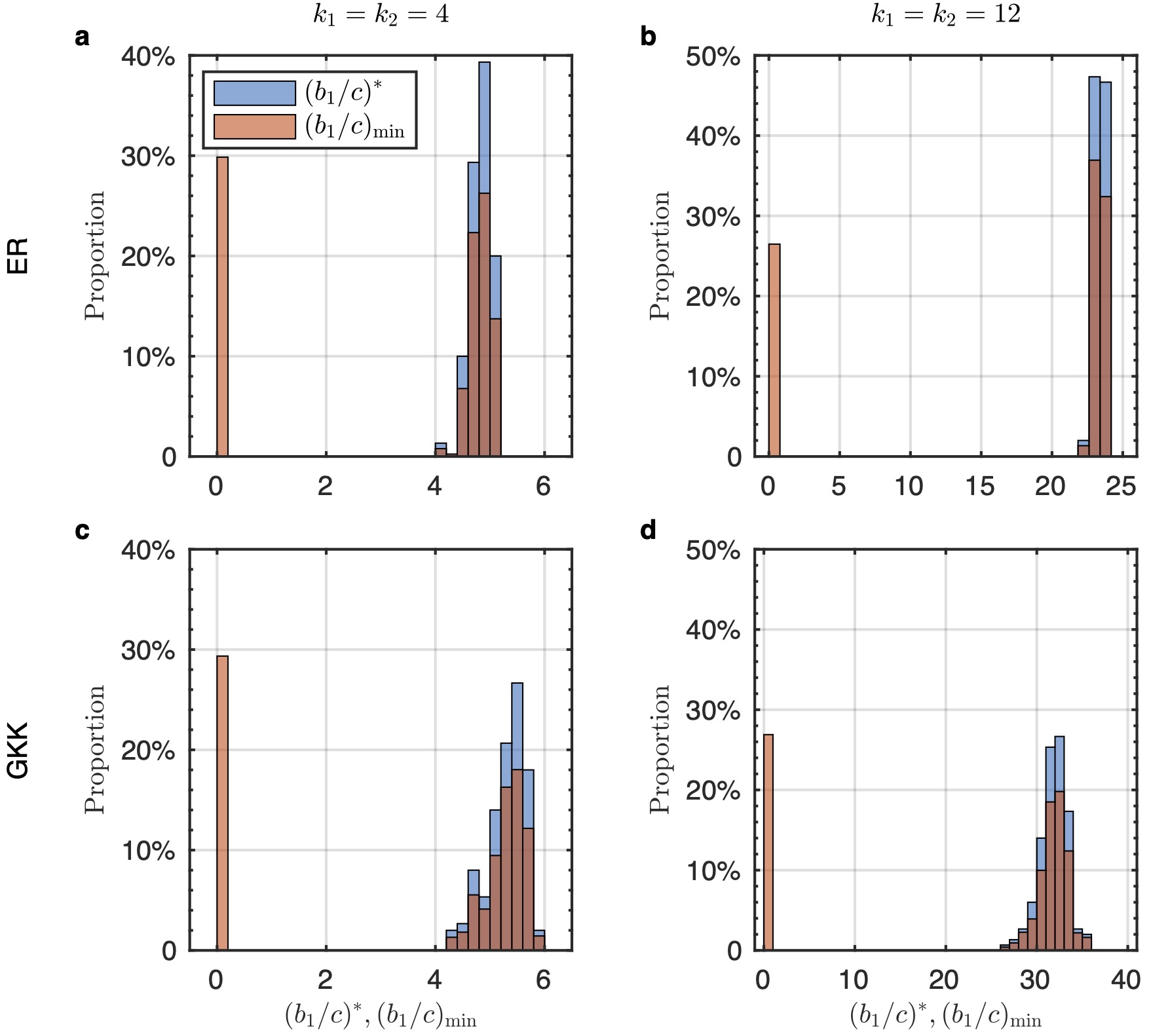}
	\caption{\label{fig:S10} \textbf{Distribution of minimal benefit-to-cost ratios required for cooperation to be favored in two-layer random networks.}
		We investigate 100 two-layer ER networks with average degree $k_1=k_2=4$ (\textbf{a}) and $k_1=k_2=12$ (\textbf{b}), and 100 two-layer GKK networks with average degree $k_1=k_2=4$ (\textbf{c}) and $k_1=k_2=12$ (\textbf{d}).
		The size for each network is $N=50$.
		In each two-layer network, we analyze all initial configurations with a single mutant cooperator in each layer, which means there are $50\times 50$ configurations.
		Let $(b_1/c)^*$ denote the critical benefit-to-cost ratio required for cooperation to be favored in layer one when layers evolve separately.
		The blue histogram shows the distribution of $(b_1/c)^*$ for all initial configurations.
		When layers are coupled, let $(b_1/c)_\text{min}$ denote the minimum ratio required to favor cooperation in layer one, for all $b_2/c>0$ in layer two.
		The red histogram presents the distribution of $(b_1/c)_\text{min}$ for all initial configurations.
		Coupling layers tends to permit cooperation to fix selectively for smaller benefit-to-cost ratio in layer one, including many cases where cooperation is favored despite providing no immediate benefit in layer one at all, i.e. $(b_1/c)_\text{min}=0$.
	}
\end{figure}

\newpage

\begin{figure}
	\centering
	\includegraphics[width=0.7\textwidth]{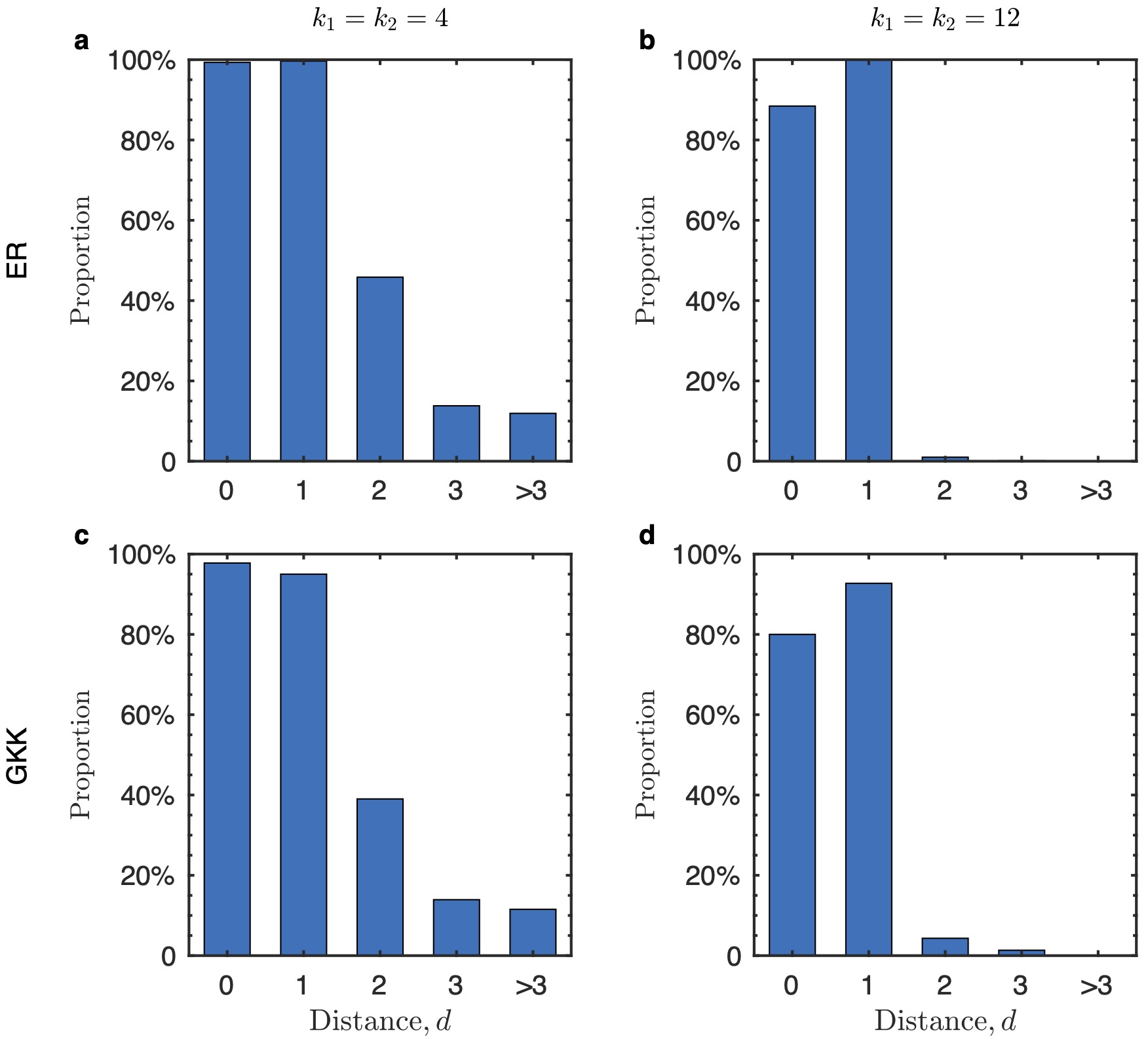}
	\caption{\label{fig:S11} \textbf{Mutants' positions in both layers decides if coupling layers can reduce $(b_1/c)_\text{min}$ to $0$}.
		We investigate ER networks with average degree $k_1=k_2=4$ (\textbf{a}) and $k_1=k_2=12$ (\textbf{b}), and GKK networks with average degree $k_1=k_2=4$ (\textbf{c}) and $k_1=k_2=12$ (\textbf{d}).
		Analogous to our analysis in Supplementary Fig.~\ref{fig:S8}, the closer mutants are in layer two, the more likely coupling layers can reduce the benefit-to-cost ratio required for cooperation to be favored in layer one to zero.
	}
\end{figure}

\newpage

\begin{figure}
	\centering
	\includegraphics[width=0.7\textwidth]{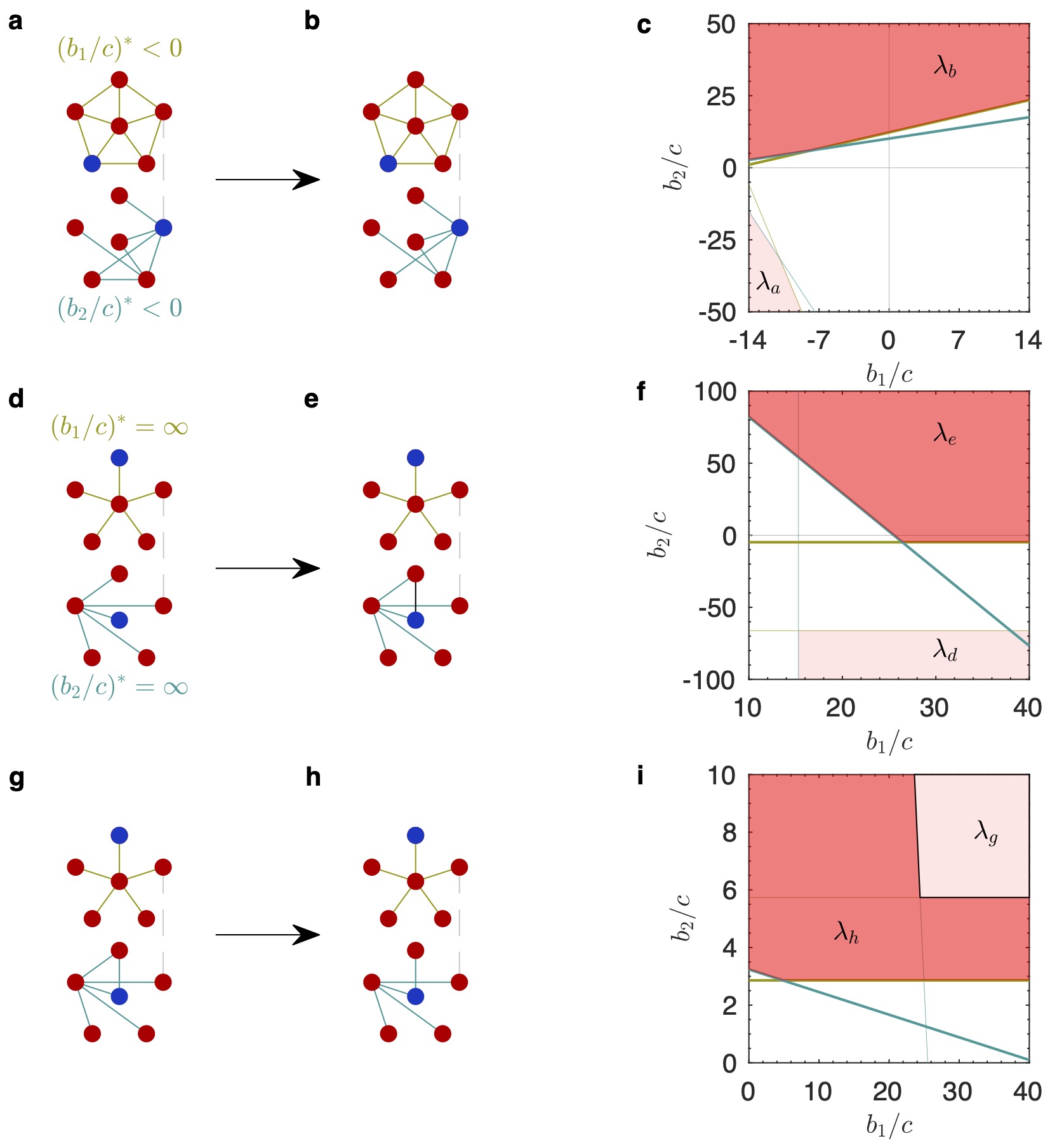}
	\caption{\label{fig:S12} \textbf{A slight modification of one layer can promote cooperation in both layers}.
		For a given initial strategy configuration, coupling layers is not always conducive to cooperation in both layers.
		In such cases, a slight modification of one layer, if managed properly, can make cooperation favored in both layers.
		In panel \textbf{a}, we show a two-player population that each individual layer favors spite, i.e. $(b_1/c)^*<0$ and $(b_2/c)^*<0$.
		Coupling the two layers does not enable the evolution of cooperation in any layer under any positive values of $b_1/c$ and $b_2/c$ (see region $\lambda_{\mathrm{A}}$ in panel \textbf{c}).
		Severing an edge in layer two, as shown in panel \textbf{b}, makes cooperation evolve in both layers possible (see region $\lambda_{\mathrm{B}}$ in panel \textbf{c}).
		In panel \textbf{d}, we present a two-layer population that each individual layer disfavors cooperation replacing defection for any benefit and cost, i.e. $(b_1/c)^*=\infty$ and $(b_2/c)^*=\infty$.
		Coupling the two layers still disfavor cooperation since it requires a negative value of $b_2/c$ (see region $\lambda_{\mathrm{D}}$ in panel \textbf{f}).
		Adding an edge in layer two, as shown in panel \textbf{e}, makes cooperation favored in both layers (see region $\lambda_{\mathrm{E}}$ in panel \textbf{f}).
		In panels \textbf{g}-\textbf{i}, we present that a slight modification to layer two can further enhance the cooperation-promoting effects of multilayer games.
		For some positive values of $b_1/c$ and $b_2/c$, cooperation can evolve in both layers in the two-layer population as presented in \textbf{g} (see region $\lambda_{\mathrm{G}}$).
		Severing an edge in layer two expands the region to $\lambda_{\mathrm{H}}$.
		Here we summarize two intuitions. Let nodes $i$ and $j$ respectively denote the cooperator mutants in layer one and two. 
		(1) connecting $i$'s ($j's$) associated node to $j$ ($i$), as shown in (\textbf{d}-\textbf{f}). Such connections enable $i$'s ($j$'s) associated node to obtain benefits from cooperator $j$ ($i$), which accordingly strengthens cooperator $i$ ($j$). 
		(2) if $i$'s associated node and $j$ have the common neighbor $k$ in layer two, severing the connection between $i$'s associated node and $k$, as shown in (\textbf{a}-\textbf{c}) and (\textbf{g}-\textbf{i}). Such a modification aims to weaken defector $k$'s advantages in exploiting cooperator $i$, $j$, and their associated nodes in the early stage.
	}
\end{figure}

\newpage

\begin{figure}
	\centering
	\includegraphics[width=0.7\textwidth]{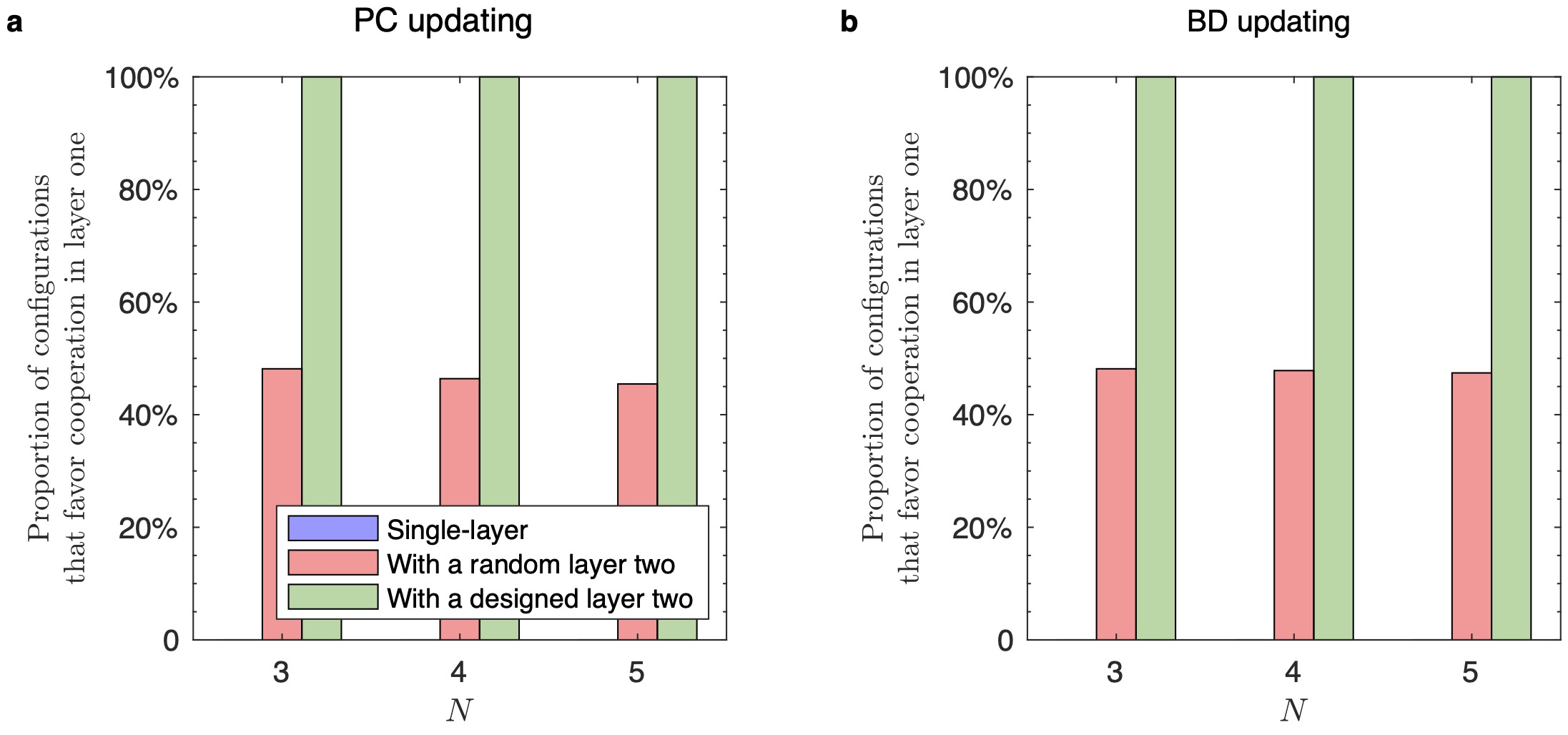}
	\caption{\label{fig:S13} \textbf{Proportion of networks and mutant configurations making  cooperation favored possible under pairwise-comparison updating (PC) and birth-death (BD) updating.} 
		We systematically analyzed all networks of $N=3$, 4, or 5, including all configurations of a single mutant cooperator in each layer.
		Blue bars indicate the proportion of single-layer networks and mutant configurations in which selection can favor cooperation for some benefit-to-cost ratio, i.e.\ $(b_1/c)^*>0$. 
		Note that under PC and BD updating, cooperation is never favored in a single-layer population.
		Coupling layer one with a randomly chosen network and strategy configuration in layer two increases the frequency of selection for cooperation (i.e.~selection favors cooperation in layer one for some choice of $b_1/c>0$ and $b_2/c>0$, red).  Coupling layer one with a deliberately designed network and strategy configuration in layer two further increases the frequency of cooperation in layer one (green).
	}
\end{figure}

\newpage

\begin{figure}
	\centering
	\includegraphics[width=0.7\textwidth]{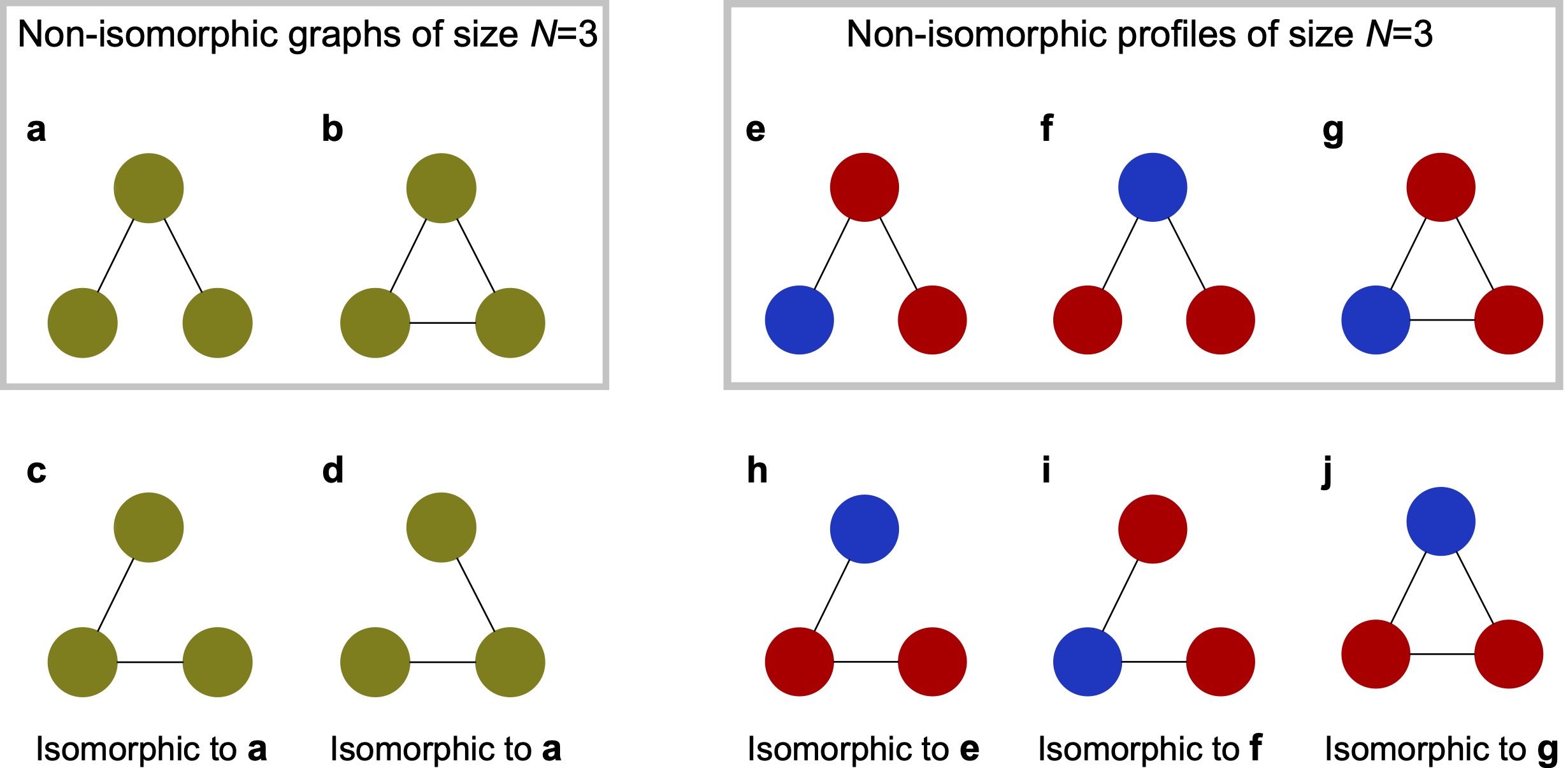}
	\caption{\label{fig:S14} \textbf{Non-isomorphic graphs (\textbf{a},\textbf{b}) and profiles (\textbf{e}-\textbf{g}) of a single-layer graph with $N=3$ nodes}.
		\textbf{a},\textbf{b}, There are two non-isomorphic connected graphs, without considering the configuration of strategies.
		\textbf{c},\textbf{d}, Examples of graphs isomorphic to \textbf{a}.
		\textbf{e}-\textbf{g}, There are three non-isomorphic profiles, which include consideration of the configuration of strategies.
		\textbf{h}-\textbf{j}, Examples of profiles isomorphic to \textbf{e}-\textbf{g} separately.}
\end{figure}

\newpage

\begin{figure}
	\centering
	\includegraphics[width=0.3\textwidth]{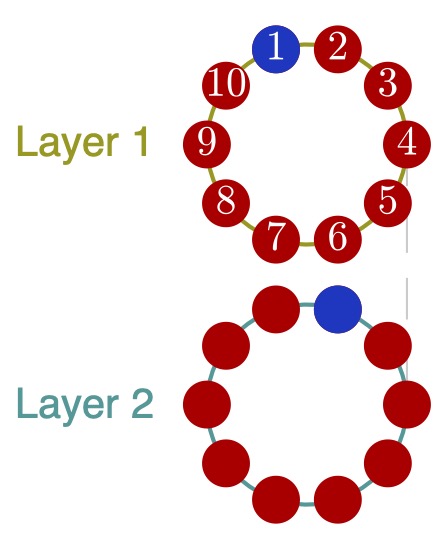}
	\caption{\label{fig:S15} \textbf{A two-layer ring network with a given strategy configuration}.
		The ring in each layer has 10 nodes and the two rings are symmetrical.
		Blue means $A$-strategy and red $B$-strategy.
		In the configuration illustrated, the distance between $A$-strategies in layer one and two is $d=1$.
	}
\end{figure}

\newpage

\begin{figure}
	\centering
	\includegraphics[width=0.3\textwidth]{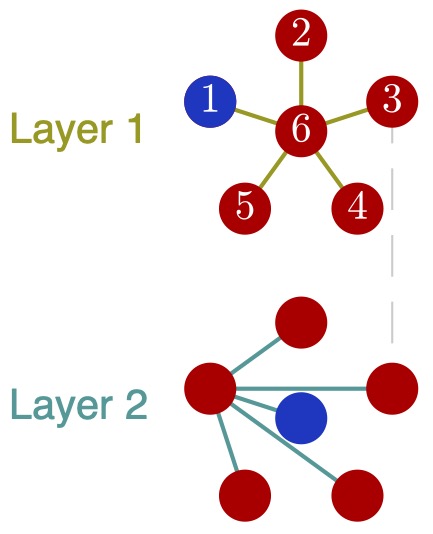}
	\caption{\label{fig:S16} \textbf{A two-layer star network with a given strategy configuration}.
		The star in each layer has $6$ nodes.
		Node $6$ and $1$ are separately the hub in layer one and two.
		Blue means $A$-strategy and red $B$-strategy.
	}
\end{figure}

\newpage

\begin{figure}
	\centering
	\includegraphics[width=0.7\textwidth]{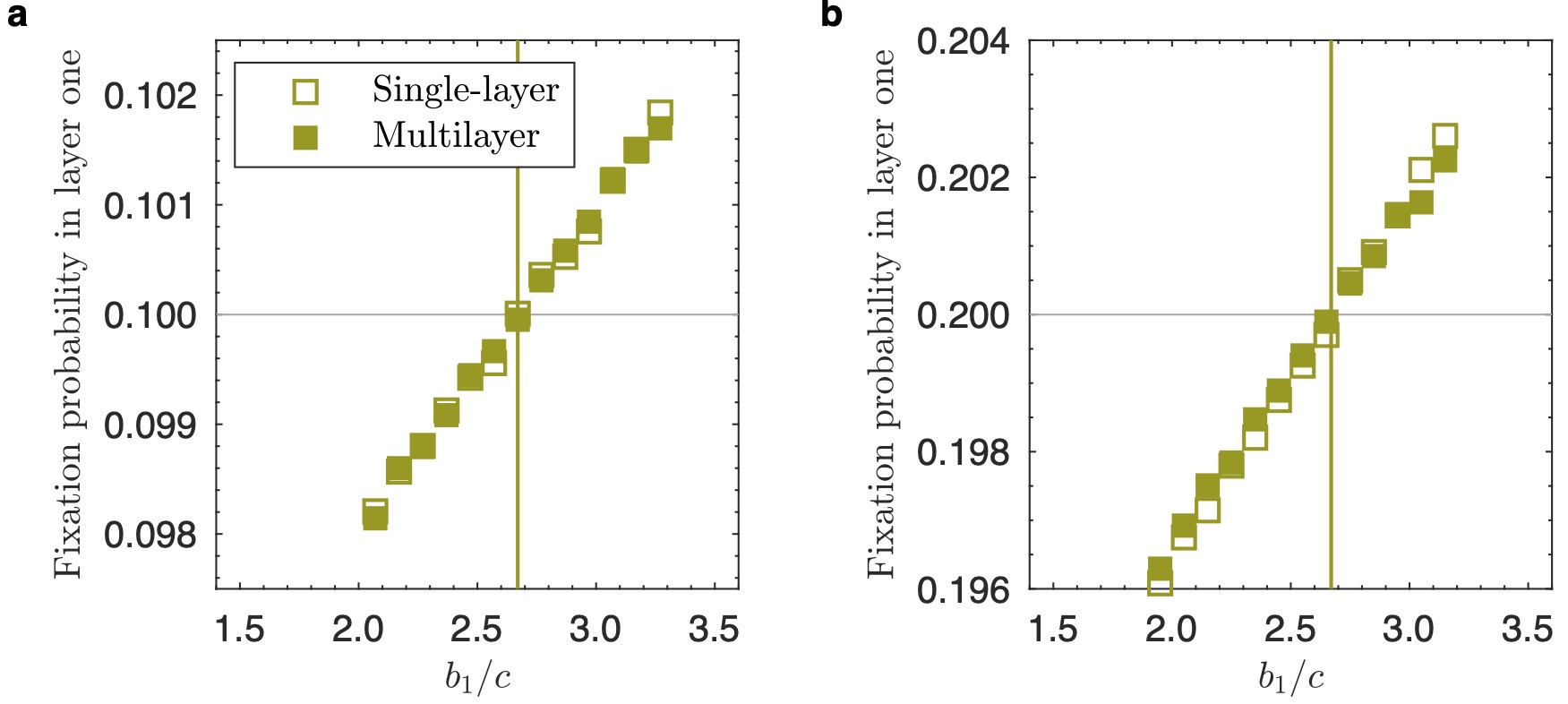}
	\caption{\label{fig:S17} \textbf{The independence property of coupling layers under a uniform distribution of mutants}.
		We investigate a two-layer circle with size $N=10$.
		\textbf{a},\textbf{b}, Fixation probability of a mutant cooperator.
		In Panel \textbf{a}, initially, a cooperator is randomly and uniformly distributed to a node in layer one, and a cooperator is designated to a fixed node in layer two (Fig.~3\textbf{a} in the main text is a specific case).
		In panel \textbf{b}, initially, two cooperators are randomly and uniformly distributed to two nodes in layer
		one, and three cooperators are designated to three fixed nodes in layer two (Supplementary Fig.~\ref{fig:S16}\textbf{a} is a specific case).
		When initial cooperators are uniformly distributed, in single-layer and multilayer games, the fixation probabilities of cooperators in layer one are identical.
		Introducing layer two does not affect the evolutionary dynamics in layer one at all.
		We take $b_2=10$ and $c=1$.
	}
\end{figure}

\end{document}